\documentclass[structabstract]{aa}
\usepackage{graphicx}
\usepackage{txfonts}
\usepackage{aas_macros}
\usepackage[sort]{natbib}
\usepackage{rotating}
\usepackage{subfig}
\usepackage{longtable}

\begin{document}

\title{Wide-field VLBA Observations of the Chandra Deep Field South}

\author{
Enno Middelberg\inst{1}
\and Adam Deller\inst{2}
\and John Morgan\inst{3,4,5}
\and Helge Rottmann\inst{5}
\and Walter Alef\inst{5}
\and Steven Tingay\inst{6}
\and Ray Norris\inst{7}
\and Uwe Bach\inst{5}
\and Walter Brisken\inst{3}
\and Emil Lenc\inst{7}
}

\institute{Astronomisches Institut, Ruhr-Universit\"at Bochum, Universit\"atsstr. 150, 44801 Bochum, Germany\\ \email{middelberg@astro.rub.de}\and
National Radio Astronomy Observatory, PO Box 0, Socorro, NM, 87801, USA                                                                       \and
Istituto di Radioastronomia -- INAF, Via Gobetti 101, I-40129 Bologna, Italy                                                                  \and
Dipartimento di Astronomia, Universit\`{a} degli Studi, via Ranzani 1, I–40127 Bologna, Italy                                                 \and
Max-Planck-Institut f\"ur Radioastronomie, Auf dem H\"ugel 69, 53121 Bonn, Germany                                                            \and
Curtin University of Technology, GPO BOX U1987, Perth, WA 6845, Australia                                                                     \and
Australia Telescope National Facility, PO Box 76, Epping NSW 1710, Australia                                                                  
}

\date{Received...}

\abstract {Wide-field surveys are a commonly-used method for studying
  thousands of objects simultaneously, to investigate, e.g., the joint
  evolution of star-forming galaxies and active galactic nuclei. Very
  long baseline interferometry (VLBI) observations can yield valuable
  input to such studies because they are able to identify AGN
  unambiguously in the moderate to high-redshift Universe. However,
  VLBI observations of large swaths of the sky are impractical using
  standard methods, because the fields of view of VLBI observations
  are of the order of $10''$ or less, and have therefore so far played
  only a minor role in galaxy evolution studies.}
{We have embarked on a project to carry out Very Long Baseline Array
  (VLBA) observations of all 96 known radio sources in one of the
  best-studied areas in the sky, the Chandra Deep Field South
  (CDFS). The challenge was to develop methods which could
  significantly reduce the amount of observing (and post-processing)
  time, making such a project feasible.}
{We have developed an extension to the DiFX software correlator which
  allows one to efficiently correlate up to hundreds of positions
  within the primary beams of the interferometer antennas. This
  extension enabled us to target many sources simultaneously, at full
  resolution and high sensitivity, using only a small amount of
  observing time. The combination of wide fields-of-view and high
  sensitivity across the field in this survey is unprecedented.}
{We have observed with the VLBA a single pointing containing the
  Chandra Deep Field South, in which 96 radio sources were known from
  previous observations with the Australia Telescope Compact Array
  (ATCA). From our input sample of 96 sources, 20 were detected with
  the VLBA, and one more source was tentatively detected. The majority
  of objects have flux densities in agreement with arcsec-scale
  observations, implying that their radio emission comes from very
  small regions. Two objects are visibly resolved. One VLBI-detected
  object had earlier been classified as a star-forming
  galaxy. Comparing the VLBI detections to sources found in sensitive,
  co-located X-ray observations we find that X-ray detections are not a
  good indicator for VLBI detections.}
{We have successfully demonstrated a new extension to the DiFX
  software correlator, allowing one to observe hundreds of fields of
  view simultaneously. In a sensitive observation of the CDFS we
  detect 21\,\% of the sources and were able to re-classify 7 sources
  as AGN which had not been identified as such before. Wide-field VLBI
  survey science is now coming of age.}


\maketitle

\section{Introduction}
\label{sec-1}

VLBI observations provide the highest resolution in observational
astronomy, but at the cost of tiny fields of view. Until recently,
only objects with brightness temperatures of the order of 10$^{6}$\,K
and greater could be observed, and since such objects are distributed
sparsely on the sky, the small field of view has not been much of a
limitation in the past. However, recent advances in technology allow
the use of much larger bandwidths in the observations, allowing much
fainter objects to be detected. Their density on the sky is so high
that many can be found in the antenna primary beams no matter where
one looks.

Unfortunately, the long baselines of VLBI observations traditionally
allow only objects within a few arcseconds of the phase centre to be
observed, before effects arising from time and frequency averaging
wash out the signal. The bottleneck has been correlator capacity (and
post-processing power). Software correlators, and in particular DiFX
(\citealt{Deller2007}), are much more flexible than traditional
correlators, giving observers the possibility to image objects
anywhere in the antenna primary beams. We have developed the necessary
software, expertise and techniques to image all known radio sources
within the primary beam of the interferometer elements using a single
VLBI observation, the results of which we present in this paper.

There are numerous scientific motivations for carrying out wide-field
VLBI surveys. One of the most pressing is to identify active galactic
nuclei (AGN). For example, \cite{Heisler1998} obtained VLBI detection
rates of 0\,\% for starbursts compared to more than 90\,\% for AGNs in
their sample of infrared-selected galaxies. However, as instrumental
sensitivities increase, VLBI observations have become more sensitive
to detection of radio supernovae in the relatively nearby Universe,
such as those in Arp\,220 \citep{Smith1998}, and so a VLBI detection
no longer unambiguously implies an AGN. For example, \cite{Kewley2000}
observed a sample of 61 luminous infrared galaxies and found compact
radio cores in 37\,\% of starburst galaxies and 80\,\% of
AGNs. However, the radio luminosity distribution of the Kewley et
al. compact cores showed a bimodal distribution. Most higher radio
luminosity cores ($>2\times 10^{21}$\,W\,Hz$^{-1}$) are AGNs, while
most lower radio luminosity cores ($<2\times 10^{21}$\,W\,Hz$^{-1}$)
are starbursts. We conclude that high luminosity radio cores ($\gg
2\times 10^{21}$\,W\,Hz$^{-1}$) VLBI detections are almost certainly
AGN, whereas lower luminosity cores may be caused either by AGN or by
supernova activity (also \citealt{Smith1998b}). For our detection
limit of $\sim 0.5$\,mJy, we can be confident that VLBI detections above
$z>0.1$ represent AGNs. Below this redshift, a VLBI detection is an
ambiguous indicator of AGN activity. Furthermore, the amount of radio
emission from an AGN does not correlate well with emission at other
wavelengths; in particular there appears to be no correlation between
a galaxy's nuclear radio and its large-scale FIR emission
(\citealt{Corbett2002}).

It is now recognised that AGN play an important role in star formation
and galaxy evolution. They are not only found in powerful quasars and
radio galaxies, but also in the local universe, with radio
luminosities as small as $10^{20}\,{\rm W\,Hz^{-1}}$ (e.g.,
\citealt{Filho2006}). AGN feedback can either push back and heat the
gas, reducing the formation of stars in the galaxy, or compress the
gas clouds, triggering the same process. Particularly at high
redshift, both AGN and star formation processes are likely to be
important in a large fraction of galaxies (\citealt{Bower2006}), but
we can not yet measure the fraction of the luminosity (bolometric and
radio) generated by each process, nor how they are influenced by
feedback. Whilst observations at other wavelengths, in particular in
the optical and X-ray regime, can identify accretion onto black holes,
these data cannot reliably indicate whether or not an AGN also
produces radio jets which interact with their surroundings.

Furthermore, VLBI observations of a substantial number of galaxies
yield information about specific classes of object. For example, the
compact steep-spectrum (CSS) and gigahertz-peaked spectrum (GPS)
sources make up a significant fraction (10\,\% to 30\,\%,
\citealt{Odea1998}) of radio sources. CSS and GPS sources are very
small, yet strong radio sources with either steep spectra
($\alpha<-0.5$, $S\propto\nu^\alpha$) or spectra that peak between
500\,MHz and 10\,GHz. Since only bright members of this class have yet
been studied with VLBI observations, a wide-field VLBI survey can
yield data in a previously inaccessible flux density regime. Another
example is the infrared-faint radio sources (IFRS), a mysterious class
of object characterised by strong (up to 20\,mJy) 1.4\,GHz radio
emission but a flux density of less than $\sim 1\,\mu$Jy in the
near-IR at 3.6\,$\mu$m (\citealt{Norris2006a}). It is now established
that they host AGN (\citealt{Norris2010}, \citealt{Middelberg2010} and
references therein), but their relation to other classes of object is
unclear.

The connection between radio and X-ray emission has recently been
investigated by \cite{Dunn2010}, using a sample of nearby
($D<100$\,Mpc) X-ray bright elliptical galaxies. They find that almost
all sources are also radio emitters and most display a likely
interaction between the X-ray and radio emitting plasmas. On the other
hand, they find no correlation between radio and X-ray luminosity. For
a study of AGN which are not recognized as such in optical
observations (``elusive AGN''), \cite{Maiolino2003} used a
VLBI-selected sample of 18 galaxies to maximise the probability of
X-ray detections. They find that these objects are heavily obscured
(column densities exceeding $10^{24}\,{\rm cm^{-2}}$) and that their
space density is comparable to, or exceeding that of, optically
classified Seyferts. We note however, that both works are difficult to
compare to what we present here. \cite{Dunn2010} used VLA observations
of nearby elliptical galaxies, and even though the sample by
\cite{Maiolino2003} was VLBI-selected, the parent samples were chosen
for high IR luminosity and detectability with VLBI. In contrast, our
sample is only flux density-limited, and we make no further selection.

Throughout the paper, we adopt a flat $\Lambda$CDM cosmology with
$H_0=71.0$, $\Omega_{\rm M}=0.27$, $\Omega_{\rm vac}=0.73$.

\section{The target sources}
\label{sec-2}

\begin{figure}[htpb]
\includegraphics[width=\linewidth]{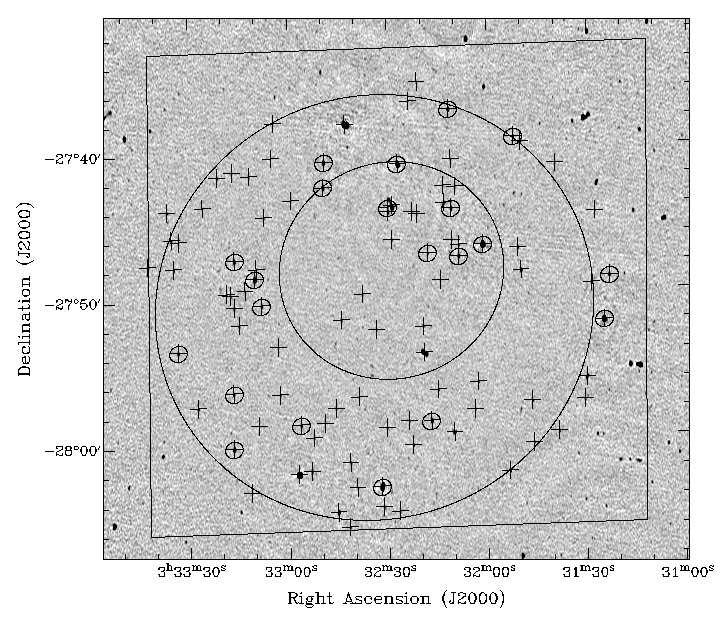}
\caption{An overview of the observed area. The background image is a
  radio image of the CDFS made with the Australia Telescope Compact
  Array (\citealt{Norris2006a}). The rms of the image is around
  20\,$\mu$Jy\,beam$^{-1}$, and the faintest sources have flux
  densities of 100\,$\mu$Jy. The square indicates the ECDFS area
  observed with the Chandra satellite by \cite{Lehmer2005} with an
  integration time of 240\,ks; the large circle indicates a typical
  VLBA antenna's primary beam size at the half-power level at
  1.4\,GHz; and the medium circle is the region covered uniformly with
  a 2\,Ms exposure with Chandra by \cite{Luo2008} (centred on the
  average aim point with a radius of 7.5', see their Figure~2).
  Crosses are drawn at the locations of the 96 targets taken from
  \cite{Norris2006a} and small circles those targets which were
  detected with the VLBA.}
\label{fig:overview}
\end{figure}

Executing a pilot widefield VLBI observation such as the one described
below required the development of new processing techniques, their
implementation in software and the purchase of necessary computing
infrastructure.  A perfect test-bed for these techniques was the
Chandra Deep Field South (CDFS). It has rich complementary coverage at
many wavelengths, particularly in the radio regime. However, even with
novel processing techniques it was not possible to image the entire
area contained within the antennas' primary beams, nor would this be a
promising exercise, since any flux detected in the VLBI observations
must have been detected in existing compact-array interferometry data
because of their sensitivity to lower-brightness temperature
sources\footnote{This reasoning ignores variability and transient
  events, see \cite{Lenc2008} for an example}.  We therefore used the
ATCA observations published by \cite{Norris2006a} as an input
catalogue, and aimed at imaging all 96 targets from that catalogue in
a region centred on the GOODS/CDFS and contained within the VLBA
primary beam area. An overview of the observed field is given in
Figure~\ref{fig:overview}, and the details of the target sources are
listed in Table~\ref{tab:input}.

\begin{table*}[htbp]
\centering
\caption{The 96 sources in our sample taken from
  \cite{Norris2006a}. Columns are the ID we use in this paper, the IAU
  designation, right ascension and declination, and ATCA flux density
  in mJy.}
\begin{tabular}{|l|l|l|l|r|}
\hline
\hline
ID   & IAU designation             & RA (J2000)        & Dec (J2000)  & $S_{\rm 1.4\,GHz}$ / mJy\\
\hline
S329 &   ATCDFS\_J033123.30$-$274905.6 &  03:31:23.305 &   -27:49:05.630 &  1.1  \\
S331 &   ATCDFS\_J033124.89$-$275208.3 &  03:31:24.892 &   -27:52:08.320 &  35.5 \\
S339 &   ATCDFS\_J033127.51$-$274440.1 &  03:31:27.519 &   -27:44:40.130 &  0.1  \\
S340 &   ATCDFS\_J033128.59$-$274934.9 &  03:31:28.594 &   -27:49:34.930 &  0.5  \\
S343 &   ATCDFS\_J033130.07$-$275602.5 &  03:31:30.077 &   -27:56:02.520 &  0.9  \\
S347 &   ATCDFS\_J033130.72$-$275734.0 &  03:31:30.720 &   -27:57:34.050 &  0.3  \\
S359 &   ATCDFS\_J033138.47$-$275942.0 &  03:31:38.473 &   -27:59:42.060 &  0.4  \\
S360 &   ATCDFS\_J033139.57$-$274119.4 &  03:31:39.572 &   -27:41:19.400 &  0.2  \\
S367 &   ATCDFS\_J033146.08$-$280027.2 &  03:31:46.080 &   -28:00:27.260 &  0.2  \\
S368 &   ATCDFS\_J033146.57$-$275734.8 &  03:31:46.572 &   -27:57:34.840 &  0.3  \\
S374 &   ATCDFS\_J033149.87$-$274838.8 &  03:31:49.876 &   -27:48:38.800 &  2.1  \\
S376 &   ATCDFS\_J033150.07$-$273947.1 &  03:31:50.076 &   -27:39:47.190 &  0.6  \\
S377 &   ATCDFS\_J033150.78$-$274704.0 &  03:31:50.788 &   -27:47:04.030 &  0.5  \\
S380 &   ATCDFS\_J033152.12$-$273926.4 &  03:31:52.122 &   -27:39:26.450 &  0.8  \\
S382 &   ATCDFS\_J033153.41$-$280221.1 &  03:31:53.412 &   -28:02:21.130 &  0.5  \\
S393 &   ATCDFS\_J033201.44$-$274647.5 &  03:32:01.440 &   -27:46:47.590 &  49.1 \\
S395 &   ATCDFS\_J033202.83$-$275612.6 &  03:32:02.836 &   -27:56:12.610 &  0.2  \\
S396 &   ATCDFS\_J033203.84$-$275804.6 &  03:32:03.841 &   -27:58:04.610 &  0.2  \\
S403 &   ATCDFS\_J033208.54$-$274647.7 &  03:32:08.548 &   -27:46:47.750 &  0.2  \\
S404 &   ATCDFS\_J033208.67$-$274734.4 &  03:32:08.670 &   -27:47:34.430 &  1.7  \\
S405 &   ATCDFS\_J033209.75$-$274247.4 &  03:32:09.753 &   -27:42:47.420 &  0.2  \\
S407 &   ATCDFS\_J033210.13$-$275936.8 &  03:32:10.132 &   -27:59:36.830 &  0.9  \\
S410 &   ATCDFS\_J033210.79$-$274627.8 &  03:32:10.793 &   -27:46:27.850 &  0.2  \\
S411 &   ATCDFS\_J033210.91$-$274415.2 &  03:32:10.915 &   -27:44:15.210 &  2.7  \\
S412 &   ATCDFS\_J033211.00$-$274053.8 &  03:32:11.002 &   -27:40:53.860 &  0.3  \\
S414 &   ATCDFS\_J033211.64$-$273726.1 &  03:32:11.645 &   -27:37:26.160 &  3.6  \\
S415 &   ATCDFS\_J033213.07$-$274351.0 &  03:32:13.077 &   -27:43:51.070 &  1.2  \\
S416 &   ATCDFS\_J033213.27$-$274241.3 &  03:32:13.277 &   -27:42:41.300 &  0.1  \\
S417 &   ATCDFS\_J033214.17$-$274910.7 &  03:32:14.179 &   -27:49:10.780 &  0.1  \\
S418 &   ATCDFS\_J033214.89$-$275640.7 &  03:32:14.895 &   -27:56:40.790 &  0.2  \\
S421 &   ATCDFS\_J033217.05$-$275846.5 &  03:32:17.051 &   -27:58:46.570 &  2.6  \\
S423 &   ATCDFS\_J033218.01$-$274718.5 &  03:32:18.011 &   -27:47:18.570 &  0.4  \\
S425 &   ATCDFS\_J033219.17$-$275407.0 &  03:32:19.177 &   -27:54:07.040 &  8.8  \\
S427 &   ATCDFS\_J033219.46$-$275219.1 &  03:32:19.465 &   -27:52:19.120 &  0.3  \\
S429 &   ATCDFS\_J033221.07$-$273529.7 &  03:32:21.074 &   -27:35:29.720 &  0.2  \\
S430 &   ATCDFS\_J033221.22$-$274435.3 &  03:32:21.223 &   -27:44:35.310 &  0.2  \\
S432 &   ATCDFS\_J033222.59$-$280023.6 &  03:32:22.590 &   -28:00:23.690 &  0.4  \\
S433 &   ATCDFS\_J033222.62$-$274426.5 &  03:32:22.627 &   -27:44:26.550 &  0.1  \\
S434 &   ATCDFS\_J033223.70$-$273649.1 &  03:32:23.701 &   -27:36:49.180 &  0.1  \\
S435 &   ATCDFS\_J033223.79$-$275844.9 &  03:32:23.793 &   -27:58:44.960 &  0.2  \\
S436 &   ATCDFS\_J033226.82$-$280453.1 &  03:32:26.828 &   -28:04:53.190 &  0.2  \\
S437 &   ATCDFS\_J033226.97$-$274106.7 &  03:32:26.975 &   -27:41:06.710 &  16.6 \\
S439 &   ATCDFS\_J033228.76$-$274619.7 &  03:32:28.763 &   -27:46:19.750 &  0.3  \\
S440 &   ATCDFS\_J033228.79$-$274356.1 &  03:32:28.790 &   -27:43:56.100 &  3.8  \\
S442 &   ATCDFS\_J033229.84$-$274423.8 &  03:32:29.847 &   -27:44:23.880 &  0.9  \\
S443 &   ATCDFS\_J033229.97$-$274405.4 &  03:32:29.974 &   -27:44:05.480 &  0.4  \\
S444 &   ATCDFS\_J033230.52$-$275911.6 &  03:32:30.529 &   -27:59:11.680 &  0.3  \\
S446 &   ATCDFS\_J033231.54$-$280433.5 &  03:32:31.540 &   -28:04:33.530 &  0.3  \\
\hline
\end{tabular}
\label{tab:input}
\end{table*}

\addtocounter{table}{-1}

\begin{table*}[htbp]
\centering
\caption{(continued)}.
\begin{tabular}{|l|l|r|r|r|}
\hline
\hline
ID   & IAU designation             & RA           & Dec          & $S_{\rm 1.4\,GHz}$ / mJy\\
\hline
S447 &   ATCDFS\_J033232.04$-$280310.2 &  03:32:32.041 &  -28:03:10.290 &  23.9  \\
S450 &   ATCDFS\_J033233.48$-$275227.2 &  03:32:33.480 &  -27:52:27.240 &  0.1   \\
S453 &   ATCDFS\_J033237.73$-$275000.3 &  03:32:37.737 &  -27:50:00.330 &  0.1   \\
S457 &   ATCDFS\_J033238.95$-$275700.6 &  03:32:38.954 &  -27:57:00.660 &  0.1   \\
S458 &   ATCDFS\_J033239.57$-$280312.5 &  03:32:39.570 &  -28:03:12.570 &  0.2   \\
S459 &   ATCDFS\_J033241.60$-$280128.0 &  03:32:41.604 &  -28:01:28.060 &  0.4   \\
S461 &   ATCDFS\_J033241.80$-$280552.5 &  03:32:41.808 &  -28:05:52.520 &  0.3   \\
S462 &   ATCDFS\_J033242.83$-$273817.6 &  03:32:42.830 &  -27:38:17.660 &  72.3  \\
S463 &   ATCDFS\_J033244.16$-$275142.5 &  03:32:44.165 &  -27:51:42.580 &  0.5   \\
S465 &   ATCDFS\_J033245.36$-$280449.7 &  03:32:45.363 &  -28:04:49.730 &  2.9   \\
S469 &   ATCDFS\_J033245.95$-$275745.2 &  03:32:45.956 &  -27:57:45.290 &  0.2   \\
S472 &   ATCDFS\_J033249.19$-$274050.7 &  03:32:49.195 &  -27:40:50.710 &  4.7   \\
S473 &   ATCDFS\_J033249.33$-$275845.1 &  03:32:49.334 &  -27:58:45.100 &  0.2   \\
S474 &   ATCDFS\_J033249.43$-$274235.3 &  03:32:49.430 &  -27:42:35.350 &  3.2   \\
S479 &   ATCDFS\_J033252.48$-$275942.1 &  03:32:52.480 &  -27:59:42.180 &  0.2   \\
S481 &   ATCDFS\_J033253.34$-$280200.1 &  03:32:53.341 &  -28:02:00.150 &  0.9   \\
S482 &   ATCDFS\_J033256.47$-$275848.4 &  03:32:56.472 &  -27:58:48.460 &  1.2   \\
S484 &   ATCDFS\_J033257.17$-$280210.0 &  03:32:57.170 &  -28:02:10.030 &  23.1  \\
S485 &   ATCDFS\_J033259.22$-$274325.6 &  03:32:59.224 &  -27:43:25.640 &  0.2   \\
S488 &   ATCDFS\_J033302.69$-$275642.4 &  03:33:02.692 &  -27:56:42.400 &  0.2   \\
S489 &   ATCDFS\_J033303.29$-$275326.4 &  03:33:03.293 &  -27:53:26.470 &  0.1   \\
S491 &   ATCDFS\_J033304.38$-$273804.3 &  03:33:04.381 &  -27:38:04.310 &  0.1   \\
S492 &   ATCDFS\_J033305.15$-$274028.9 &  03:33:05.155 &  -27:40:28.900 &  0.1   \\
S497 &   ATCDFS\_J033307.24$-$274431.2 &  03:33:07.244 &  -27:44:31.260 &  0.3   \\
S500 &   ATCDFS\_J033308.16$-$275033.2 &  03:33:08.168 &  -27:50:33.260 &  1     \\
S501 &   ATCDFS\_J033309.13$-$275846.7 &  03:33:09.139 &  -27:58:46.700 &  0.1   \\
S502 &   ATCDFS\_J033309.73$-$274802.2 &  03:33:09.731 &  -27:48:02.270 &  0.3   \\
S503 &   ATCDFS\_J033310.19$-$274842.0 &  03:33:10.191 &  -27:48:42.060 &  19.3  \\
S506 &   ATCDFS\_J033311.48$-$280319.0 &  03:33:11.486 &  -28:03:19.090 &  0.2   \\
S507 &   ATCDFS\_J033311.79$-$274138.2 &  03:33:11.796 &  -27:41:38.250 &  0.2   \\
S509 &   ATCDFS\_J033313.12$-$274930.2 &  03:33:13.123 &  -27:49:30.250 &  0.3   \\
S514 &   ATCDFS\_J033314.98$-$275151.9 &  03:33:14.982 &  -27:51:51.950 &  0.5   \\
S517 &   ATCDFS\_J033316.34$-$274725.0 &  03:33:16.340 &  -27:47:25.080 &  2.4   \\
S518 &   ATCDFS\_J033316.45$-$275038.9 &  03:33:16.456 &  -27:50:38.900 &  0.4   \\
S519 &   ATCDFS\_J033316.72$-$275630.4 &  03:33:16.725 &  -27:56:30.400 &  0.9   \\
S520 &   ATCDFS\_J033316.75$-$280016.0 &  03:33:16.754 &  -28:00:16.020 &  3.7   \\
S521 &   ATCDFS\_J033316.95$-$274121.8 &  03:33:16.954 &  -27:41:21.840 &  0.3   \\
S522 &   ATCDFS\_J033317.43$-$274947.8 &  03:33:17.434 &  -27:49:47.890 &  0.3   \\
S523 &   ATCDFS\_J033318.66$-$274940.0 &  03:33:18.668 &  -27:49:40.090 &  0.2   \\
S528 &   ATCDFS\_J033321.32$-$274138.9 &  03:33:21.328 &  -27:41:38.910 &  0.3   \\
S533 &   ATCDFS\_J033325.82$-$274342.4 &  03:33:25.824 &  -27:43:42.460 &  0.2   \\
S535 &   ATCDFS\_J033327.55$-$275725.7 &  03:33:27.557 &  -27:57:25.760 &  0.6   \\
S544 &   ATCDFS\_J033333.03$-$274600.0 &  03:33:33.030 &  -27:46:00.050 &  0.1   \\
S545 &   ATCDFS\_J033333.41$-$275333.1 &  03:33:33.413 &  -27:53:33.110 &  0.9   \\
S547 &   ATCDFS\_J033334.55$-$274750.3 &  03:33:34.554 &  -27:47:50.380 &  0.5   \\
S548 &   ATCDFS\_J033335.26$-$274549.1 &  03:33:35.260 &  -27:45:49.180 &  0.2   \\
S550 &   ATCDFS\_J033336.45$-$274354.3 &  03:33:36.457 &  -27:43:54.350 &  0.2   \\
S565 &   ATCDFS\_J033342.36$-$274736.8 &  03:33:42.368 &  -27:47:36.880 &  0.4   \\
\hline
\end{tabular}
\end{table*}

\section{Observations and data analysis}
\label{sec-3}

\subsection{Observations}
\label{sec-3.1}

We observed on 3 July 2007 the GOODS/CDFS area with the VLBA at
1383\,MHz, centred on RA~03:32:34.0392, Dec~$-27$:50:50.748
(J2000). Two polarisations were recorded across a total bandwidth of
64\,MHz, using 2-bit sampling, resulting in a recording bit rate of
512\,Mbps. The fringe finders 3C\,454.3 and 4C\,39.25 were observed in
intervals of 2.5\,h, and the phase calibrator NVSS\,J034838$-$274914
was observed for 1\,min after each 5\,min scan on the target
field. The elapsed time of the observations was 9\,h, but the low
declination of the target field and the calibrator observations
reduced the total integration time on the target field to 178
baseline-hours. According to the VLBA observational status
summary\footnote{http://www.vlba.nrao.edu/astro/obstatus/current/obssum.html}
the baseline sensitivity of the VLBA at 1.4\,GHz is 3.3\,mJy in a
2\,min observation using 256\,Mbps recording, or 0.43\,mJy per
baseline-hour using 512\,Mbps. This sensitivity results in a thermal
noise limit of our images of 32\,$\mu$Jy\,beam$^{-1}$. Accounting for
the loss of about 40\,\% of the data in self-calibration on our
in-beam calibrator S503 (see Table~\ref{tab:input}), the estimate has
to be increased to about 41\,$\mu$Jy\,beam$^{-1}$. However, these
estimates are typical for observations near the zenith, but our
observations were carried out predominantly at low elevations (at an
average elevation of around 20$^\circ$), where the system temperatures
were significantly higher, resulting in a lower expected system
sensitivity. For example, the average $T_{\rm sys}$ of the Fort Davis
station when observing the target field was 27\,\% higher than
compared to a fringe finder observation at 75$^\circ$ elevation. At
Kitt Peak, the increase was 32\,\%, and at Los Alamos 42\,\%. The
final on-axis noise was found to be 55\,$\mu$Jy\,beam$^{-1}$, only
27\,\% higher than at an average declination of 75$^\circ$.

\subsection{Correlation}
\label{sec-3.2}

\subsubsection{Bandwidth and time averaging effects}
\label{sec-3.2.1}

Since the targets are scattered throughout the primary beam it was
necessary to develop new correlation strategies to overcome the
effects of bandwidth and time averaging. These effects are commonly
called bandwidth smearing and time smearing, since they smear out
emission from the target into the map and reduce the observed
amplitudes of sources away from the phase centre. Bandwidth smearing
is regarded as the aperture synthesis equivalent to chromatic
aberration, whereas time smearing can be regarded as similar to motion
blur in a photograph when the shutter time is too long. An estimate of
the magnitude of these effects can be found in \cite{Thompson2001}. In
our case, with sources having separations to the phase centre of up to
15\,arcmin, and long baselines of typically 5000\,km, amplitudes would
be reduced to a few percent of their true values if a normal
correlator setup with 500\,kHz channels and 2\,s integrations was
used.

To keep the amplitudes within 5\,\% of their true values, one would
have to use a channel width of 4\,kHz and 50\,ms integrations. Such a
correlation would result in around 3\,TB of visibility data, which
would be very difficult to manage even on large general-purpose
computers. Furthermore, current hardware correlators are not able to
produce data with such high resolution. Therefore the only option was
to use a software correlator, in particular the DiFX correlator
developed by \cite{Deller2007}. We used a development version of DiFX
which included a new multiple-field capability, and which was
installed on a computer cluster at the Max-Planck-Institut f\"ur
Radioastronomie in Bonn.

\subsubsection{An extension to the DiFX correlator}
\label{sec-3.2.2}

The DiFX software correlator is widely-used and verified
(\citealt{Deller2007}). It replaced the hardware correlator of the
Australian Long Baseline Array in 2008. Subsequent to our observations
DiFX has also been adopted as the full-time correlator for the VLBA,
and the multiple-field extension capability which is discussed below
has recently been incorporated into the VLBA implementation of DiFX
and made available to the VLBA community.

The implementation of ``simultaneous multiple field centre
correlation" is described in detail by \cite{Deller2010}, and is
briefly summarised here.  The correlation is initially performed with
high frequency resolution which is sufficient to minimise bandwidth
smearing. Periodically, but still frequently enough to minimise time
smearing, the phase centre of the correlated data is shifted from its
initial location (which is usually the pointing centre) to a target
source location. This shift requires rotating the visibility phases of
each baseline by an amount equal to the difference in the geometric
delay between the final and initial source directions, multiplied by
the sky frequency. In effect, this corrects for the ``unapplied"
differential fringe rotation between the final and initial source
directions. This phase shift is repeated for each desired source
direction. After the phase shift is applied, the visibilities are
averaged in frequency and continue to be averaged in time. Eventually,
this results in an array of ``normal--sized" visibility datasets, with
one dataset per target source. The field of view of each of these
datasets is of the order of $13''$, at which point bandwidth and time
smearing would reduce the observed amplitudes by 5\,\%.

In comparison to the normal correlation operation, which requires
operations on every baseband sample, at a time scale of $\sim$ns, these
shift/average operations need only be applied relatively infrequently
(10\,ms is sufficient to shift past the edge of the primary beam with
minimal time smearing), and thus adding many phase centres to the
correlation is relatively computationally cheap.  \cite{Deller2010}
show that hundreds of phase centres can be added within the primary
beam at a cost only $\sim$3 times that of a single traditional
correlator pass. Hence each individual phase centre requires far fewer
operations than a traditional ``multi-pass" correlation where the full
correlation operation is repeated for each phase centre.

\cite{Deller2010} also demonstrate that the accuracy of the data
obtained in this manner is on par with the accuracy of data obtained
in the traditional way. Using a small test data set observed at a
frequency of 8.4\,GHz, they find the amplitude and phase differences
of the two methods to be 0.09\,\% and 0.014\,$^\circ$ on average,
which is in agreement with effects expected to arise from numerical
errors.

\subsection{Calibration}
\label{sec-3.3}

\subsubsection{Standard steps}
\label{sec-3.3.1}

The calibration followed standard procedures used in phase-referenced
VLBI observations, using the Astronomical Image Processing System,
AIPS. Amplitude calibration was carried out using $T_{\rm sys}$
measurements and known gain curves. Fringe-fitting was carried out on
all calibrators to determine delays and phase corrections. The data of
the phase calibrator were then exported to Difmap
(\citealt{Shepherd1997}) for imaging. We have written an extension to
Difmap to allow export of the complex gains determined in Difmap to
AIPS for further
processing\footnote{http://www.atnf.csiro.au/people/Emil.Lenc/tools/Tools/
  Cordump.html}, and to apply the gains to the target data
sets. Images were then made of the brighter sources to look for a
source suitable for in-beam calibration.

One of the targets, S503, was found to be bright enough for
self-calibration. We applied the phase calibrator gains to the S503
data set and used two cycles of phase-only self-calibration to refine
the gains, when convergence was reached. Amplitude self-calibration
was found to converge on less than 30\,\% of the data and resulted in
large gain fluctuations, and was therefore not used on
S503. Furthermore, S503 was found to be resolved on long baselines,
resulting in the number of useful visibilities reduced by
$\approx40\,\%$ after self-calibration. The noise in images (before
primary beam correction) was found to be of the order of
55\,$\mu$Jy\,beam$^{-1}$. This sensitivity is in reasonable agreement
with our expectations, given that (i) the number of visibilities was
much reduced in self-calibration, and (ii) the system temperatures in
our data were significantly increased compared to observations near
the zenith, an effect arising from the average antenna elevations
being of the order of only 20\,$^\circ$ (see Section~\ref{sec-3.1}).

Since the phase response of a VLBA antenna is constant within the
primary beam, and since the geometric delays had been taken care of at
the correlation stage, the amplitude, phase, and delay correction
could simply be copied from one target source to another, and an image
could be formed. However, another correction is needed to compensate
for the amplitude response arising from primary beam attenuation,
which reduces the apparent flux density of a source by up to 50\,\%.

\subsubsection{Primary beam correction}
\label{sec-3.3.2}

Unlike in compact-array interferometry, where the primary beam
correction is carried out in the image plane, we have corrected for
primary beam attenuation by calculating visibility gains. This is
possible only because the fields of view are very small in our
observations and so the attenuation due to the primary beam does not
vary significantly across the image. In general the measured
amplitude, $A'$, of a visibility is related to the intrinsic
amplitude, $A$, via

\begin{equation}
A'=A\sqrt{g_1(t,\nu)g_2(t, \nu)},
\end{equation}

where $g_1(t, \nu)$ and $g_2(t, \nu)$ are the time and
frequency-dependent gains of the two antennas of the baseline in the
direction of the target. The gains consist of many contributions from
various instrumental and environmental effects, but the attenuation of
the signal at each antenna arising from primary beam attenuation,
$b_1(t,\nu)$ and $b_2(t,\nu)$, can be separated, yielding

\begin{equation}
A'=A\sqrt{b_1(t,\nu)g_1'(t,\nu)b_2(t,\nu)g_2'(t, \nu)}.
\end{equation}

The gains $g_1'(t,\nu)$ and $g_2'(t,\nu)$ are determined using $T_{\rm
  sys}$ measurements and antenna gain curves, but the primary beam
attenuation is ignored in standard VLBI calibration procedures because
one usually observes a tiny field of view in direction of the optical
axis where these terms can be ignored.

\begin{figure}
\includegraphics[width=\linewidth]{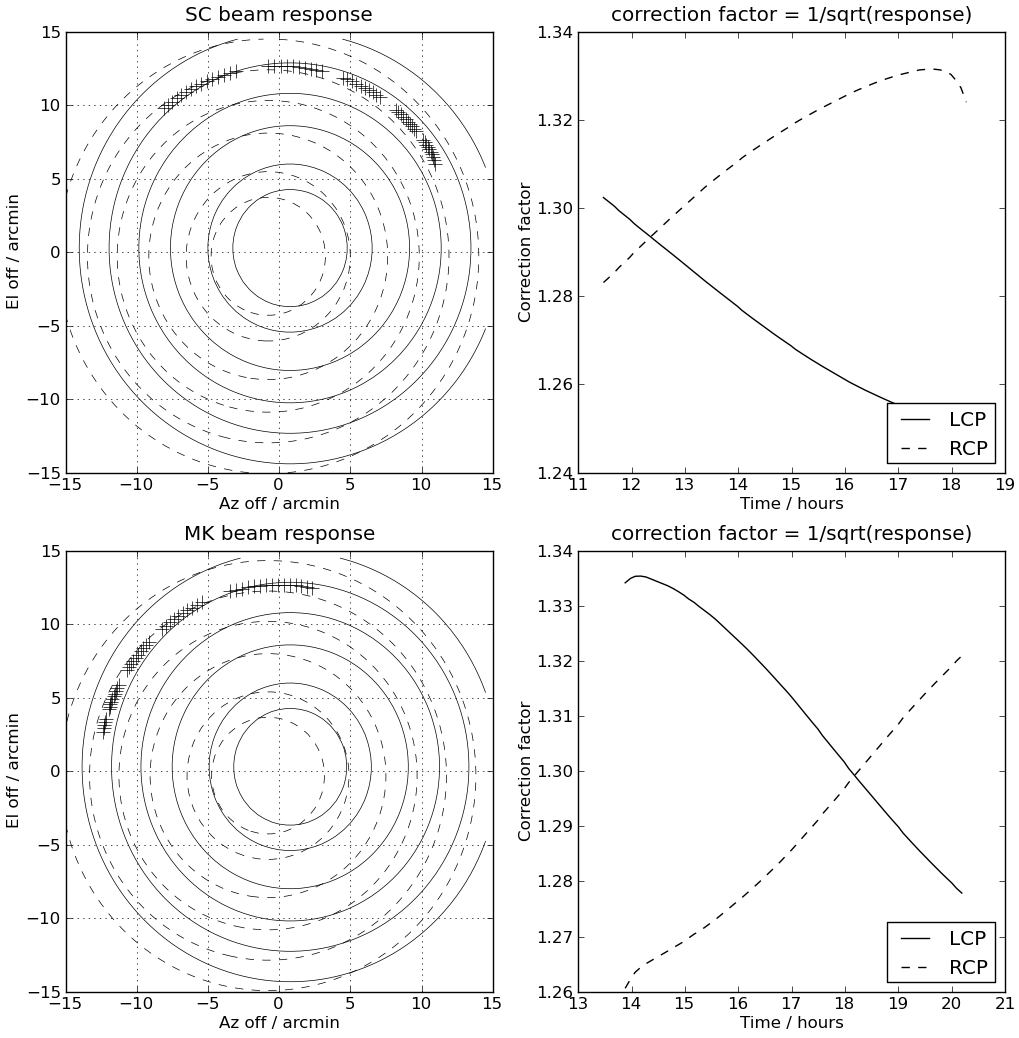}
\caption{{\it Top left:} Contour plot illustrating the relative
  orientation of the LCP (solid lines) and RCP (dashed lines) response
  of the VLBA antenna at Saint Croix. Contours are drawn at 95\,\%,
  90\,\%, 80\,\%, ..., 50\,\% of the peak response. The series of
  crosses indicates the location of S462 in the antenna's primary beam
  during the observations, in intervals of approximately 6\,min. {\it
    Top right:} The correction factor required to compensate for
  primary beam attenuation according to the location of the target,
  plotted as a function of observing time. The corrections for LCP and
  RCP can be quite different. {\it Bottom left and right:} The same
  plots for the Mauna Kea station. The series of crosses shows that
  the correction factors are markedly different from those at Saint
  Croix, because of the different parallactic angle.}
\label{fig:beams}
\end{figure}

Two effects need to be taken care of to determine $b_1(t,\nu)$ and
$b_2(t,\nu)$ at any one time. First, in the case of the VLBA, the
receivers are offset from the optical axis, and the beams of the
right-hand and left-hand circular polarisations (RCP and LCP)
therefore are offset on the sky. This effect is known as beam squint,
and it scales with frequency (see \citealt{Uson2008} for a description
of beam squint at the VLA). Second, the beamwidths also are functions
of frequency. Since the antennas are alt-azimuth antennas the sources
rotate through the two offset beams, and in general sources away from
the pointing centre have different parallactic angles, because the
separation of the antennas is large (i.e., at any given time antennas
see the targets through different portions of their primary
beams). This situation is illustrated in Figure~\ref{fig:beams}. The
angular separation (RCP-LCP) between the two beams on the sky,
averaged across all 10 antennas of the array, is $-1.42'$ in azimuth
and $-0.58'$ in elevation (R. C. Walker, priv. comm.), or a total of
$1.53'$, corresponding to 5.4\,\% of the average FWHM. This squint is
in good agreement with theoretical predictions
(\citealt{Fomalont1999}), and is small when the targets are close to
the pointing centre (which is half-way between the two
beams). However, near the FWHM point of an antenna beam the slope of
the beam response is quite steep, and the difference between the RCP
and LCP response can be up to 10\,\%.

The beamwidths and the amount and direction of beam squint have been
measured at 1438\,MHz (Walker, priv. comm.). We have scaled these
values to the centre frequency of each IF to account for their
frequency dependence. The relative locations of the antenna pointing
centre, the target coordinates and the RCP and LCP beams on the sky
were then computed in 1\,min intervals for each IF, and the
appropriate correction factor was calculated and saved to a new
calibration table. These steps were carried out for each target data
set separately.

Primary beam corrections in VLBA observations are uncommon, and this
is the first attempt to do so on a larger scale, as far as we are
aware. Even though we have taken care to account for primary beam
attenuation as accurately as possible, an amplitude error will
unavoidably remain after correction. For example, the quadrupod legs
and the subreflector blockage produce errors which have not been
accounted for, and the primary beam shape and beam squint are likely
to vary with elevation. We estimate that this residual error is of the
order of 10\,\%, which we add in quadrature to the other sources of
error.

\subsubsection{Imaging}
\label{sec-3.3.3}

We have made three images for each source, with increasing
resolution. To increase the sensitivity to partially extended sources,
we have made naturally-weighted images from data with a circular
30\,\% Gaussian taper at a $(u,v)$ distance of 10\,M$\lambda$,
yielding a resolution of 55.5$\times$22.1\,mas$^2$. For higher
resolution and maximum sensitivity, we have made naturally-weighted
images using all data, yielding a resolution of
28.6$\times$9.3\,mas$^2$. For maximum resolution we have made
uniformly-weighted images using all data, resulting in a resolution of
20.8$\times$5.9\,mas$^2$. All images were 8192$^2$\,pixels large,
showing areas of 16.4$\times$16.4\,arcsec$^2$ for the tapered image
and 8.2$\times$8.2\,arcsec$^2$ for the untapered images,
respectively. Images were centred on the radio position derived from
ATCA observations.

The naturally-weighted and uniformly-weighted images are still plagued
by significant imaging artifacts, which we were unable to remove. The
sources appear to be irregularly extended predominantly in the
north-south direction, in which the resolution is poor because of the
low declination of the field. We attribute this to the fact that the
visibilities which have the greatest resolution in the north-south
direction are preferentially produced near the rise and set of the
source. At this time, the airmass is high and these visibilities
probably have comparatively large calibration errors. Another effect
is that only relatively few long baselines are left after
self-calibration on S503, because it is slightly resolved. These
visibilities then make a disproportionately large contribution to
uniformly-weighted images, reducing their quality. The loss of long
baselines in the north-south direction is reflected in a change of the
ratio of the restoring beam axes, which increases from 2.51 (tapered
images) to 3.1 (natural weighting) and to 3.5 (uniform weighting).

\subsection{Source detection and measurements}
\label{sec-3.4}

The target source positions are relatively uncertain, in VLBI terms,
as they were determined from arcsec-scale radio observations that have
a resolution 500 times lower than the VLBI observations. To account
for this, large VLBI images have been made to increase the search
space for source emission (see Sec.~\ref{sec-3.3.3}).  We have used
three criteria to automatically search for potential detections:

\begin{enumerate}
\item We have computed the SNR of the brightest pixel in the image
  (excluding a guard band of 100\,pixels around the image edges which
  frequently have artificially high values arising from the imaging
  procedure). All images with SNR larger than 6 were treated as
  potential detections.

\item The separation of the brightest pixels in the naturally weighted
  and tapered images was calculated. If this separation was smaller
  than a beamwidth, a visual inspection was carried out.

\item We have imaged the RCP and LCP data separately, using no taper
  and natural weighting. Since the two polarisations are processed by
  different electronics in the antennas, they provide independent
  information about the sources. These images were set to zero below a
  threshold, $z/\sqrt{2}$, with $z=5\,\sigma$ and $\sigma$ the noise
  in the Stokes I image, and then multiplied. The resulting image is
  zero everywhere except at locations where both the RCP and LCP images
  exceed $z/\sqrt{2}$. Compared to using a simple $5\,\sigma$ cutoff
  in the Stokes I image this method is less susceptible to random
  noise peaks by a factor of more than 6 (see Appendix A).

\end{enumerate}

Sources matching at least one of these criteria were inspected by eye
to confirm or reject the detection.

To measure flux densities of the observed sources requires making a
choice about resolution. Many sources are significantly resolved, and
almost all sources can be reasonably imaged using heavily tapered or
untapered data, resulting in a variety of resolutions. We decided to
quote flux densities measured from images tapered at 10\,M$\lambda$,
because at this resolution all sources (with the exception of S393)
can be well approximated by a Gaussian. S393 has been fitted with two
Gaussians, and the parameters listed in Table~\ref{tab:results}
represent their combined values.

Flux densities were measured by fitting a Gaussian to the image,
starting at the location of the brightest pixel. The most accurate
results were obtained when the shape of the Gaussian was fixed, but
its position angle kept as a free parameter. In very low-SNR cases
such as ours, free fits frequently result in a gross overestimate of
the flux densities.

Flux density errors were estimated as follows. We have identified
three potential sources of error: (i) the a priori amplitudes in the
VLBI observations are commonly assumed to be to be correct to within
10\,\%; (ii) the formal errors determined by the Gaussian fits; and
(iii) the 10\,\% error estimate as a consequence of primary beam
correction (which probably is conservative, in particular for sources
near the pointing centre). These three components are added in
quadrature to yield the total flux density error.

\cite{Condon1997} gives the equations for errors of Gaussian fits. In
our case, the pixels are correlated, in fact, the area over which
pixels are correlated is approximately one resolution element in the
images, or one beam area. For this case, \cite{Condon1997} gives

\begin{equation}
\frac{\mu^2(I)}{I^2} \approx \frac{\mu^2(A)}{A^2} + \left(\frac{\theta_N^2}{\theta_M\theta_m}\right)\left[\frac{\mu^2(\theta_M)}{\theta_M^2} + \frac{\mu^2(\theta_m)}{\theta_m^2}\right]
\end{equation}

where $\mu$ denotes a variance of a parameter returned by the fitting
procedure, $I$ the volume of the Gauss (the integrated flux density),
$A$ its amplitude (the peak flux density), and $\theta_N$, $\theta_M$,
and $\theta_m$ denote the FWHM of a circular Gaussian convolving
function and the major and minor axes of the fit, respectively. In our
case, $\theta_N^2\approx\theta_M\theta_m$, hence the equation
simplifies to

\begin{equation}
\frac{\mu^2(I)}{I^2} \approx \frac{\mu^2(A)}{A^2} + \left[\frac{\mu^2(\theta_M)}{\theta_M^2} + \frac{\mu^2(\theta_m)}{\theta_m^2}\right]
\end{equation}

which we use as an estimate of the integrated flux density error
arising from the fitting procedure.

\section{Results}
\label{sec-4}

A summary of the image properties, source flux densities, and
ancilliary data is listed in Table~\ref{tab:results}.

\subsection{Images}

Contour plots of the 20 detected sources and the one candidate
detection are shown in Figure~\ref{fig:images}. The three images per
source, made with different weighting and tapering, are shown in a
row. The left panel shows a naturally-weighted image made with a
10\,M$\lambda$ tapering (restoring beam size
55.5$\times$22.1\,mas$^2$), the centre panel the untapered,
naturally-weighted image (restoring beam size
28.6$\times$9.3\,mas$^2$), and the right panel shows the
uniformly-weighted image (restoring beam
20.8$\times$5.9\,mas$^2$). Positive contours start at three times the
rms level of the images and increase by factors of $\sqrt{2}$. One
negative contour is shown at three times the rms.

\begin{figure*}
\includegraphics[width=0.3\linewidth]{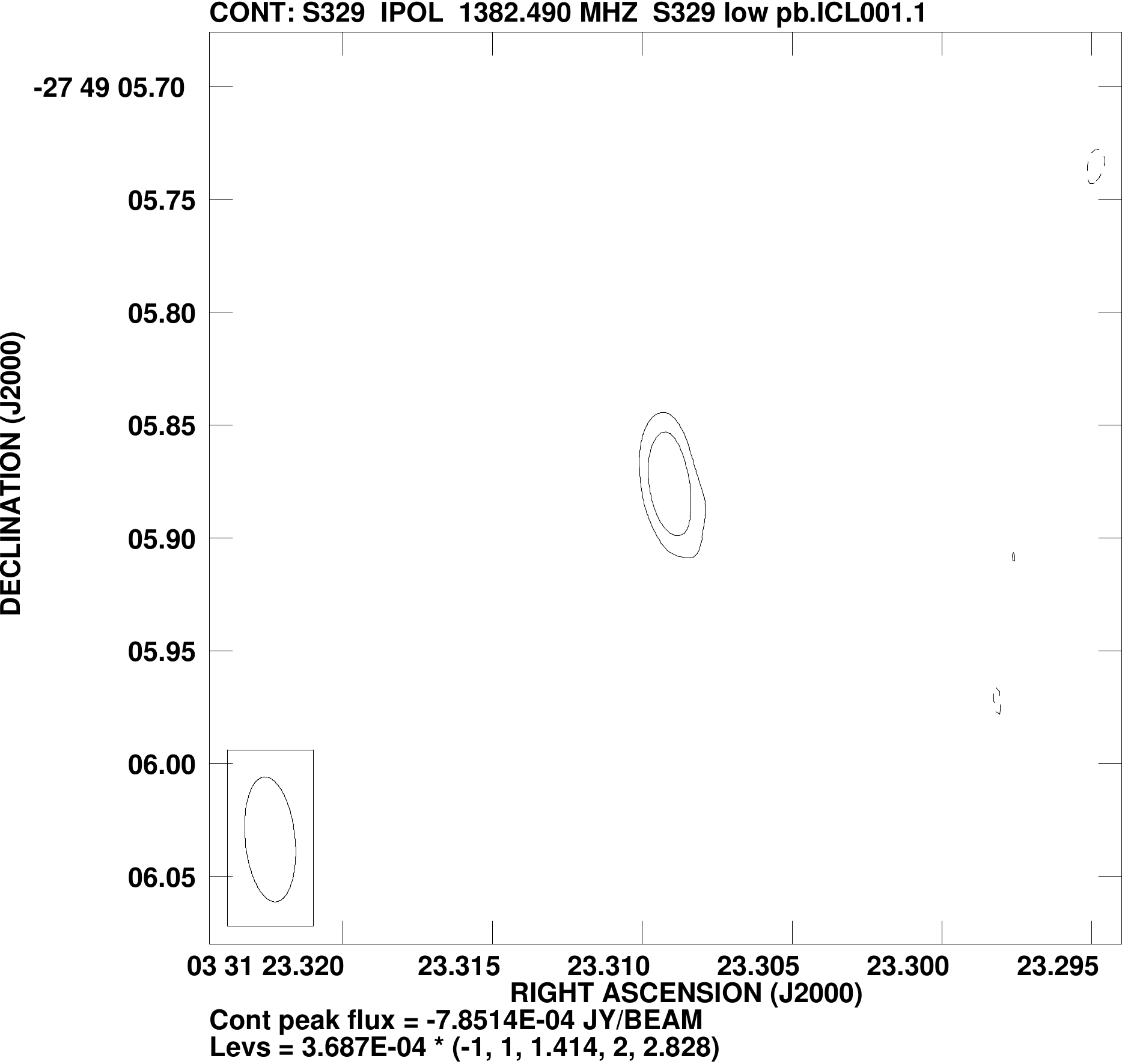} 
\includegraphics[width=0.3\linewidth]{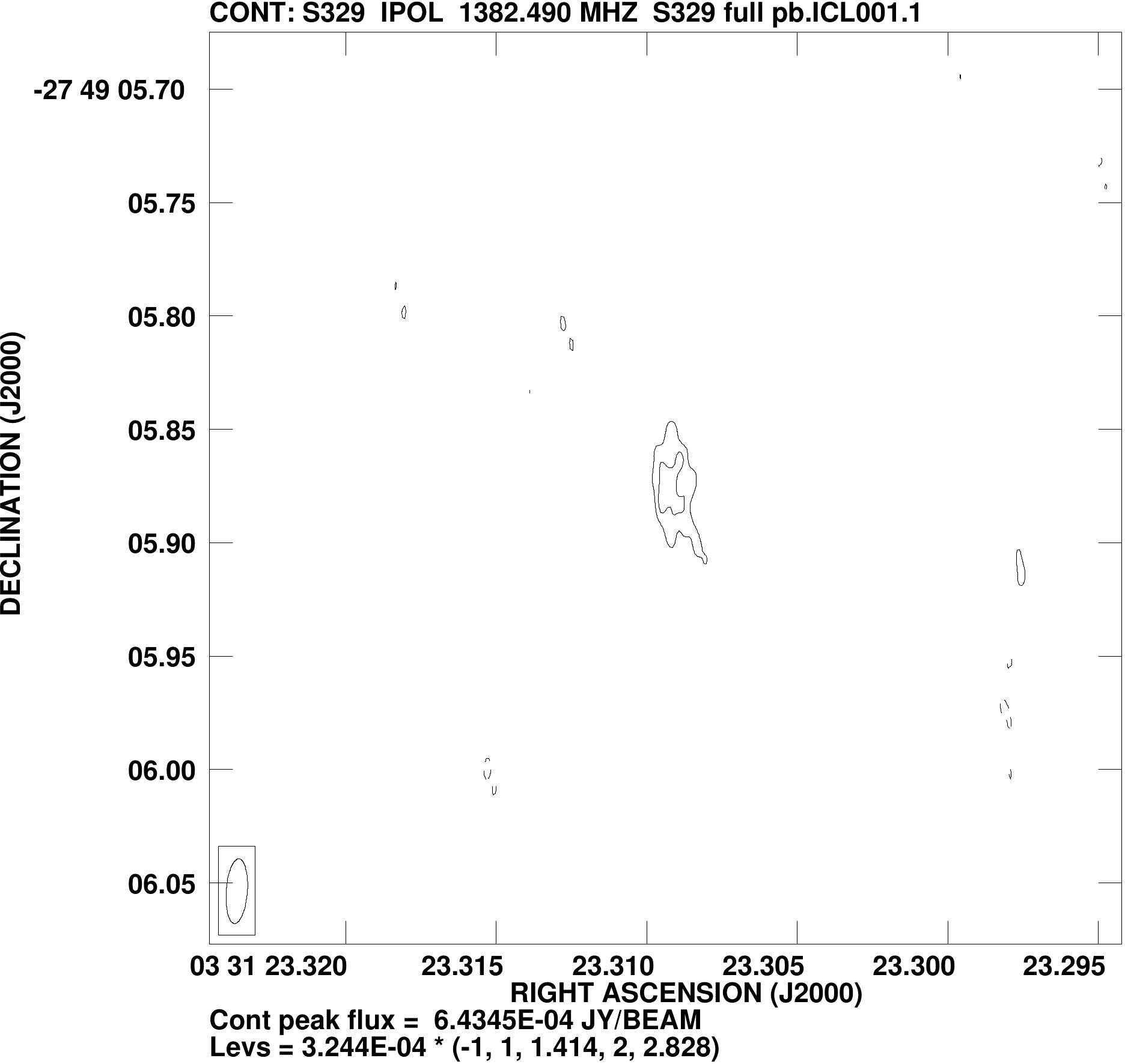}
\includegraphics[width=0.3\linewidth]{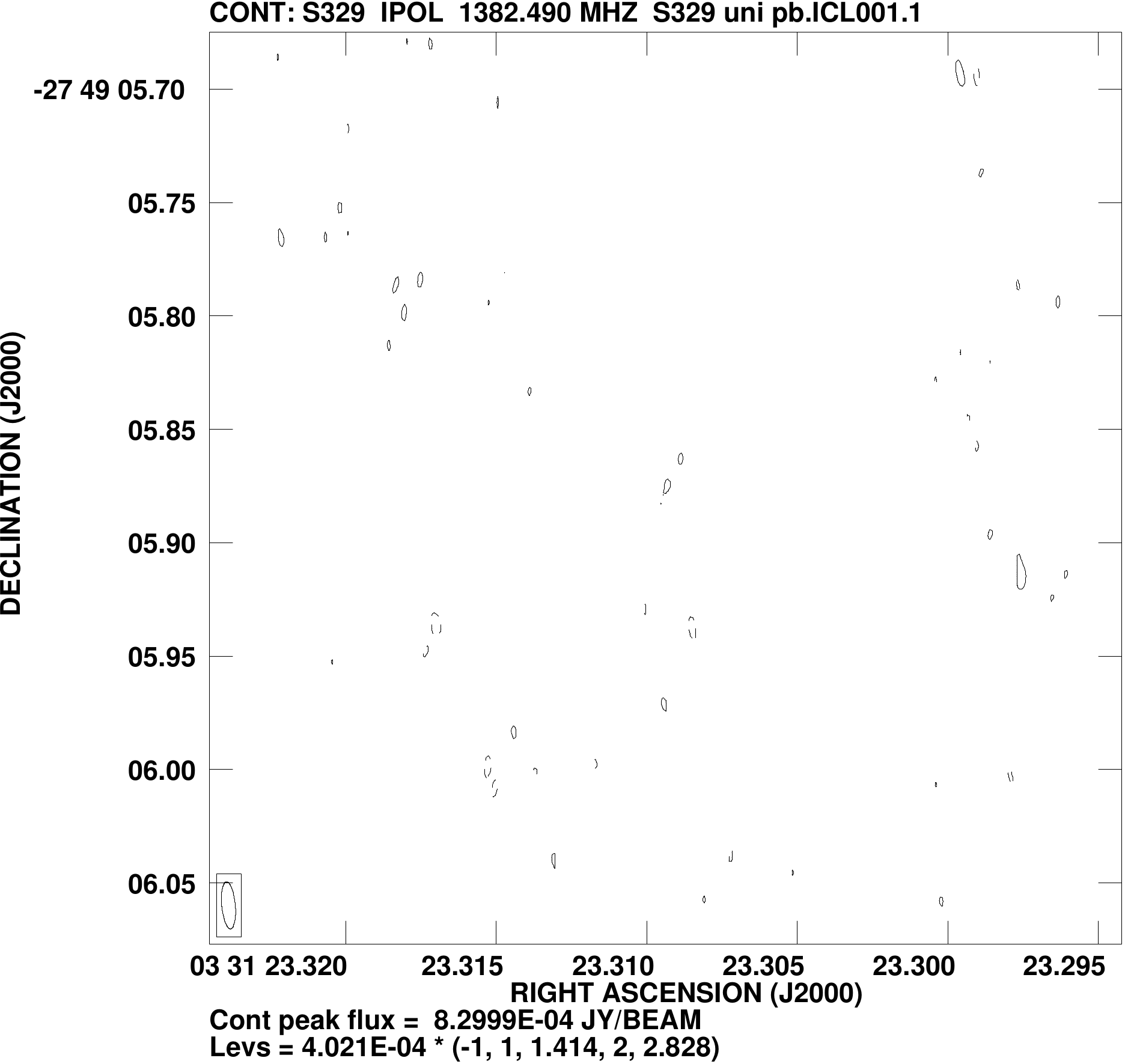} \\
\includegraphics[width=0.3\linewidth]{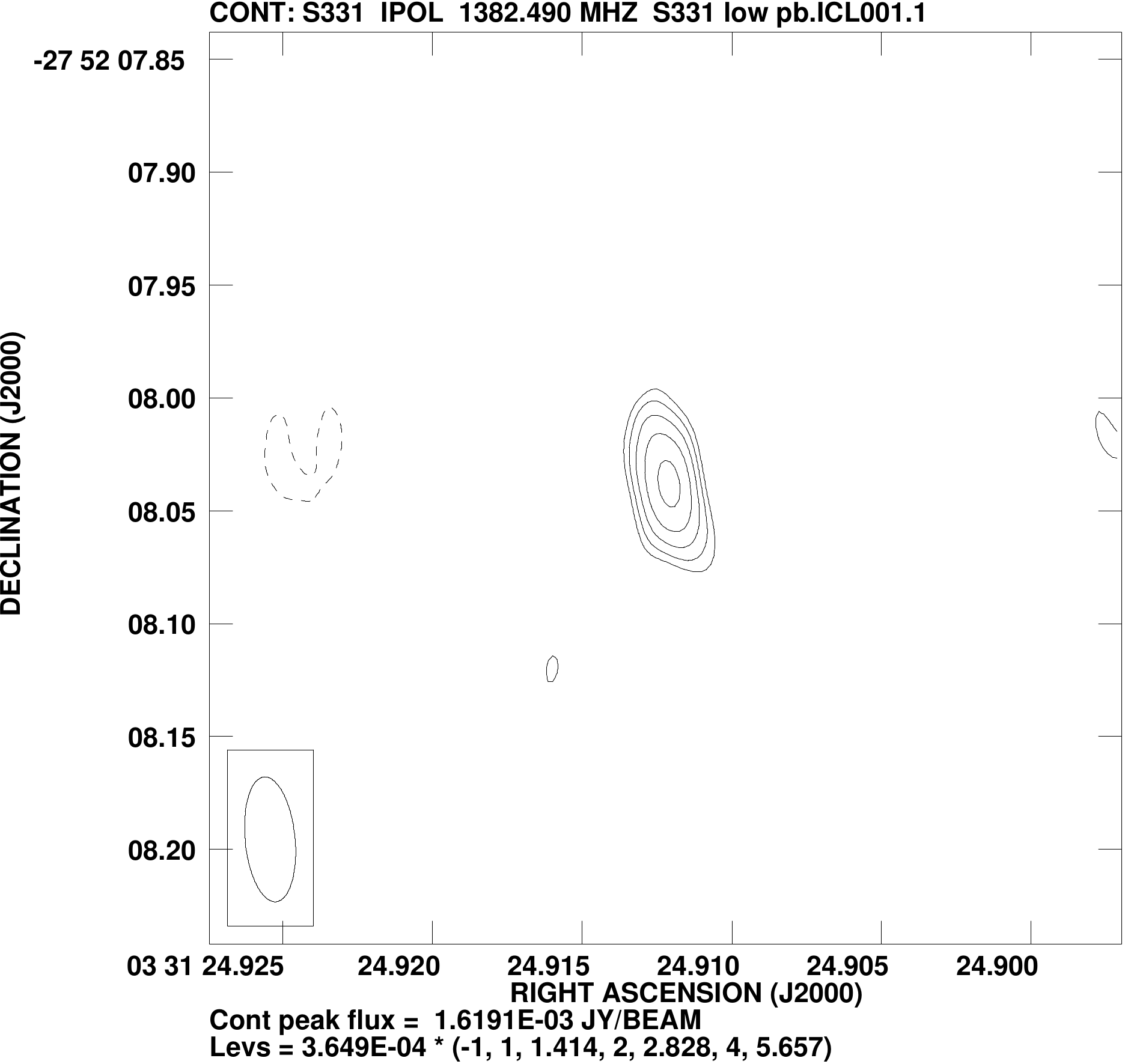} 
\includegraphics[width=0.3\linewidth]{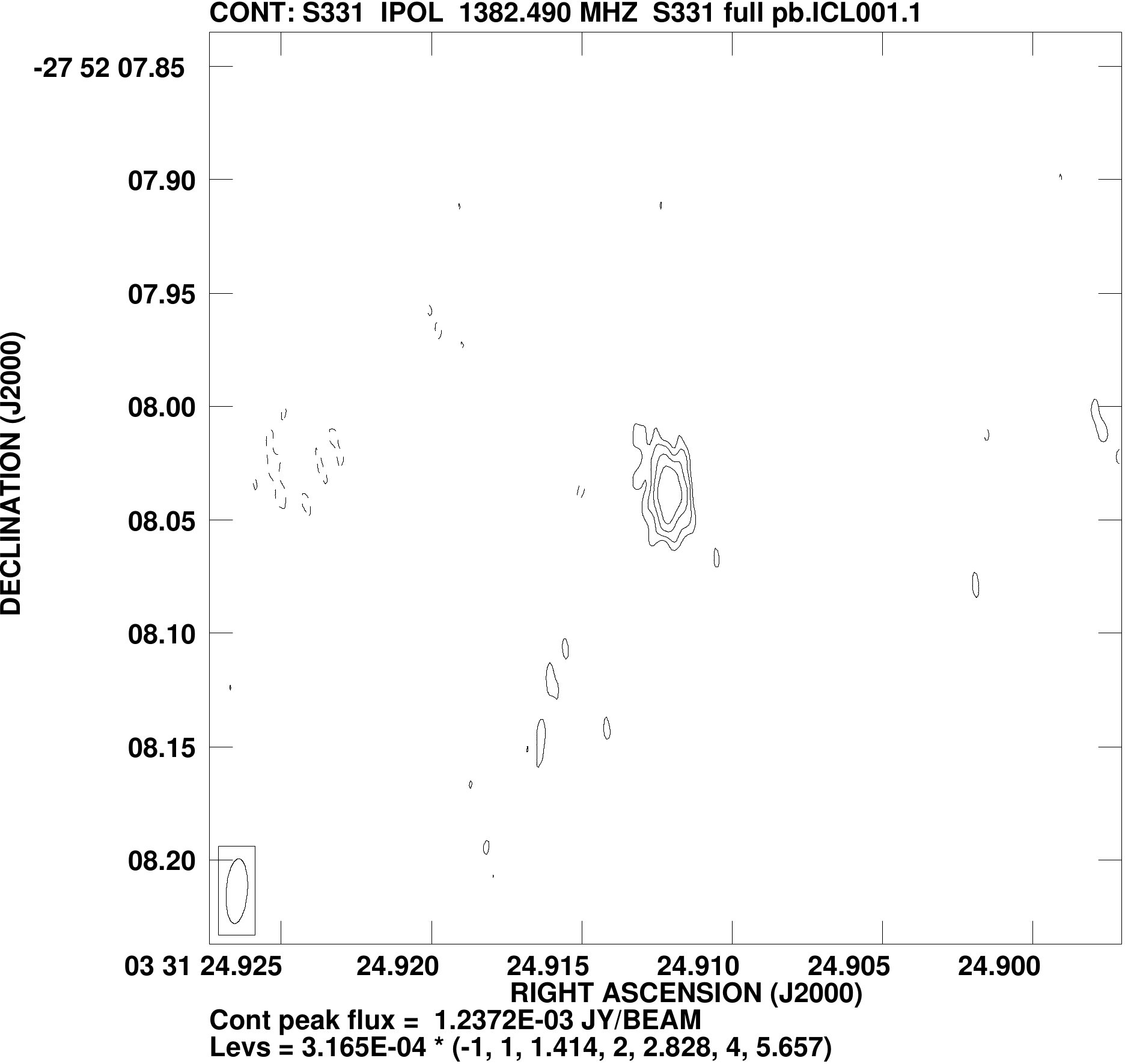}
\includegraphics[width=0.3\linewidth]{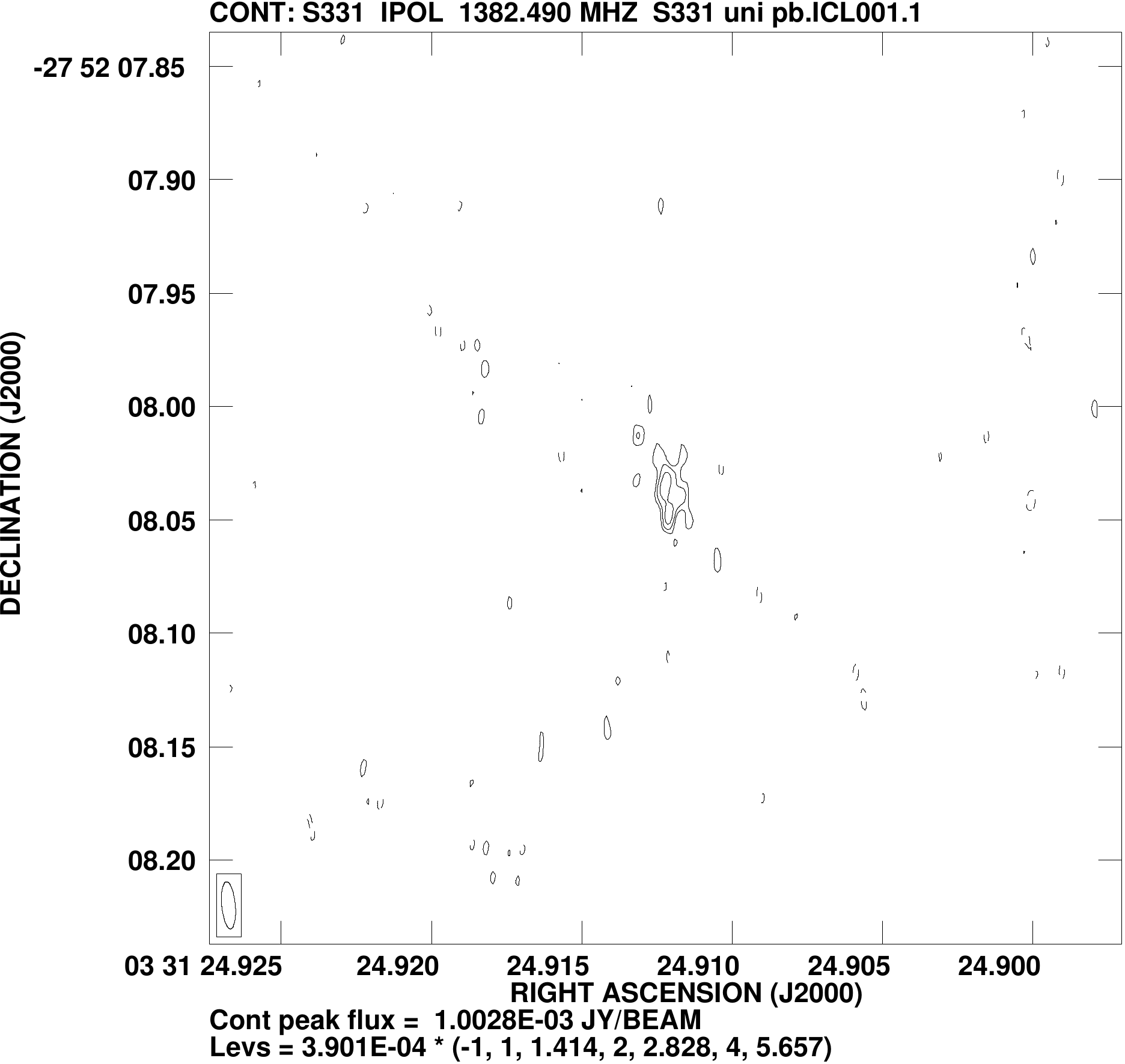} \\
\includegraphics[width=0.3\linewidth]{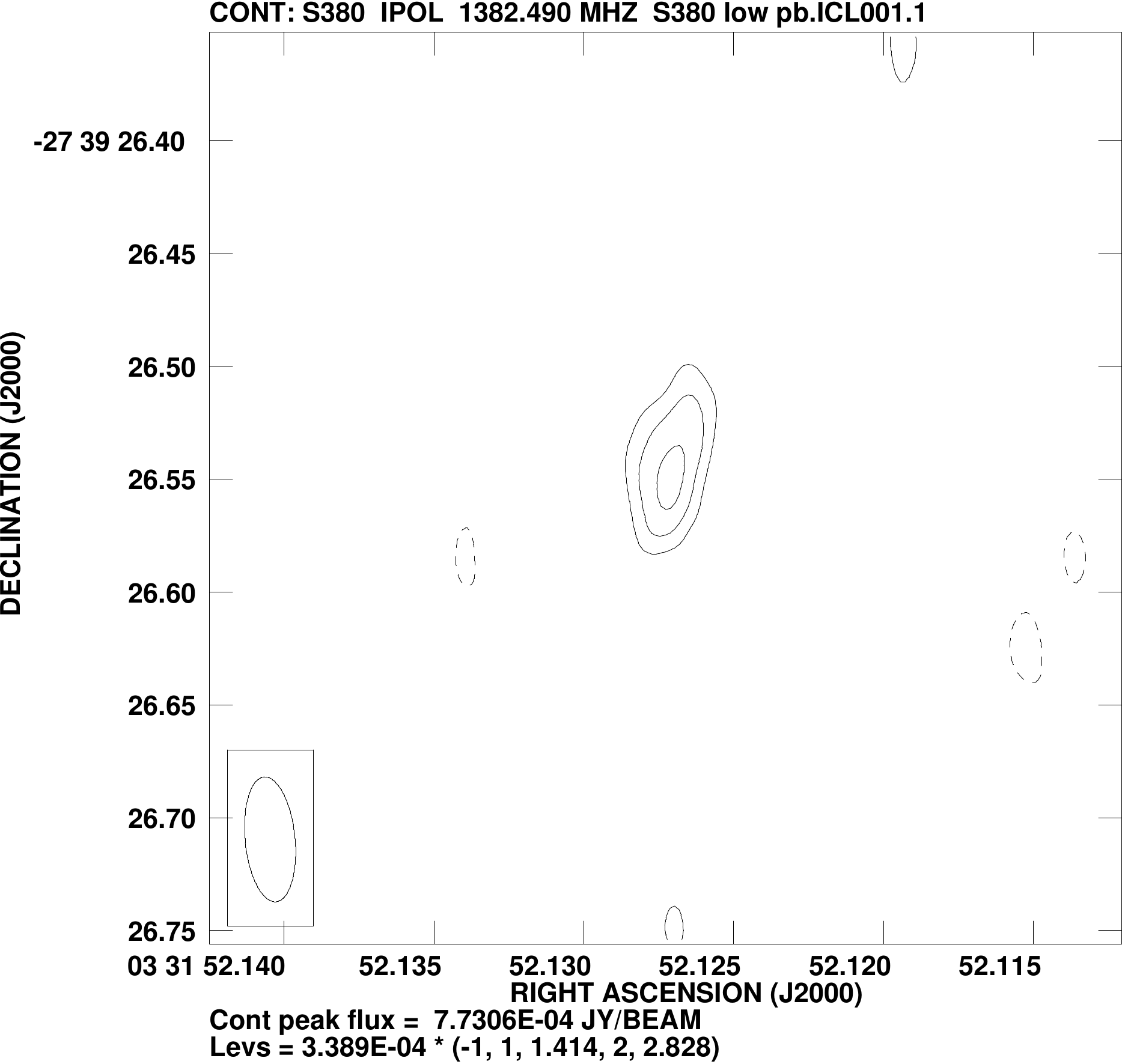} 
\includegraphics[width=0.3\linewidth]{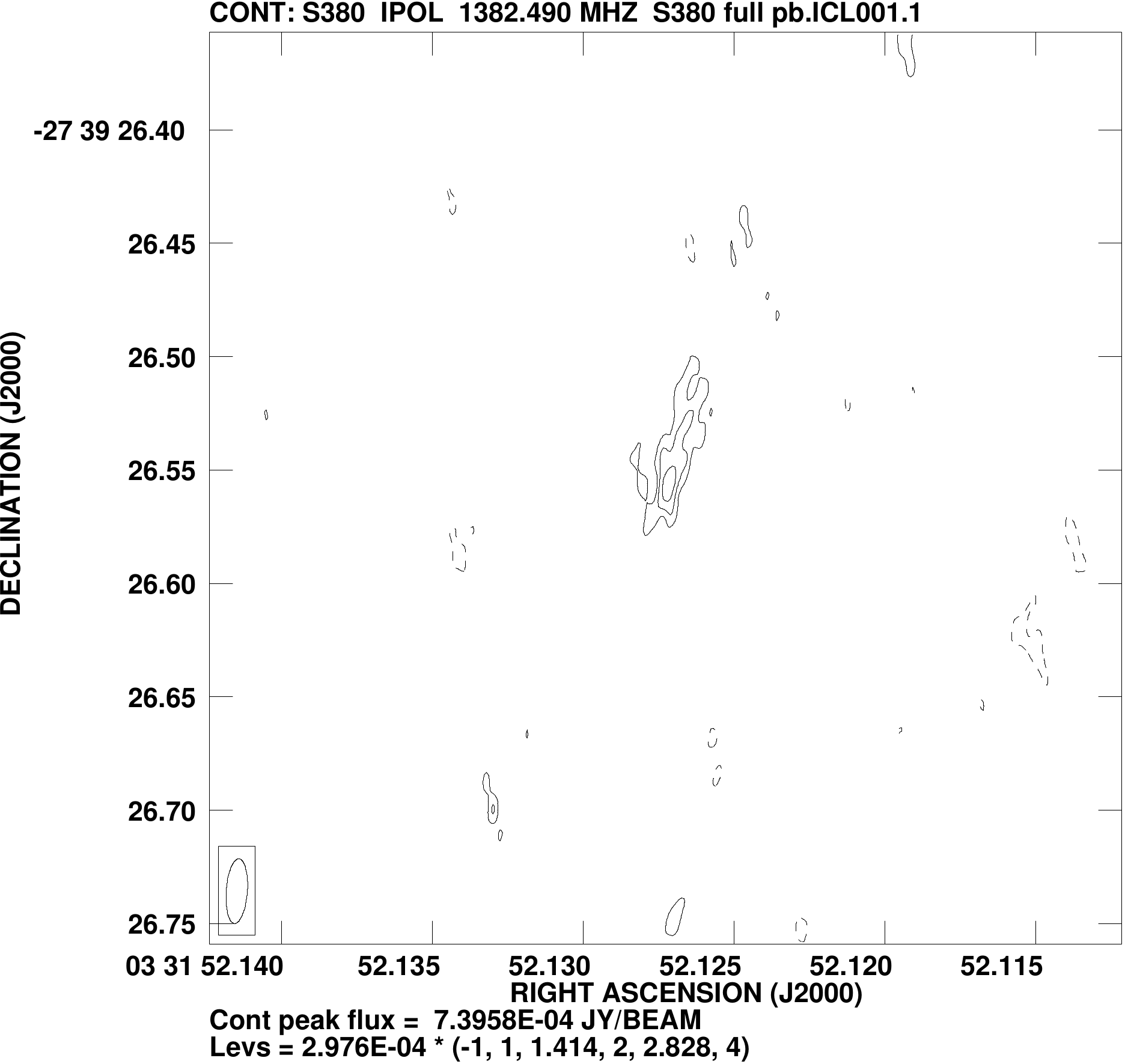}
\includegraphics[width=0.3\linewidth]{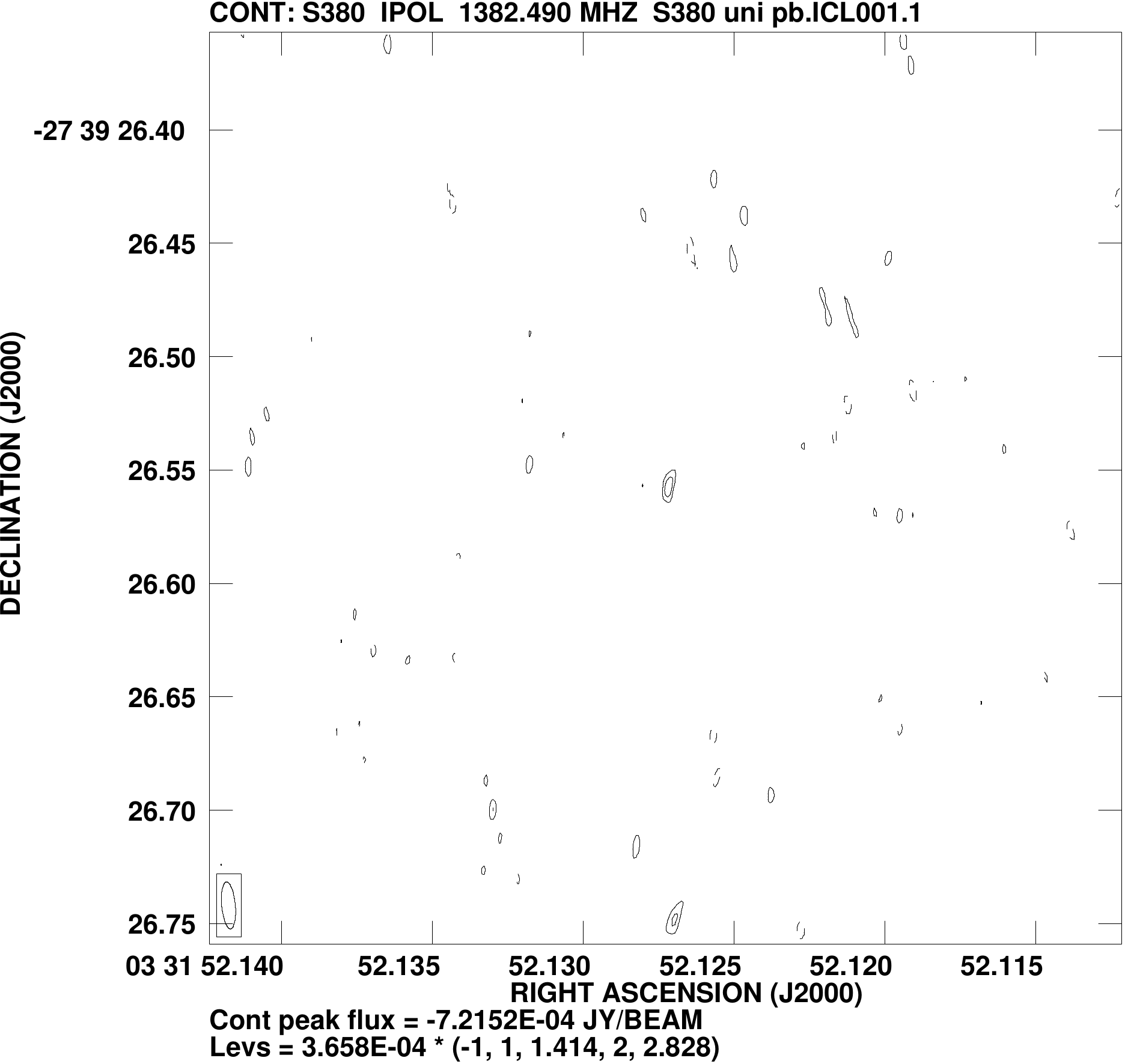} \\
\includegraphics[width=0.3\linewidth]{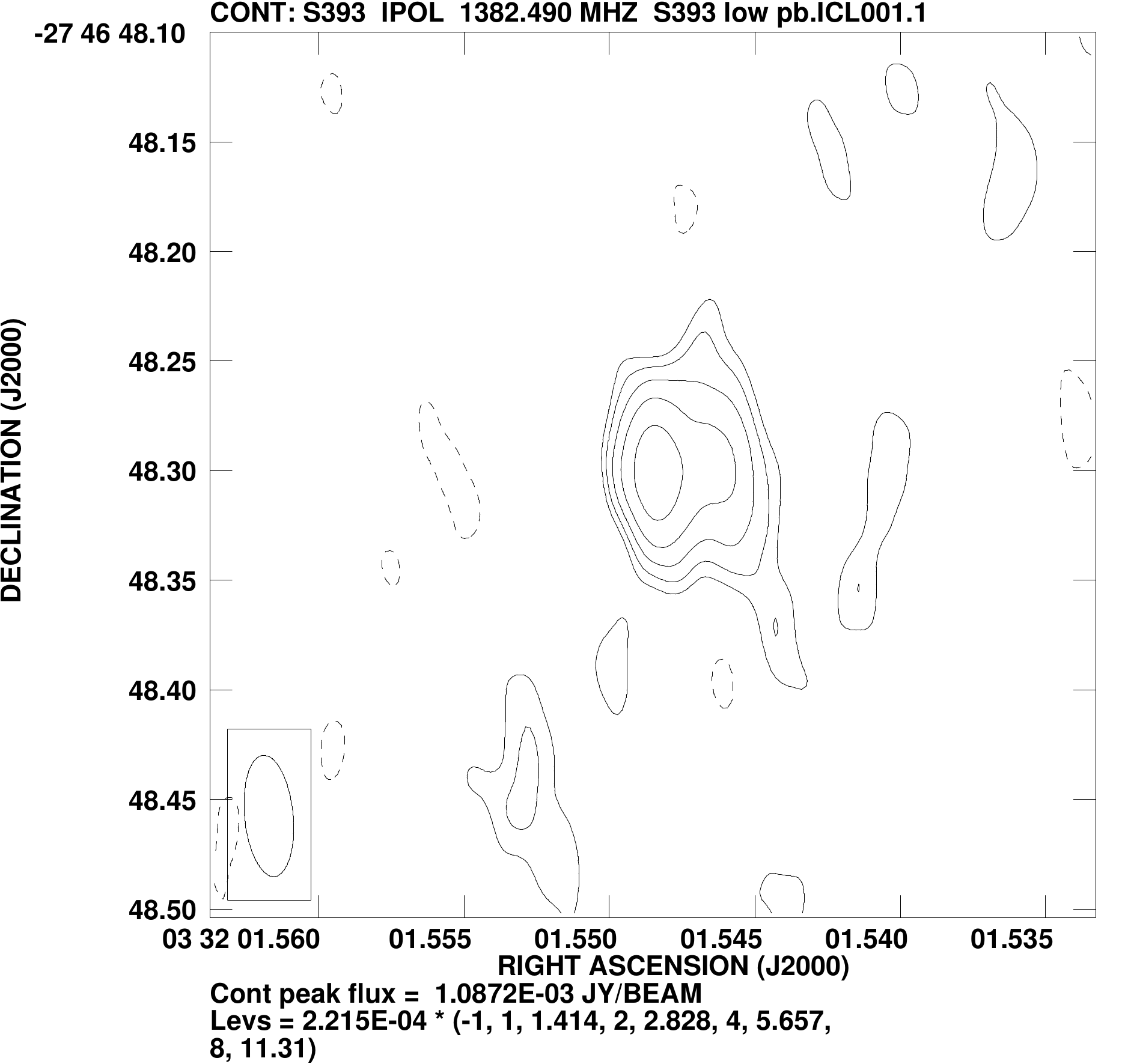} 
\includegraphics[width=0.3\linewidth]{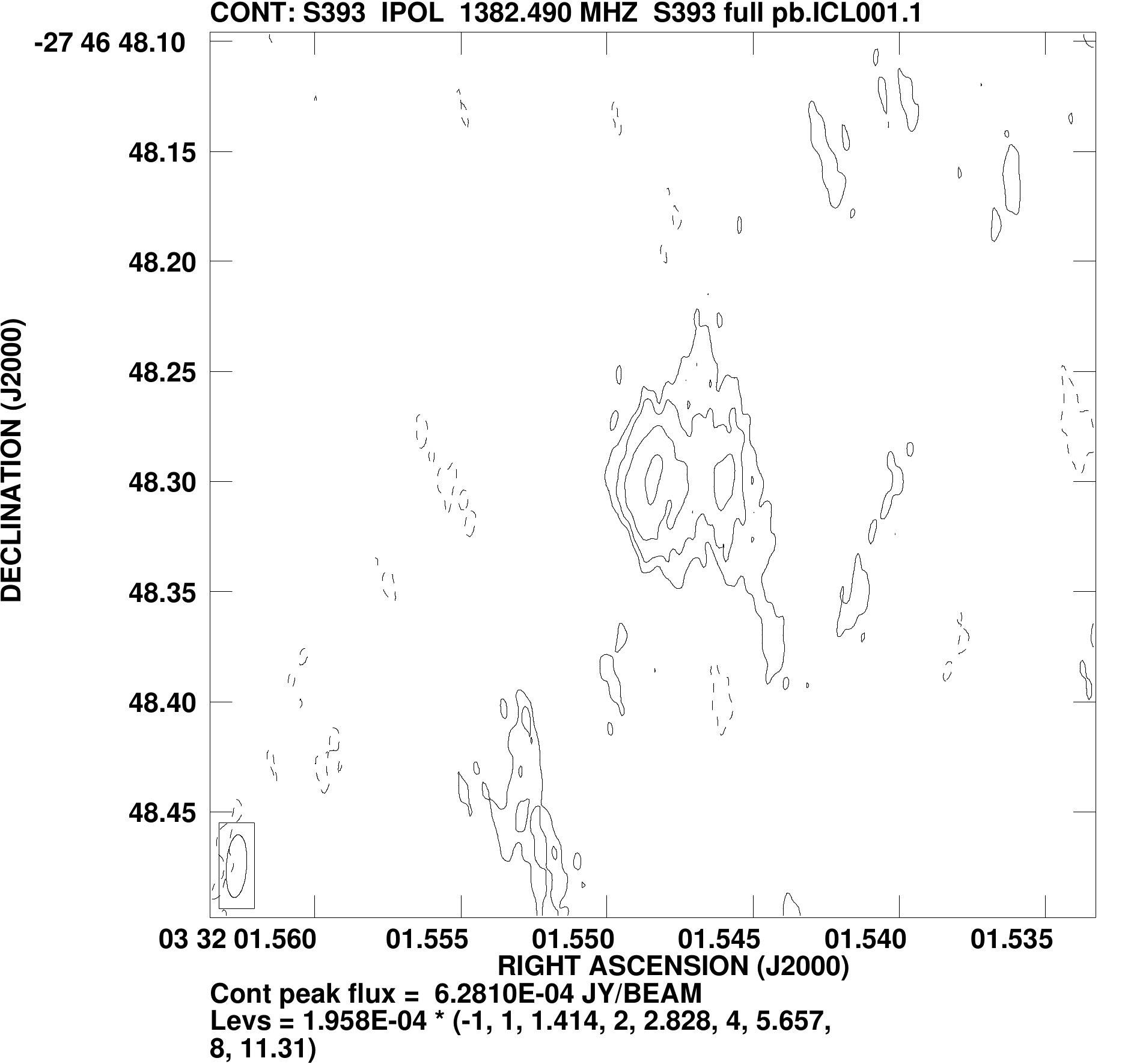}
\includegraphics[width=0.3\linewidth]{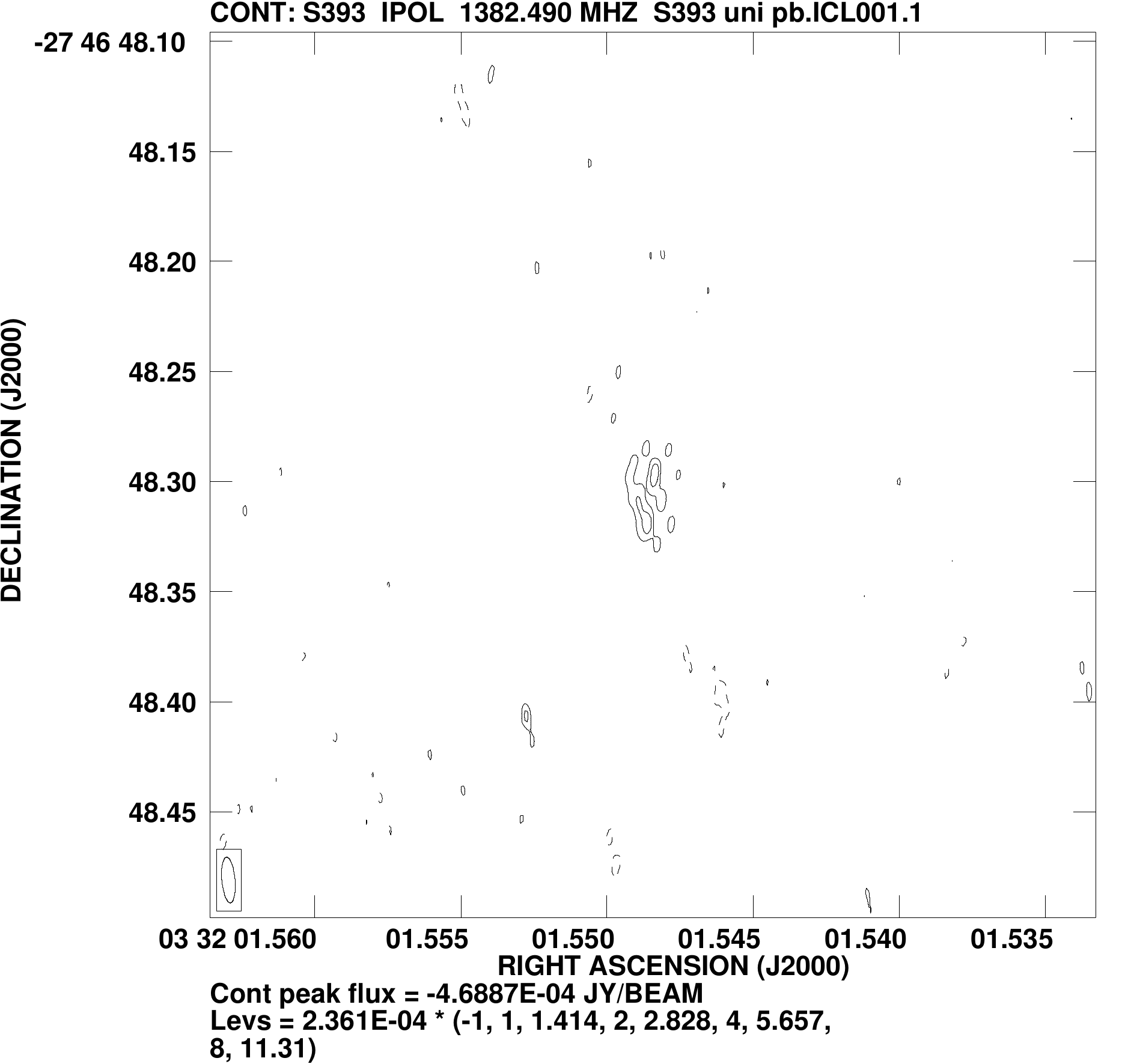} \\
\caption{Contour plots of the detected sources. Three images per
  source are shown in a row. {\it Left column:} naturally-weighted
  images made with a 10\,M$\lambda$ tapering (restoring beam size
  55.5$\times$22.1\,mas$^2$); {\it middle column:} untapered,
  naturally-weighted image (restoring beam size
  28.6$\times$9.3\,mas$^2$); {\it right column:} uniformly-weighted
  image (restoring beam 20.8$\times$5.9\,mas$^2$). Positive contours
  start at three times the rms level of the images and increase by
  factors of $\sqrt{2}$. One negative contour is shown at three times
  the rms.}
\label{fig:images}
\end{figure*}

\addtocounter{figure}{-1}

\begin{figure*}
\includegraphics[width=0.3\linewidth]{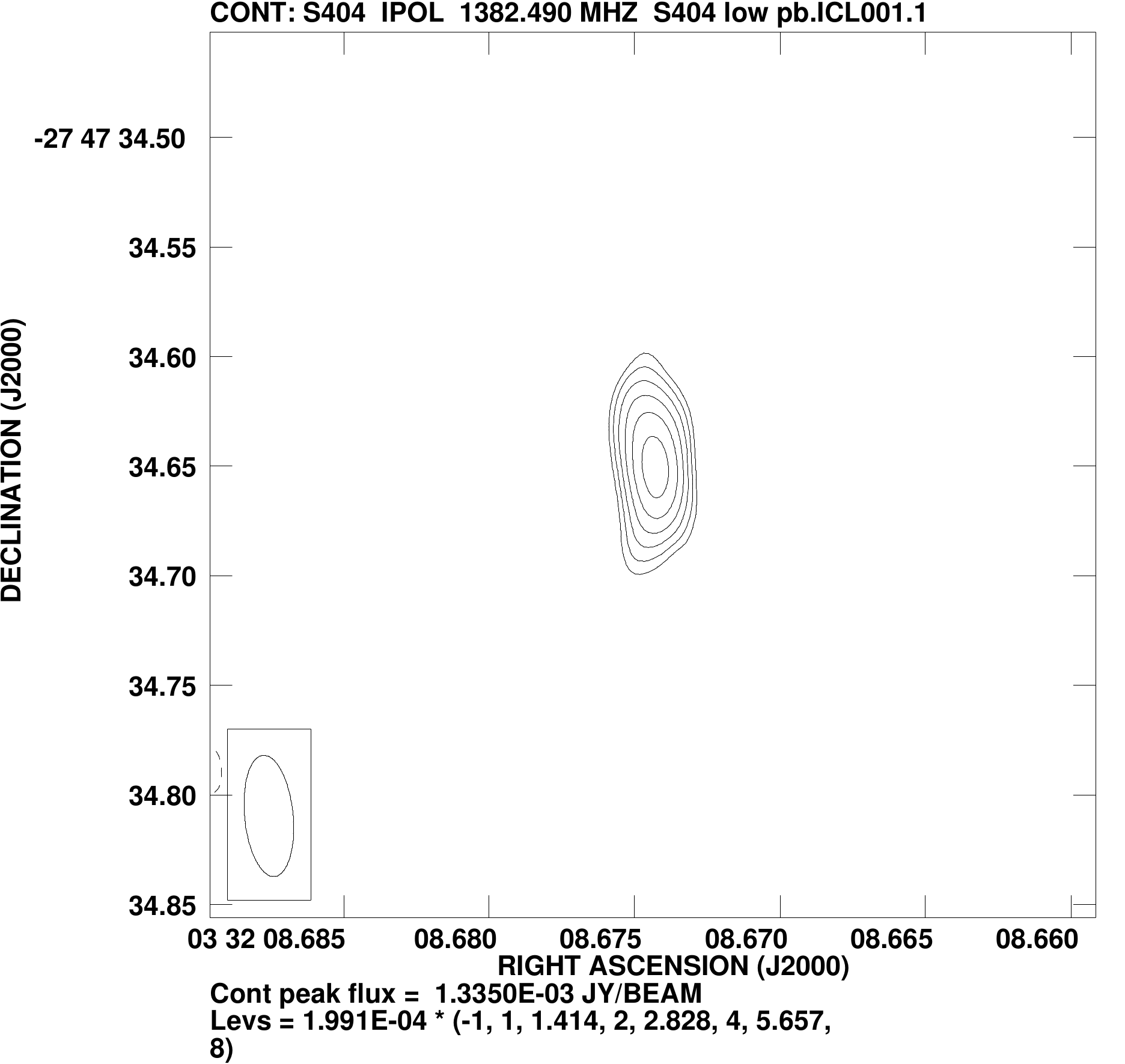} 
\includegraphics[width=0.3\linewidth]{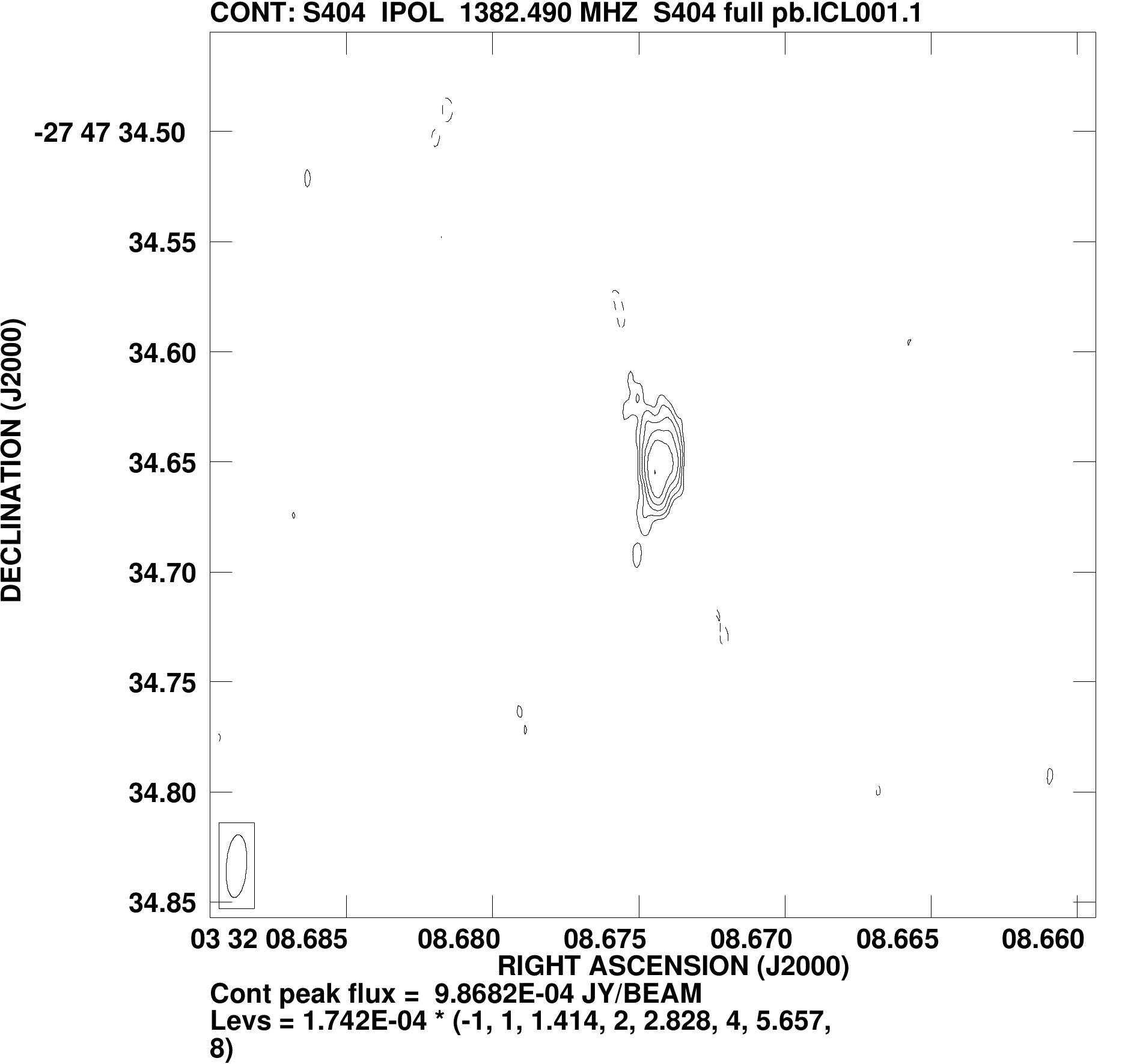}
\includegraphics[width=0.3\linewidth]{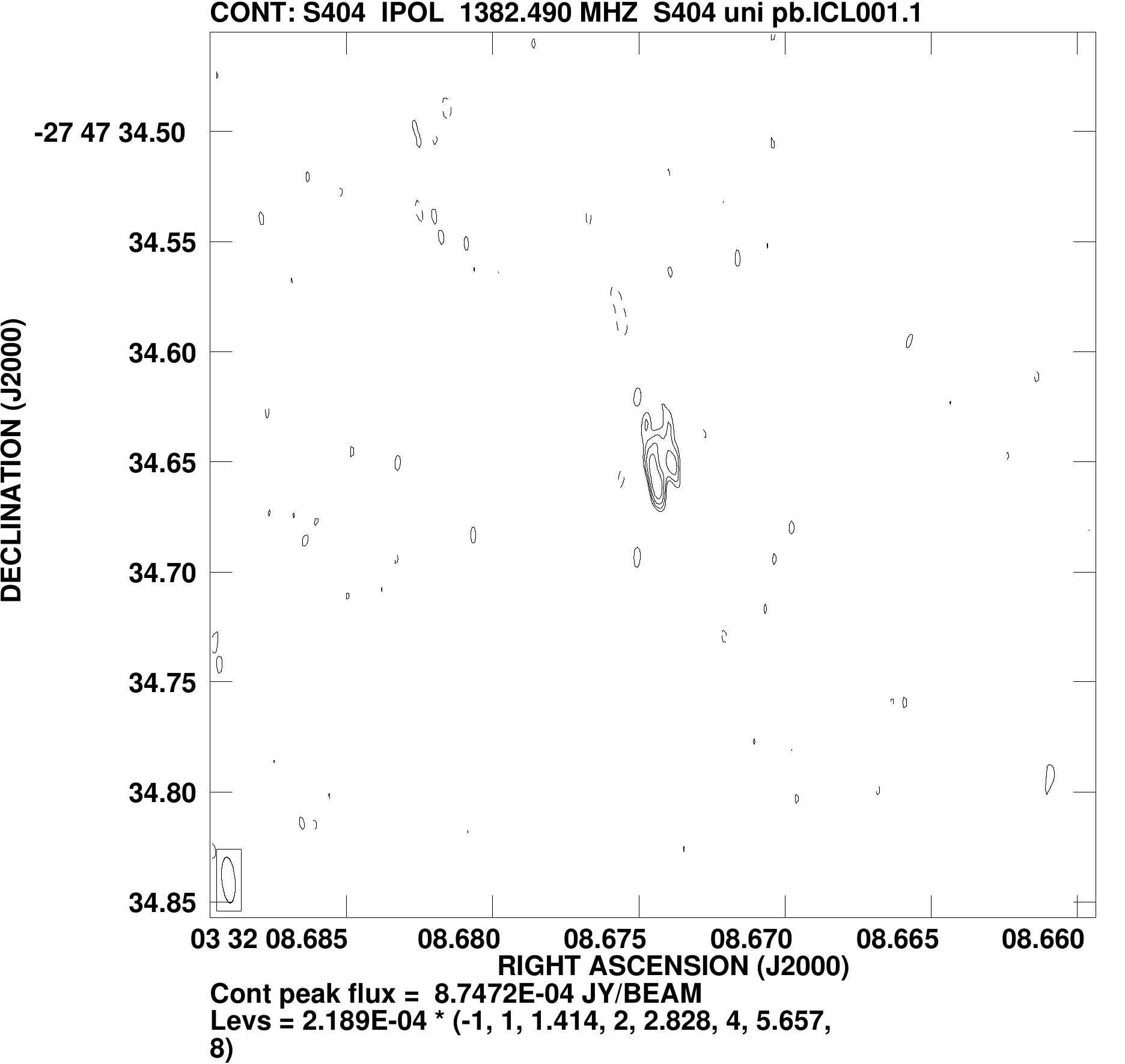} \\
\includegraphics[width=0.3\linewidth]{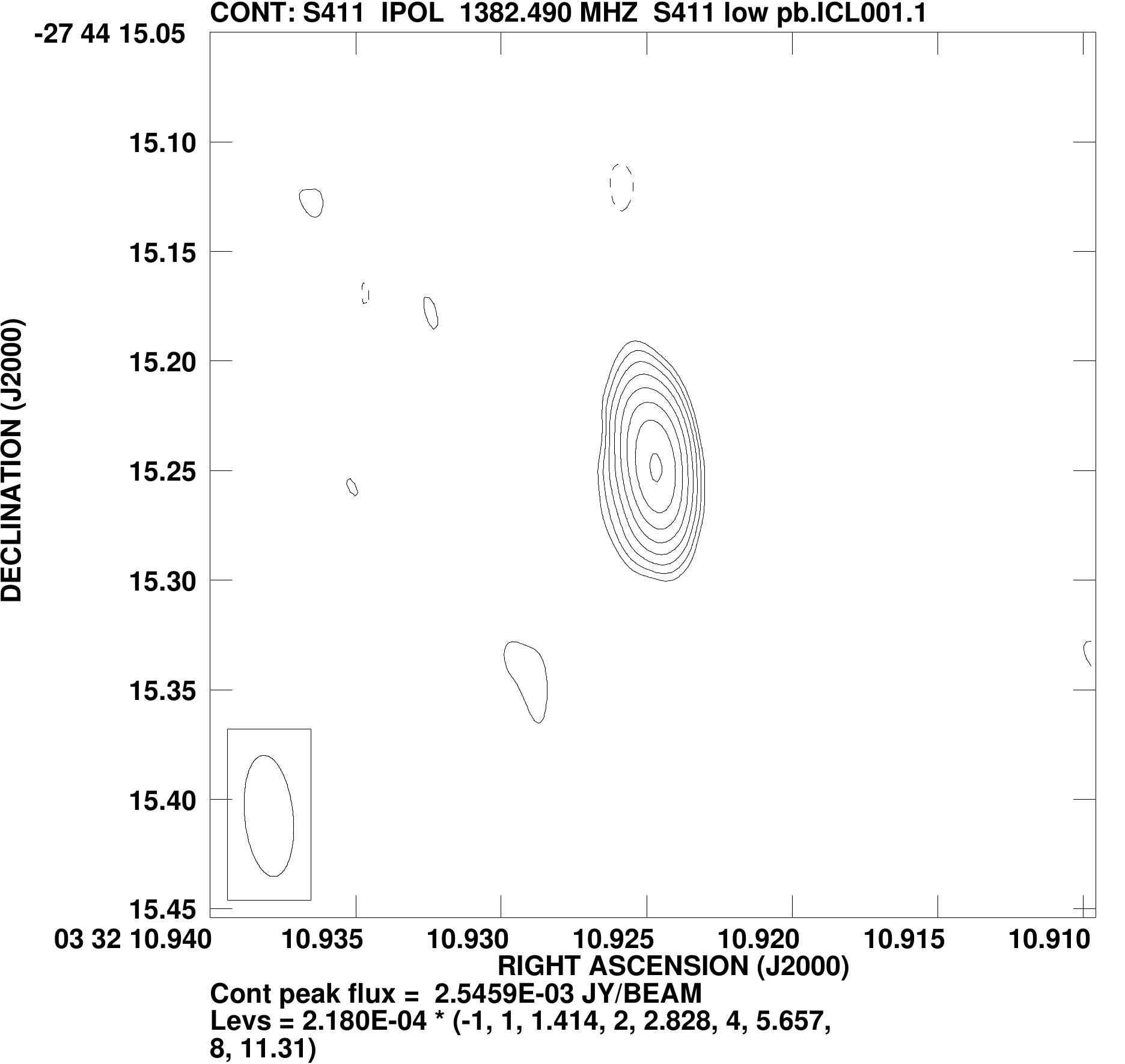} 
\includegraphics[width=0.3\linewidth]{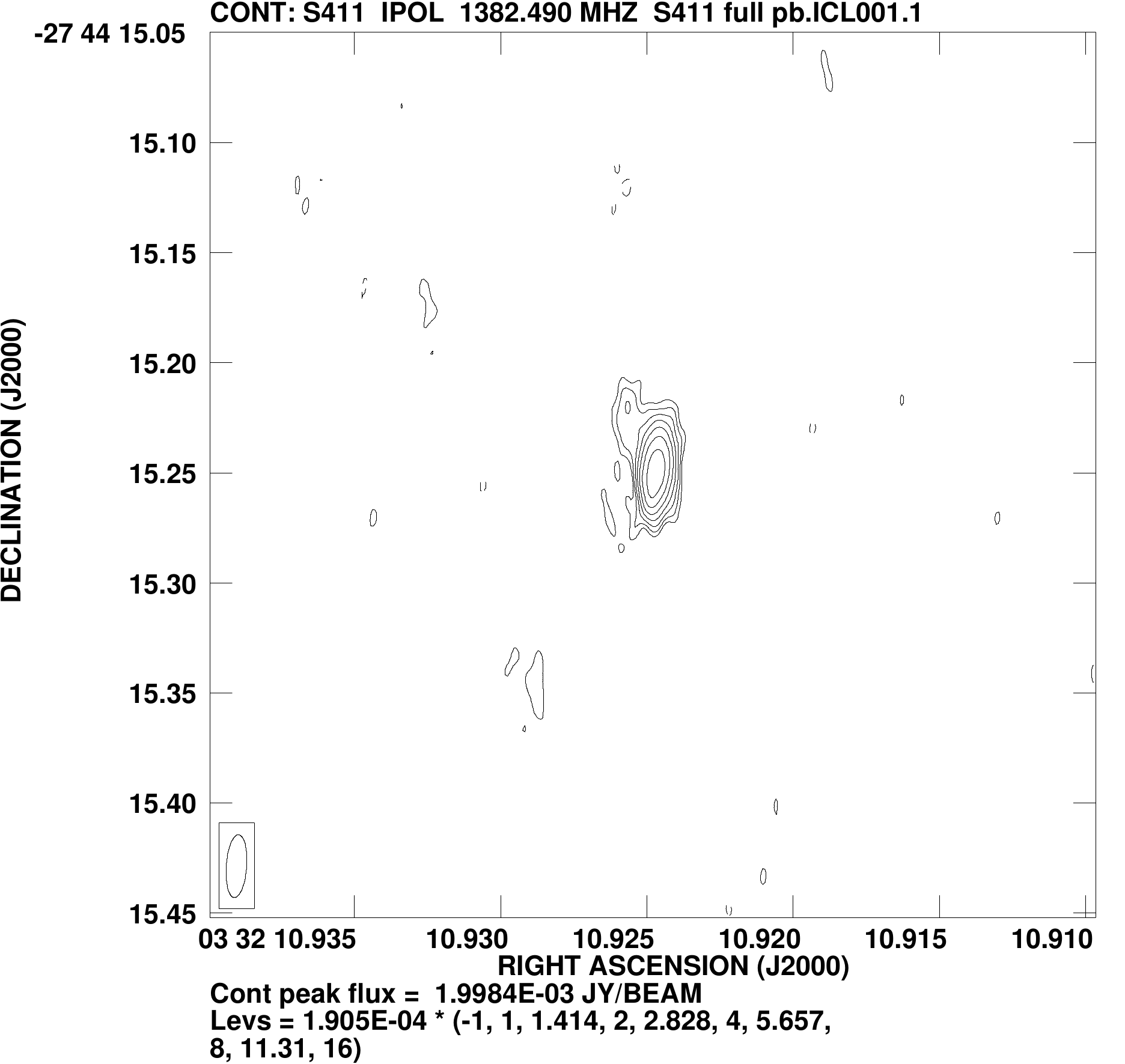}
\includegraphics[width=0.3\linewidth]{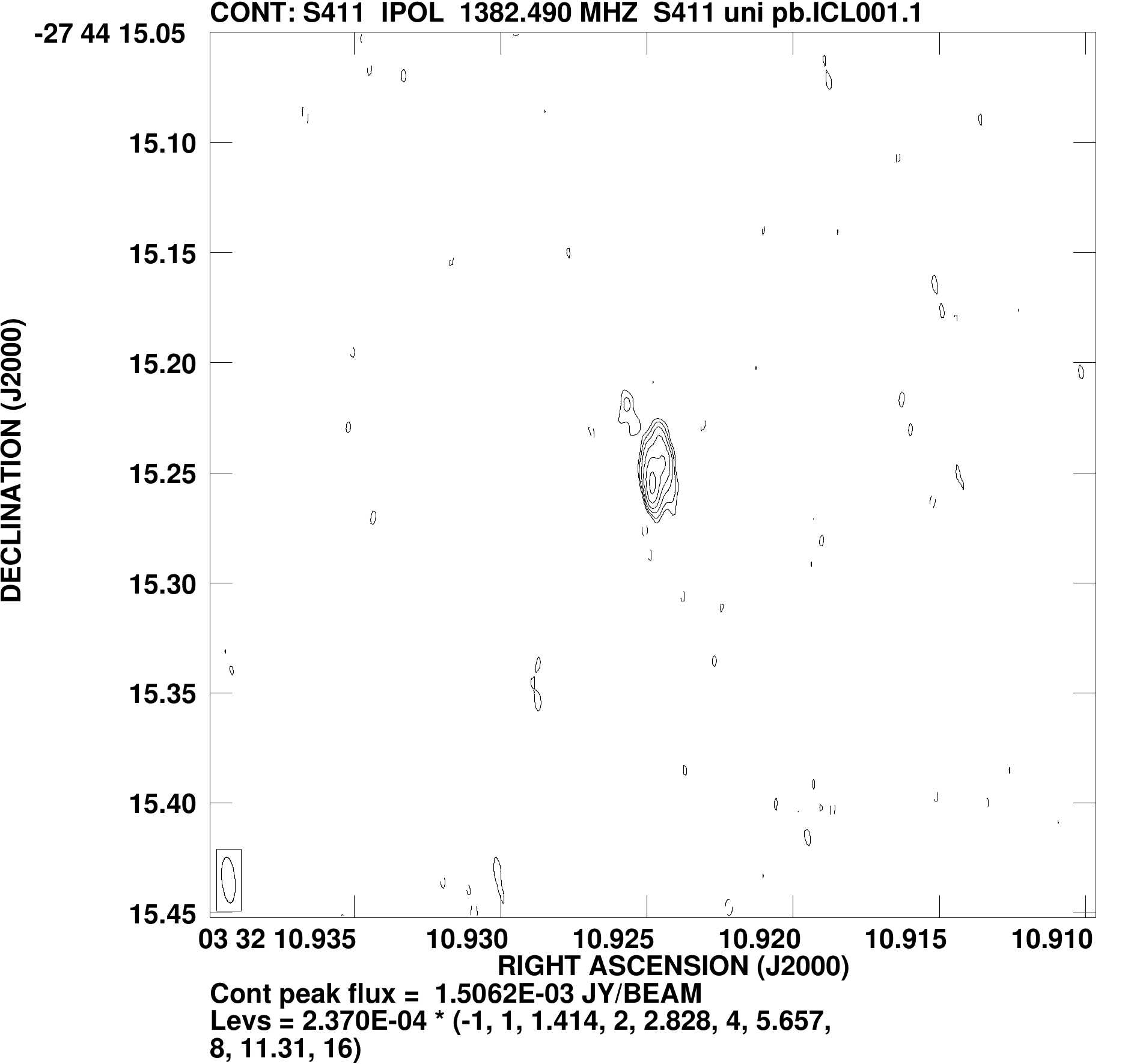} \\
\includegraphics[width=0.3\linewidth]{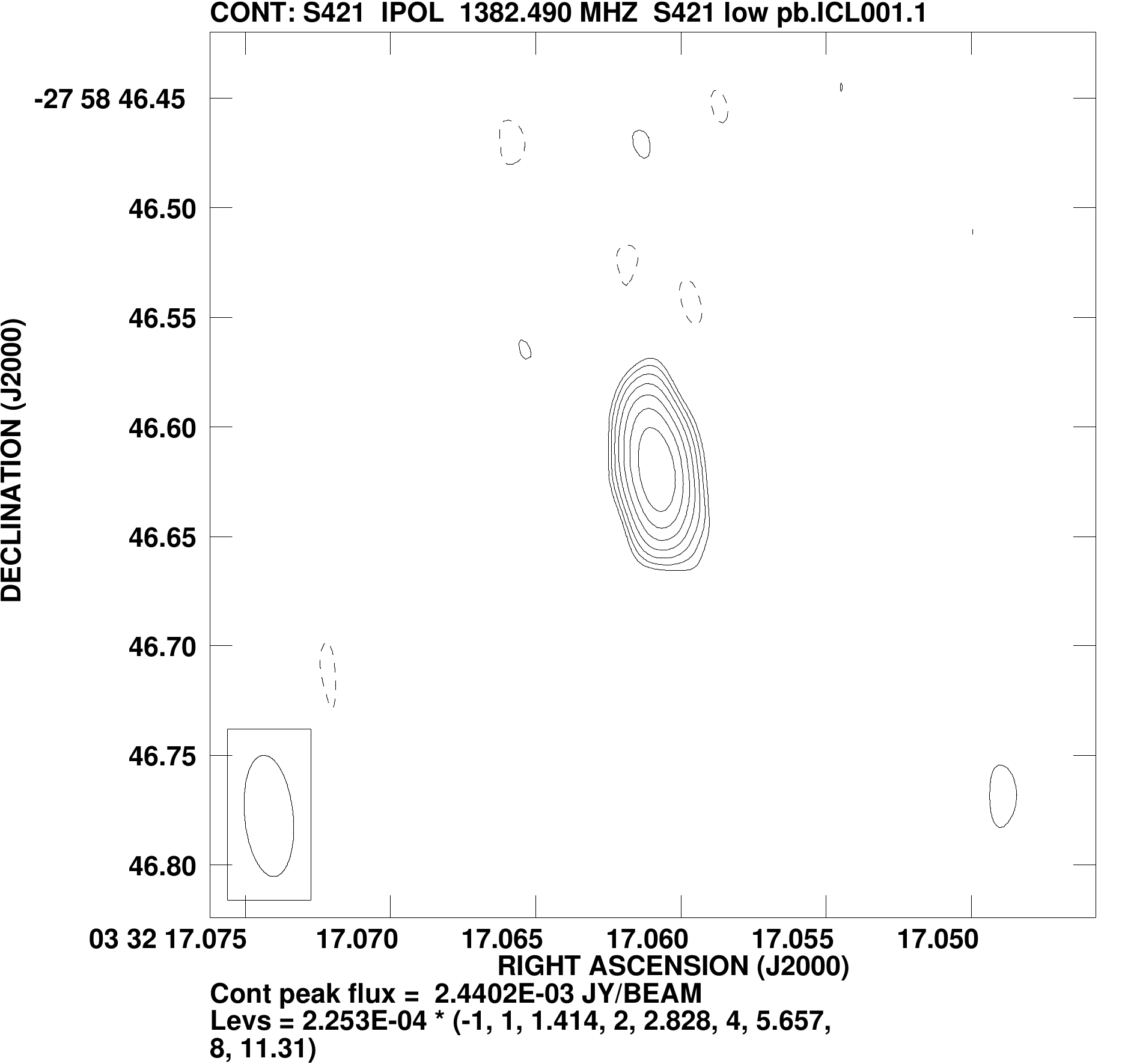} 
\includegraphics[width=0.3\linewidth]{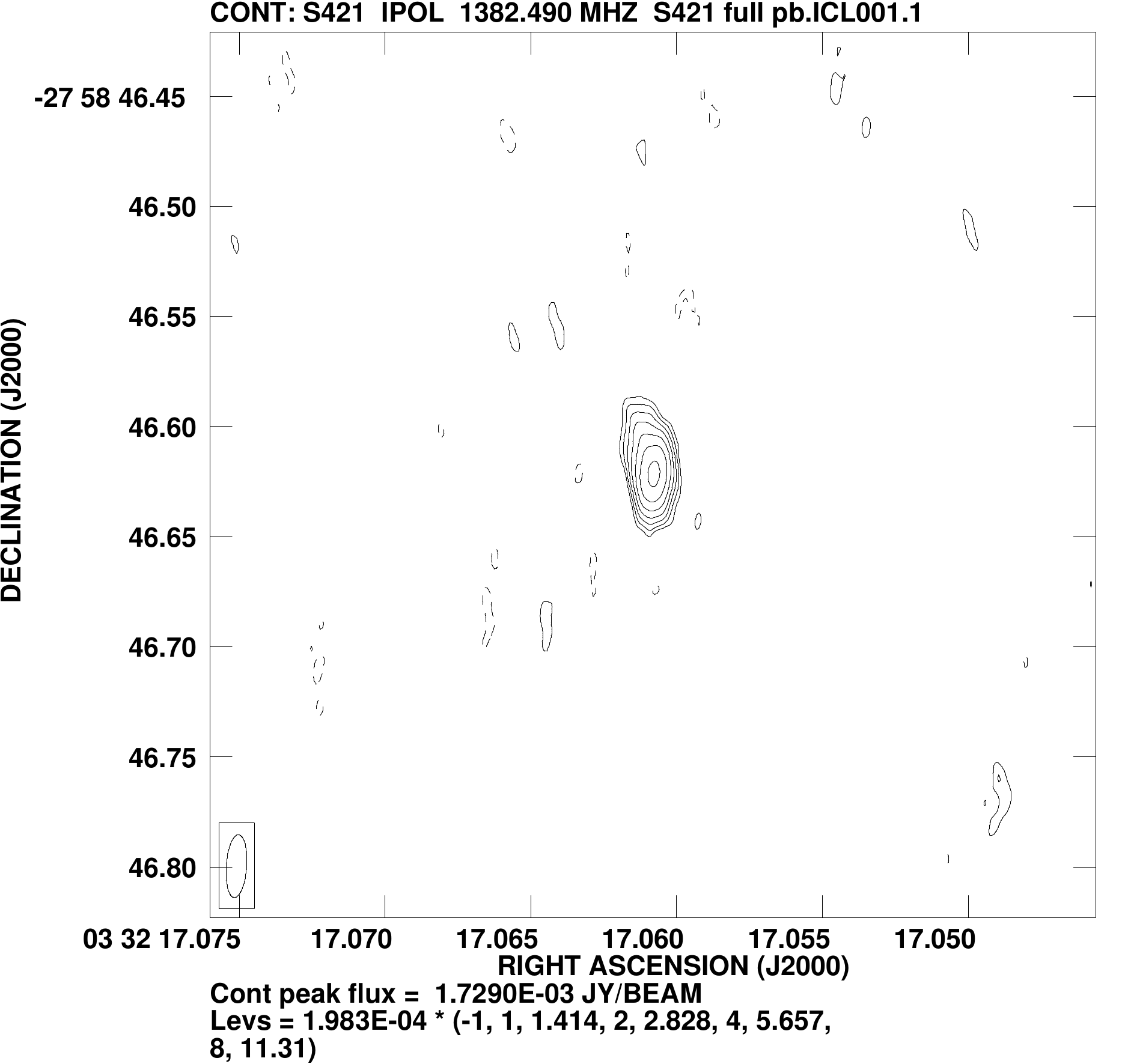}
\includegraphics[width=0.3\linewidth]{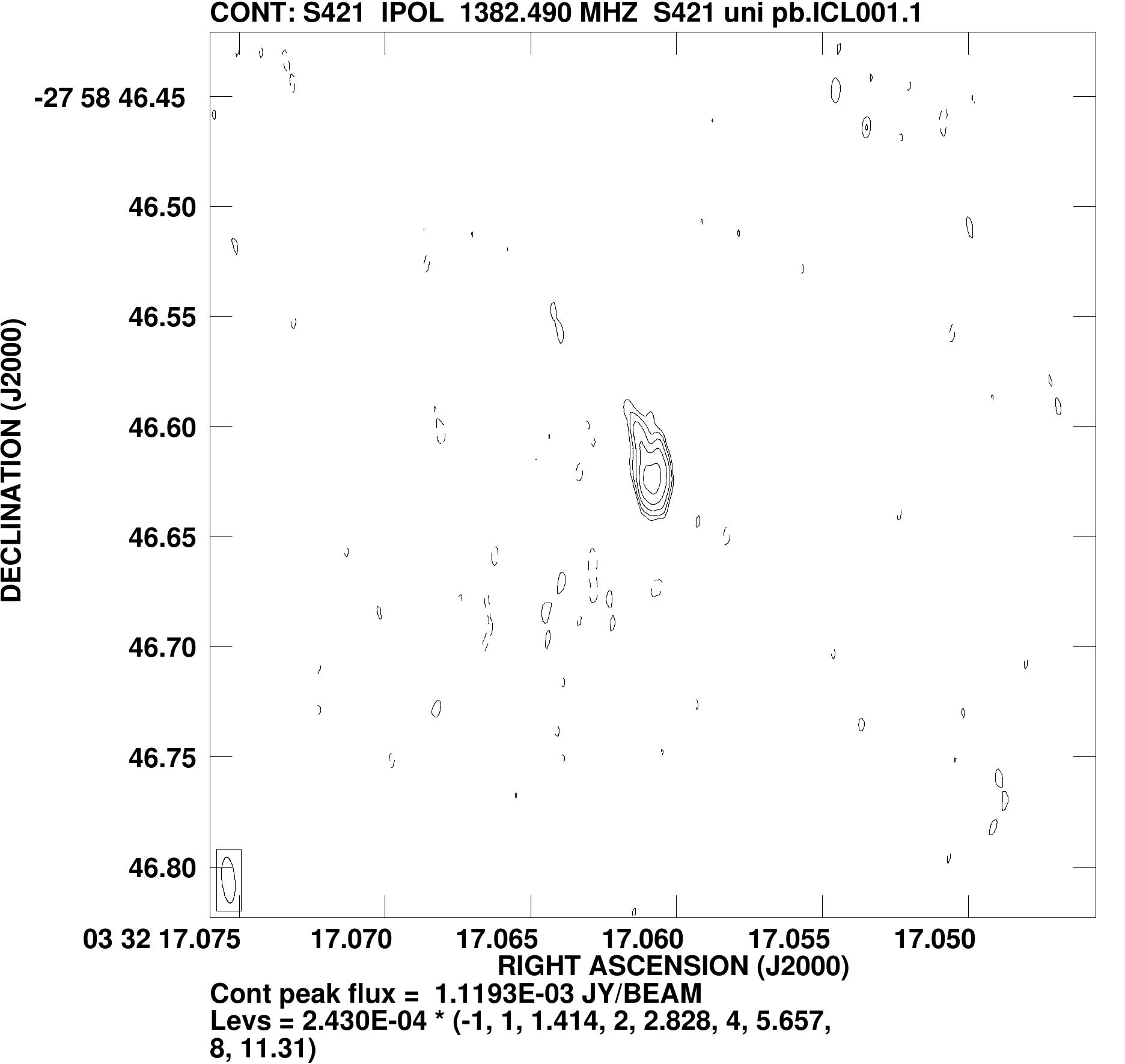} \\
\includegraphics[width=0.3\linewidth]{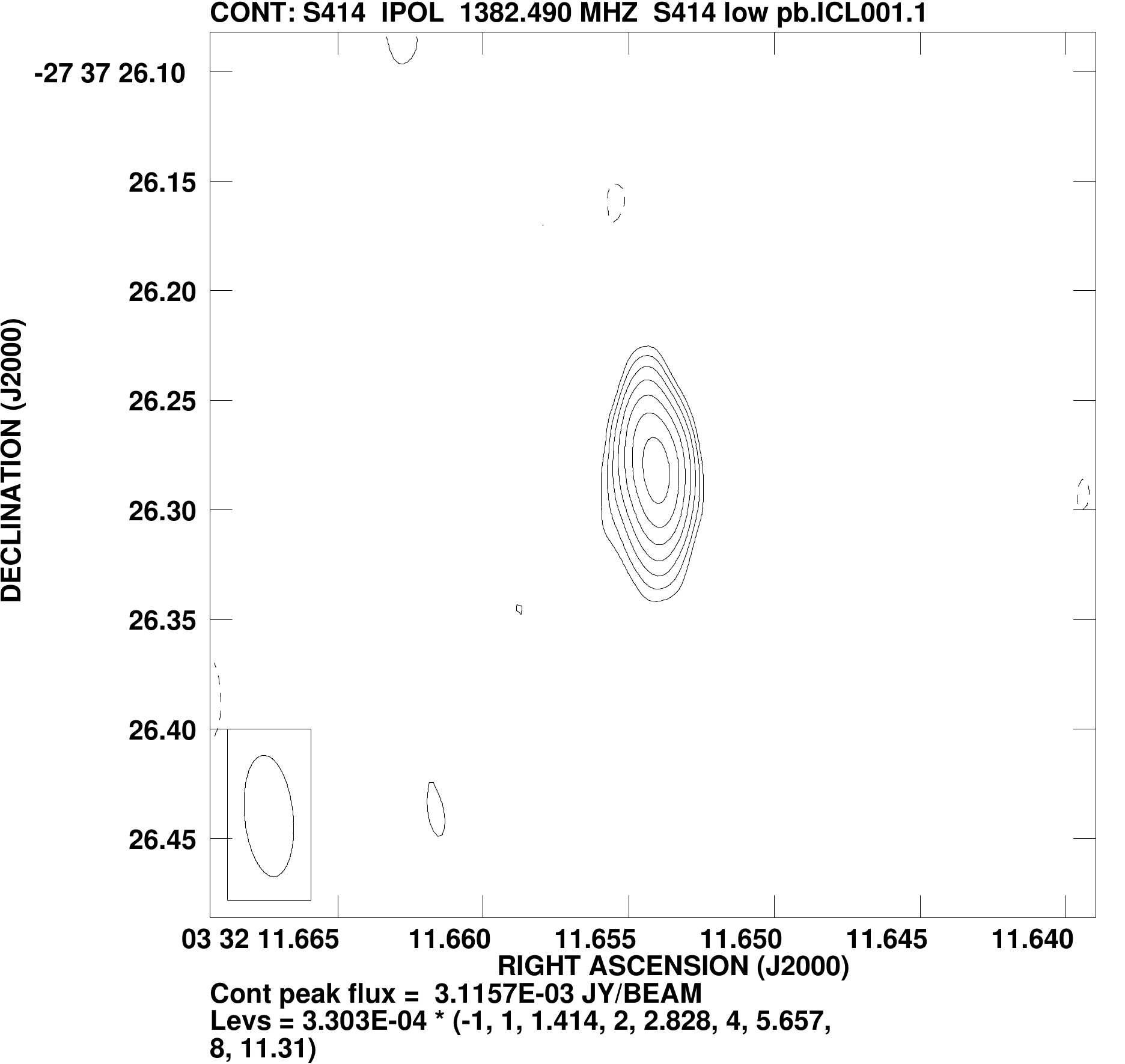} 
\includegraphics[width=0.3\linewidth]{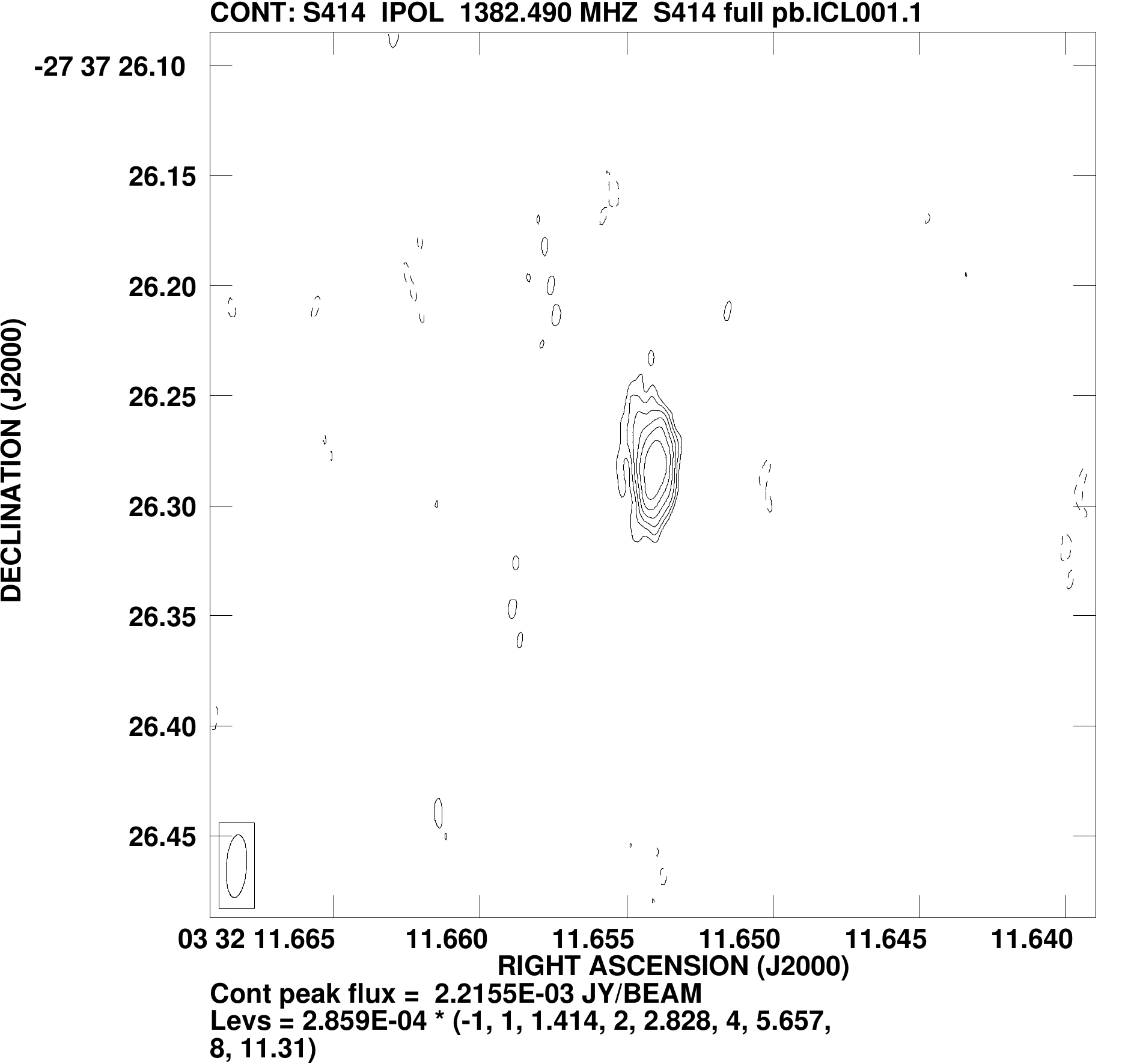}
\includegraphics[width=0.3\linewidth]{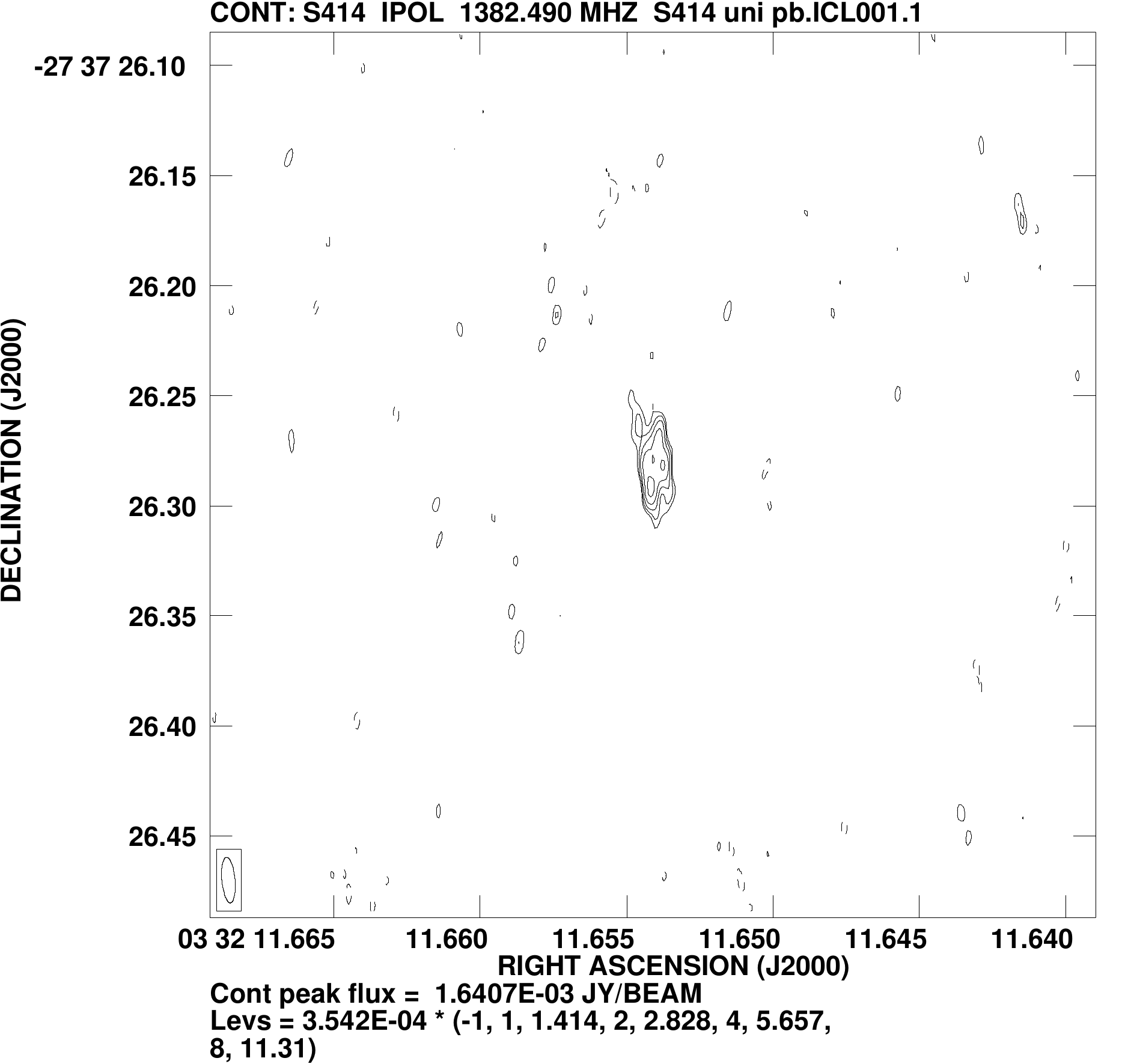} \\
\caption{Continued}
\end{figure*}

\addtocounter{figure}{-1}

\begin{figure*}
\includegraphics[width=0.3\linewidth]{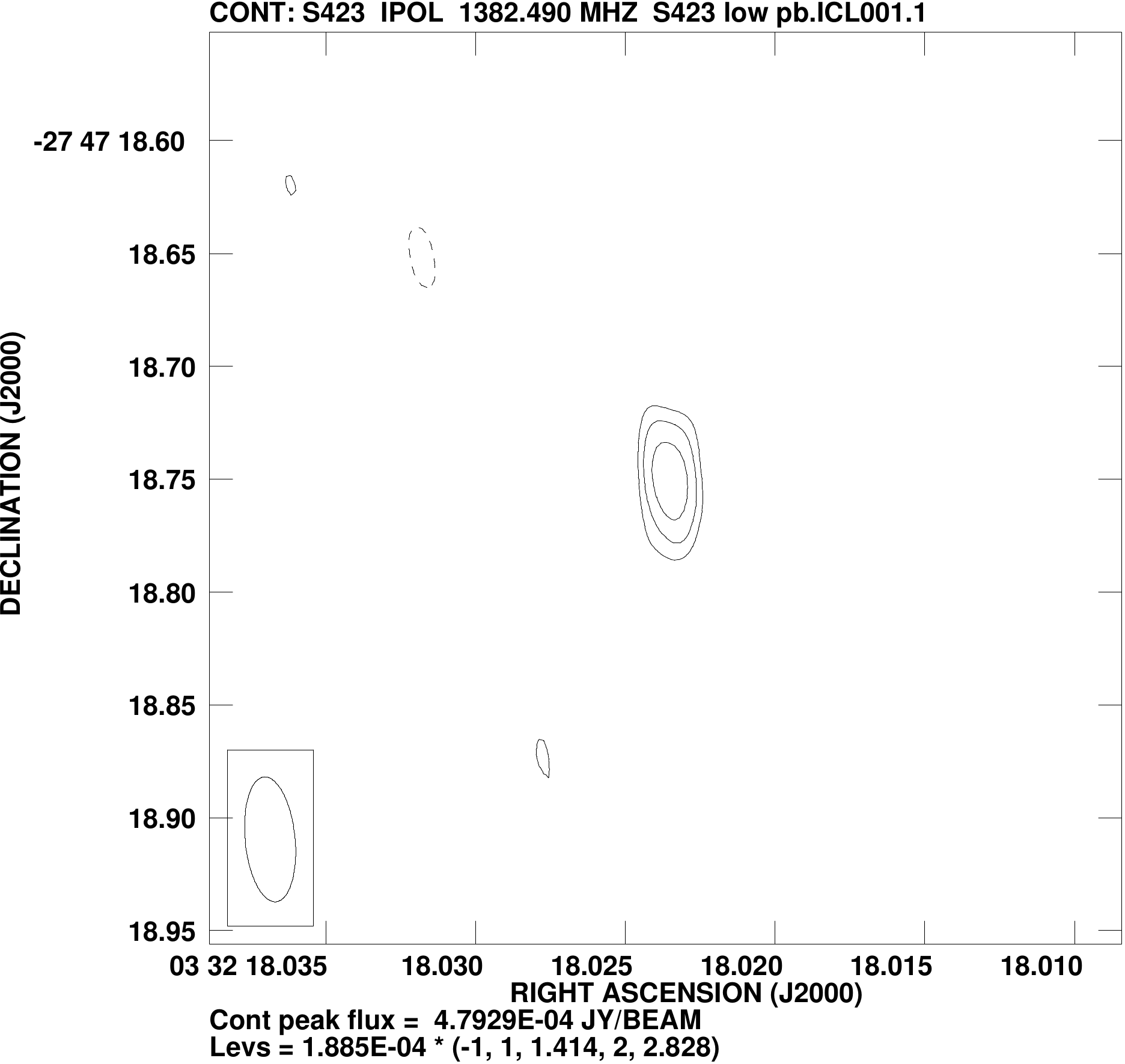} 
\includegraphics[width=0.3\linewidth]{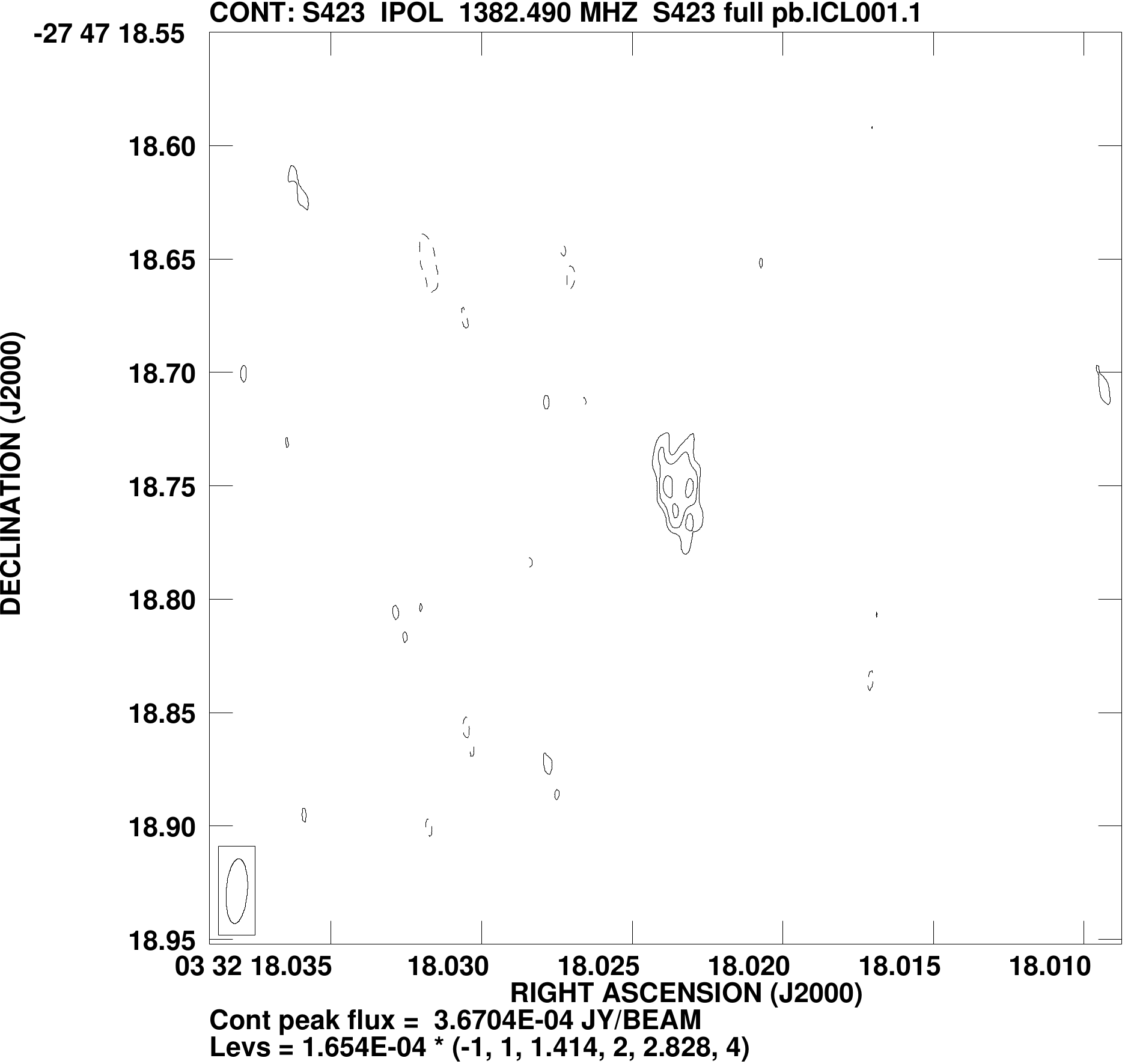}
\includegraphics[width=0.3\linewidth]{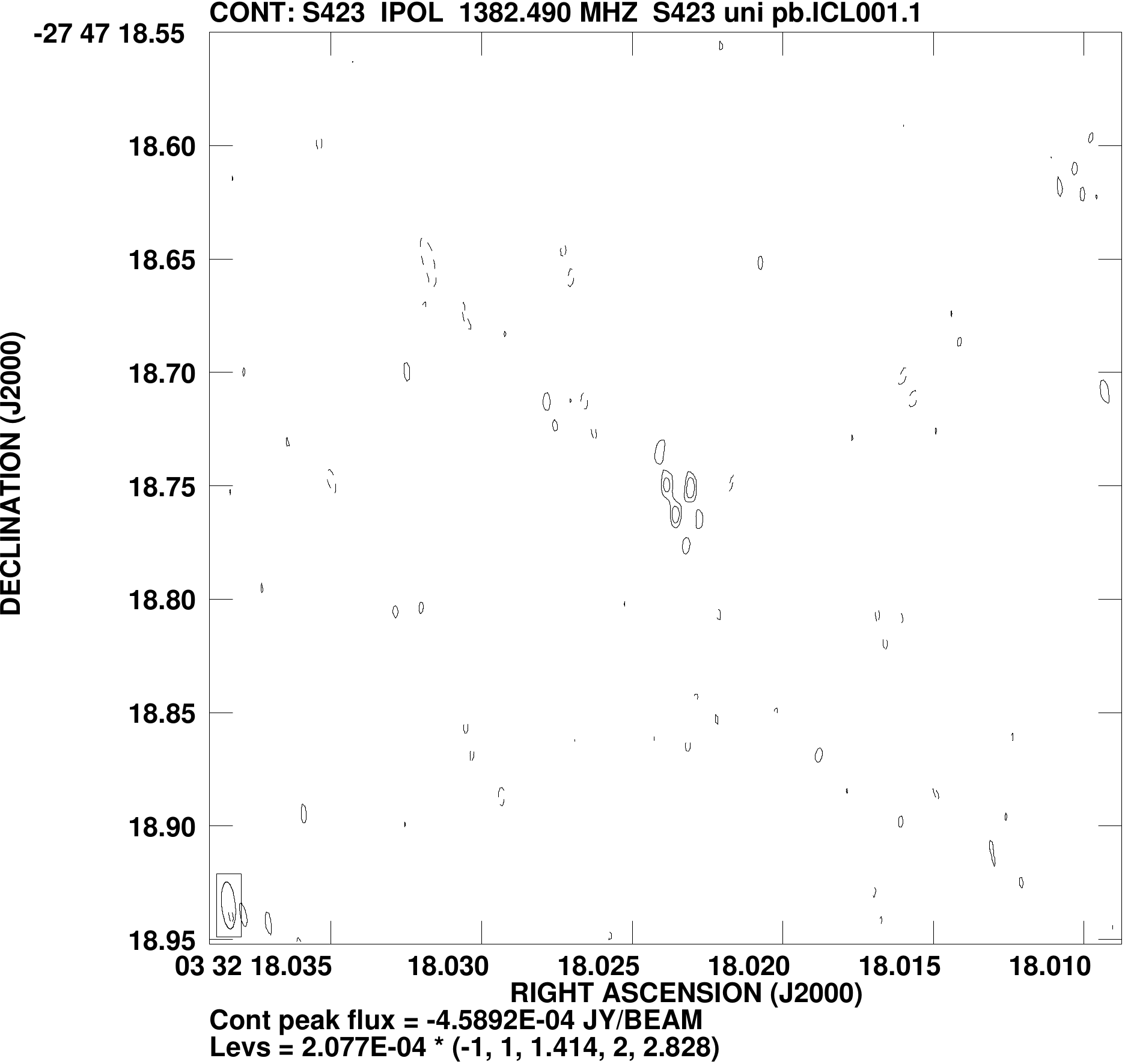} \\
\includegraphics[width=0.3\linewidth]{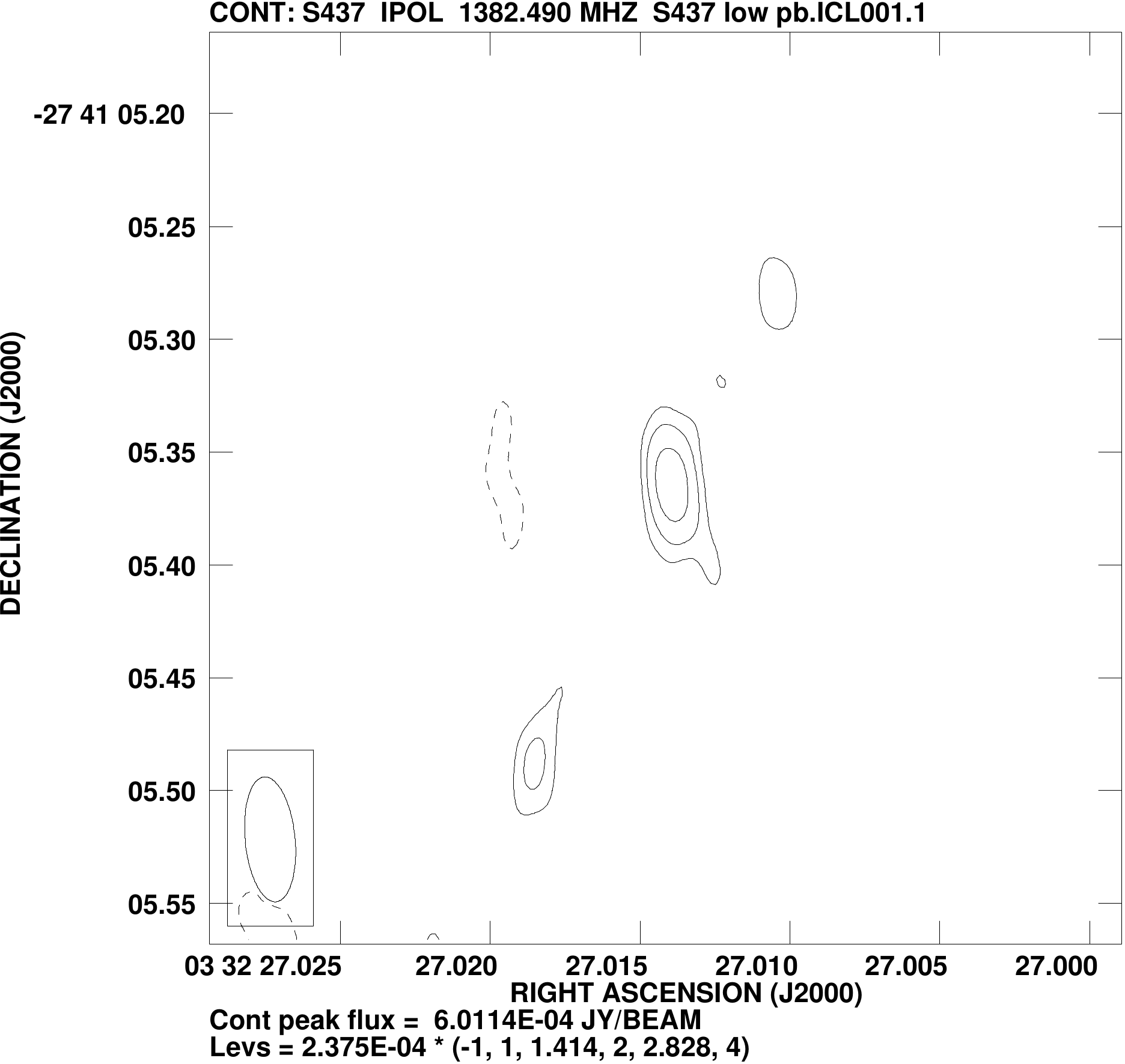} 
\includegraphics[width=0.3\linewidth]{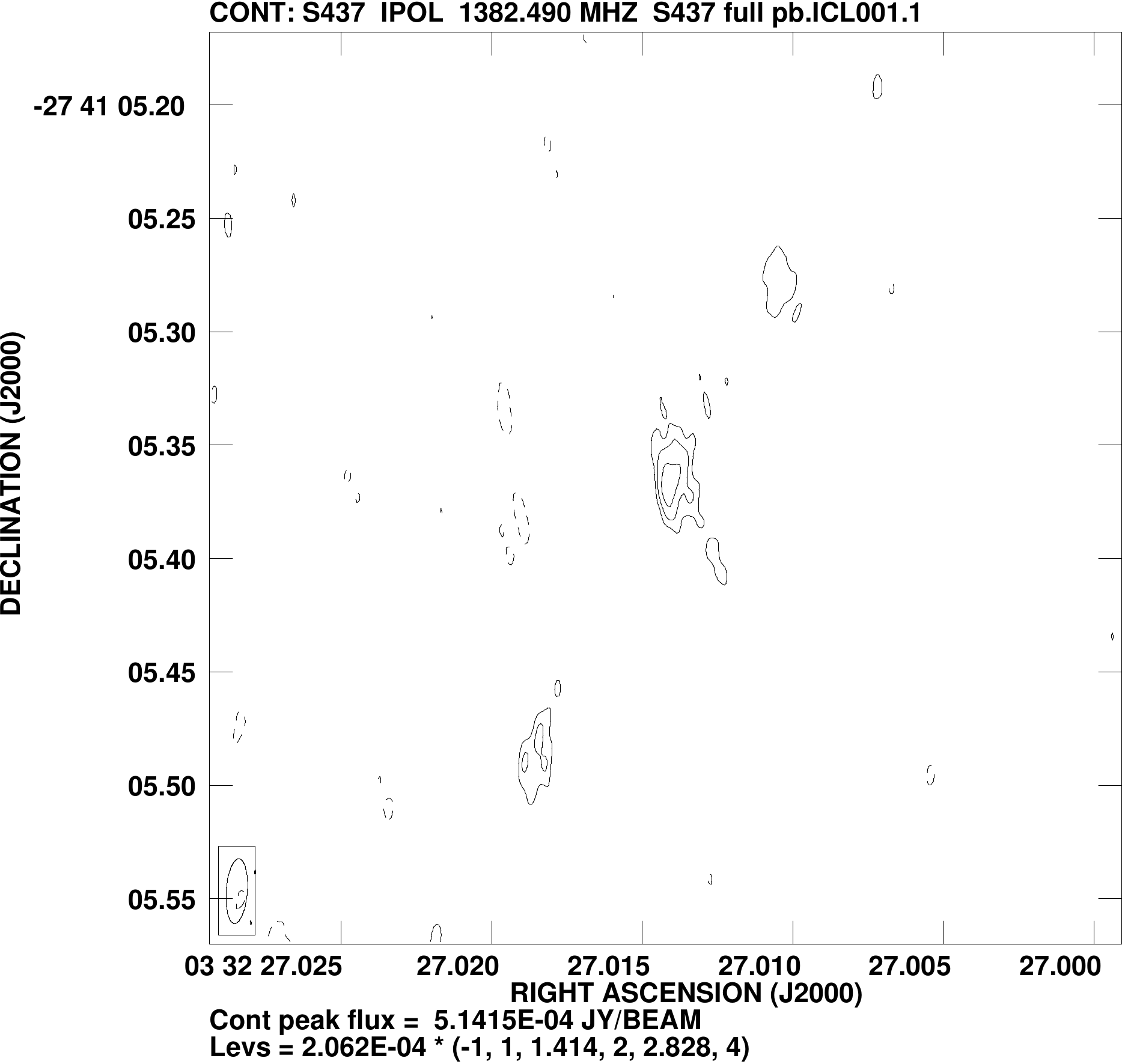}
\includegraphics[width=0.3\linewidth]{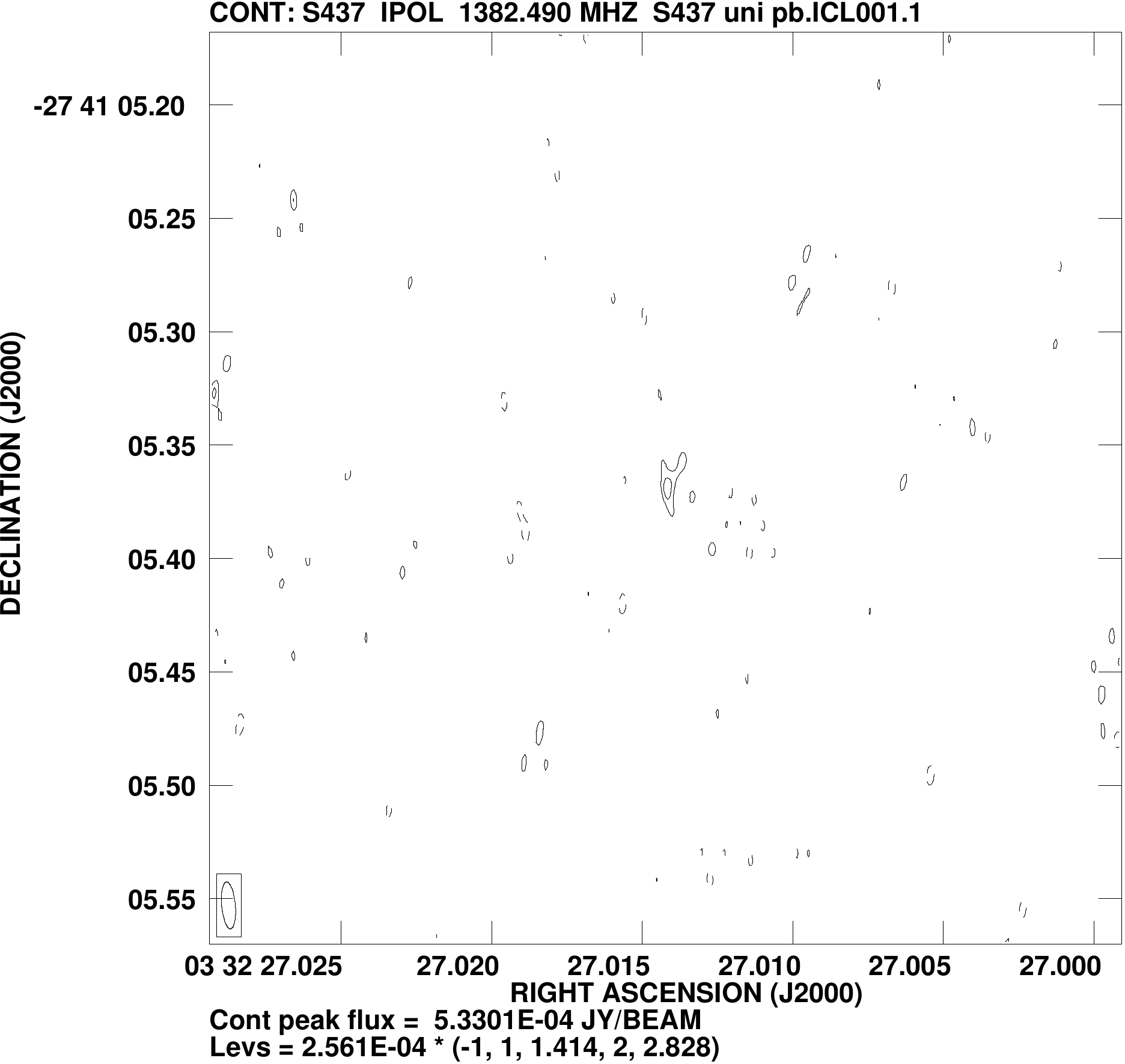} \\
\includegraphics[width=0.3\linewidth]{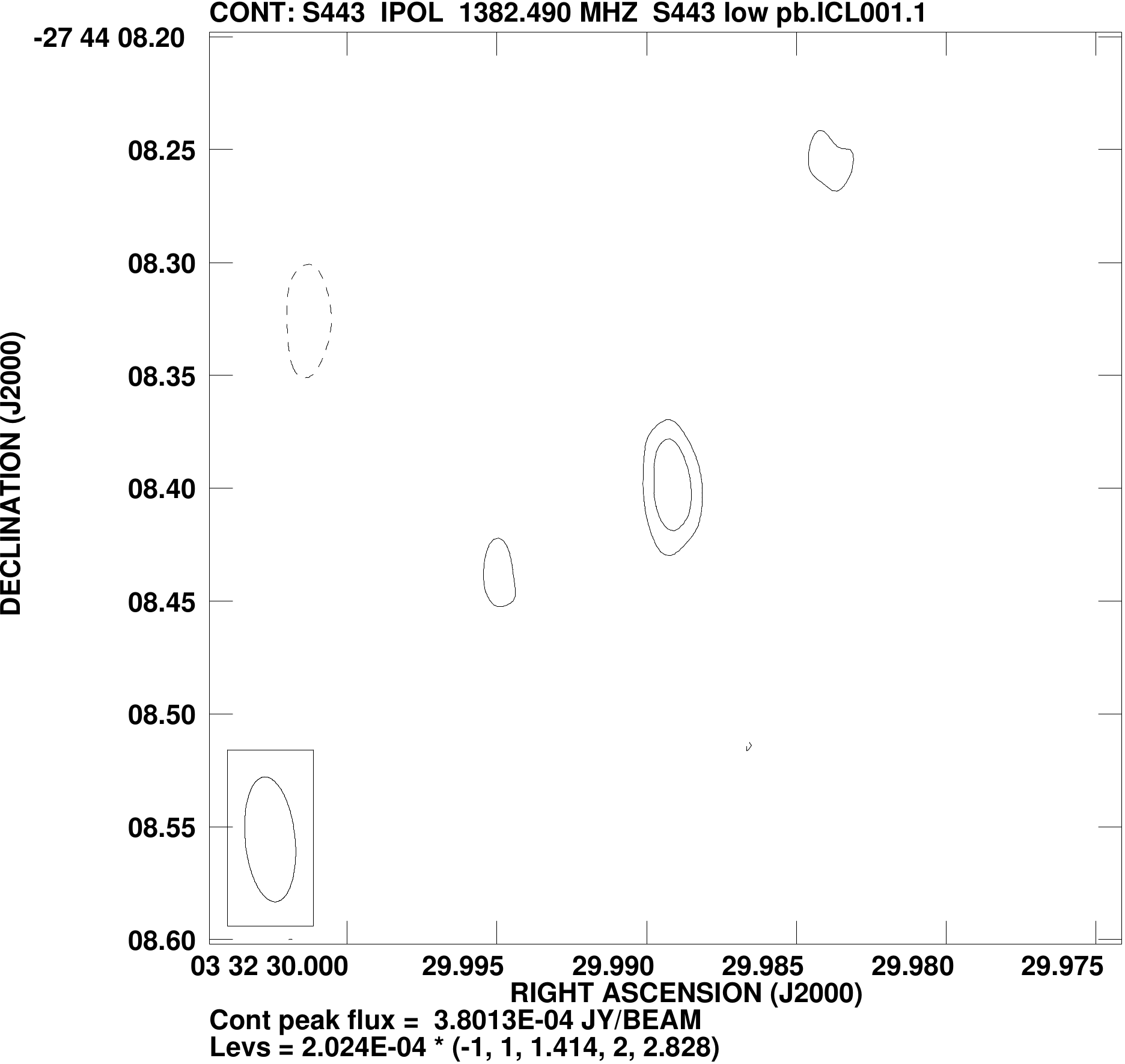} 
\includegraphics[width=0.3\linewidth]{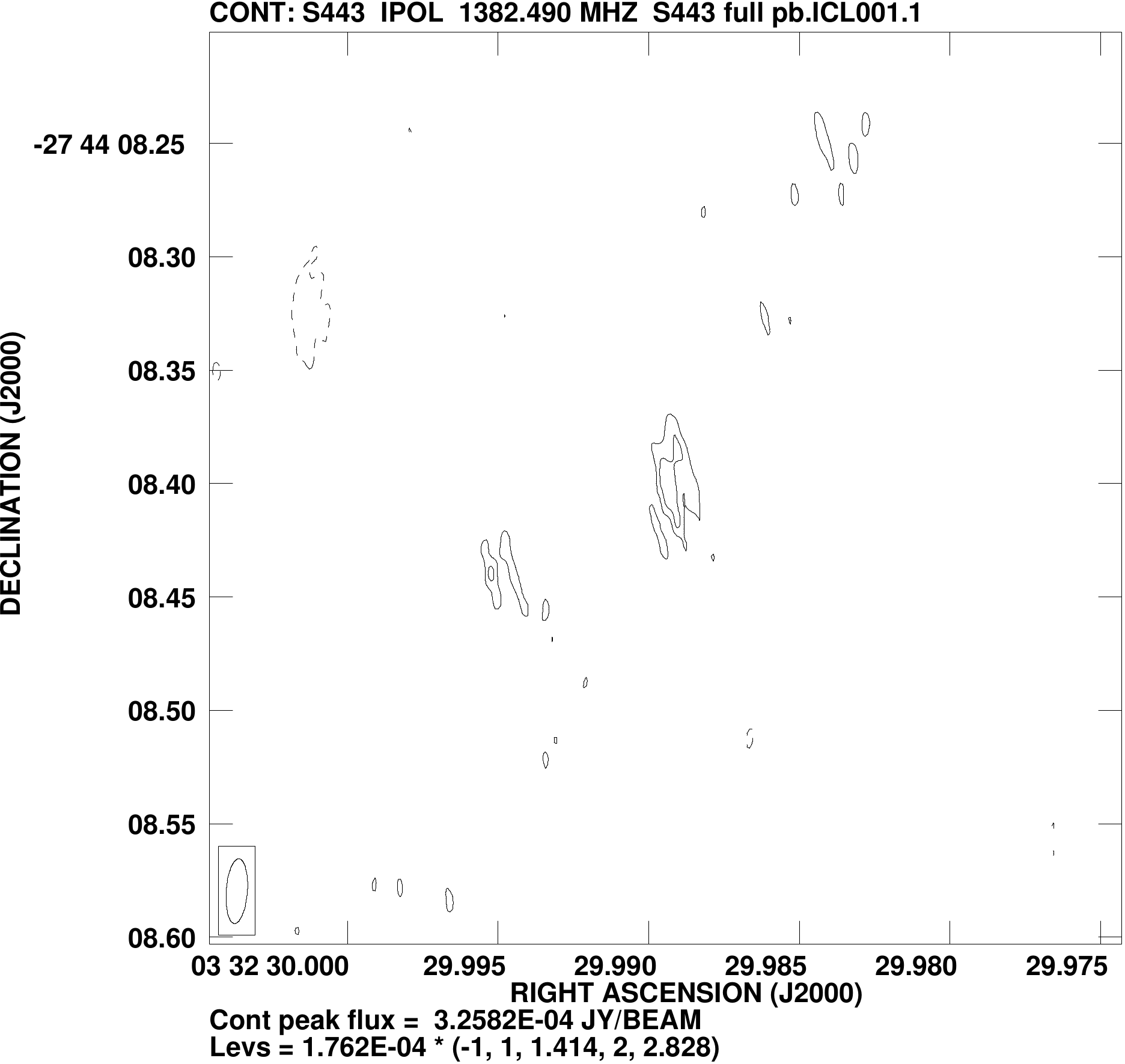}
\includegraphics[width=0.3\linewidth]{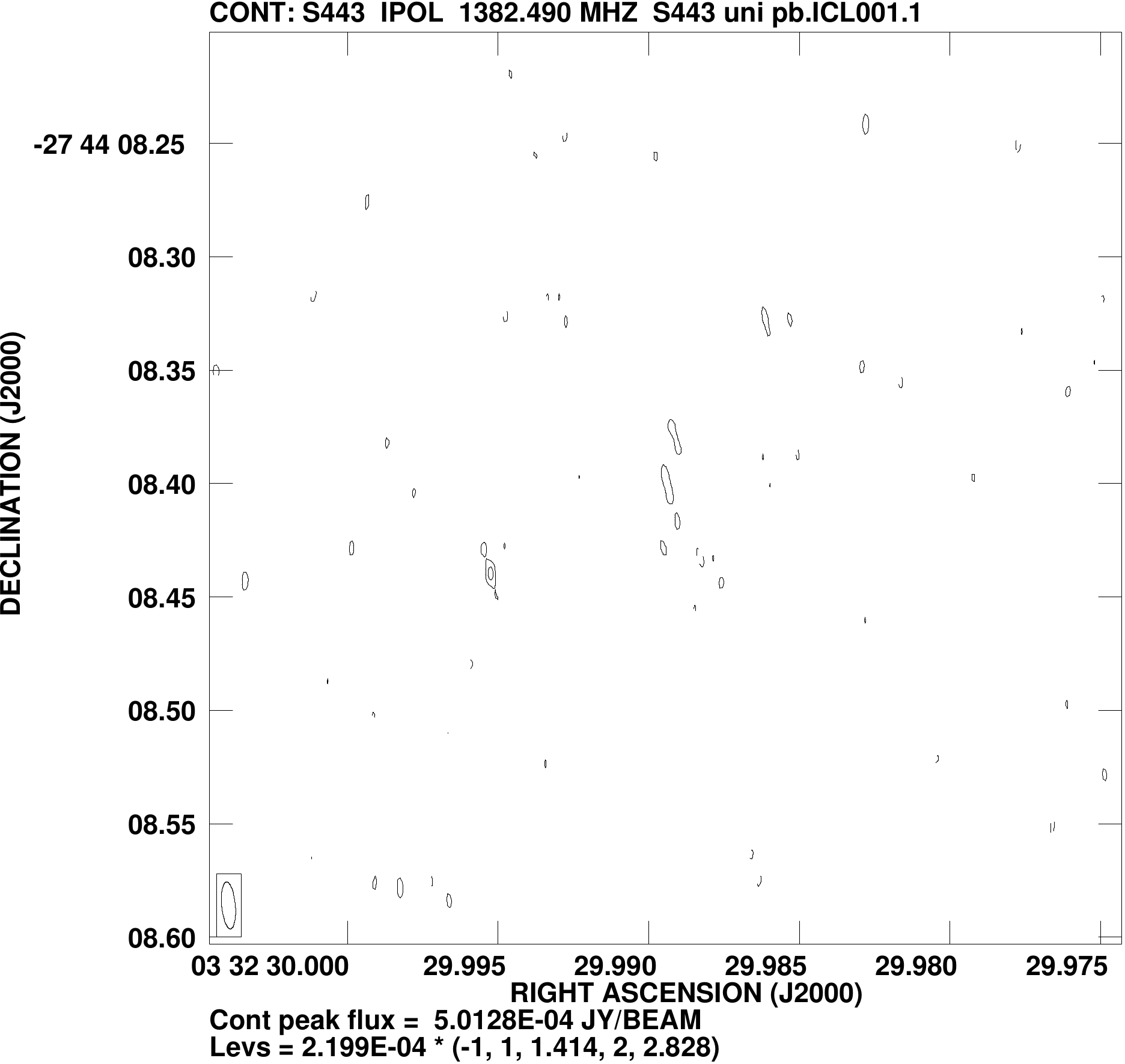} \\
\includegraphics[width=0.3\linewidth]{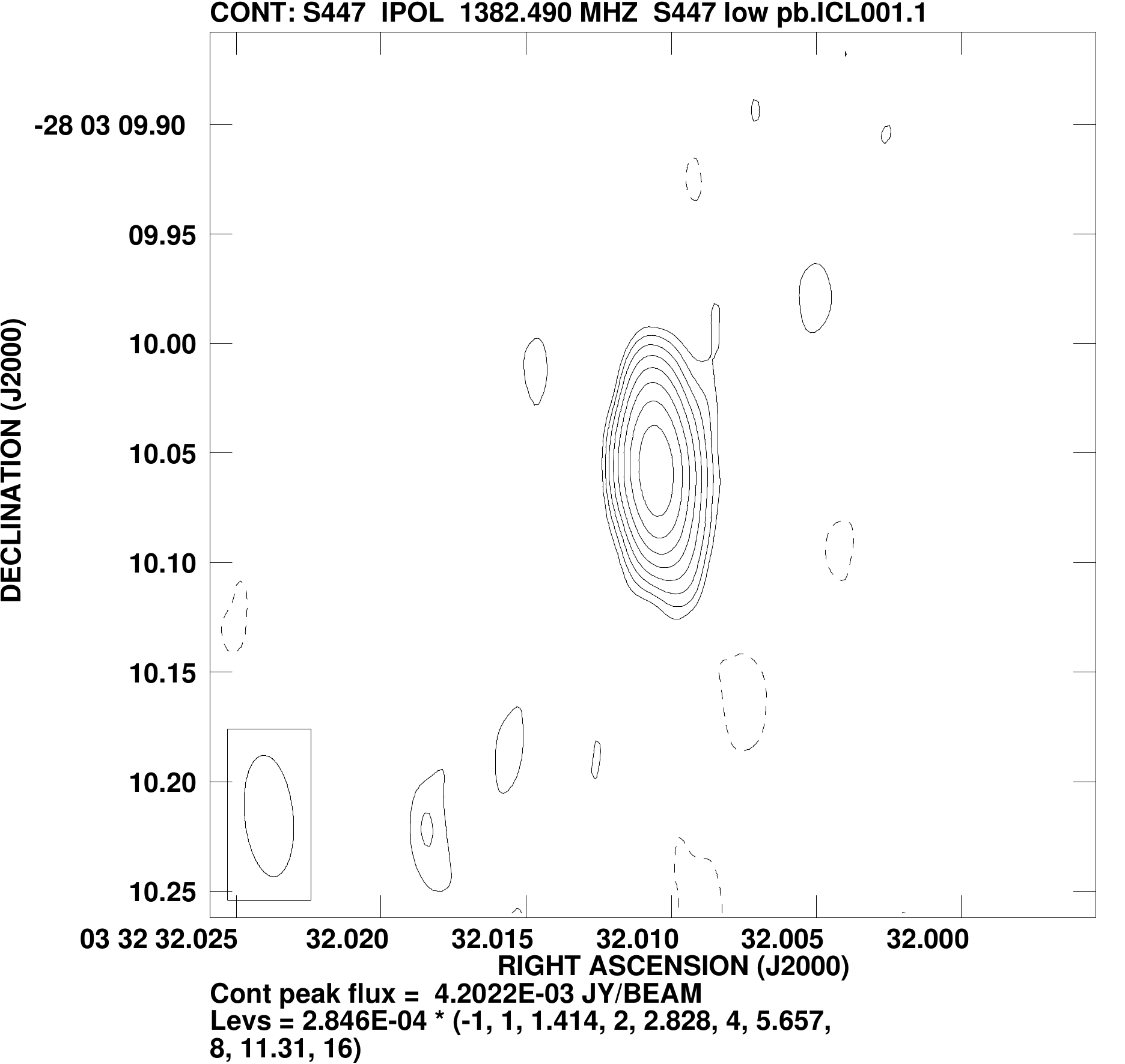} 
\includegraphics[width=0.3\linewidth]{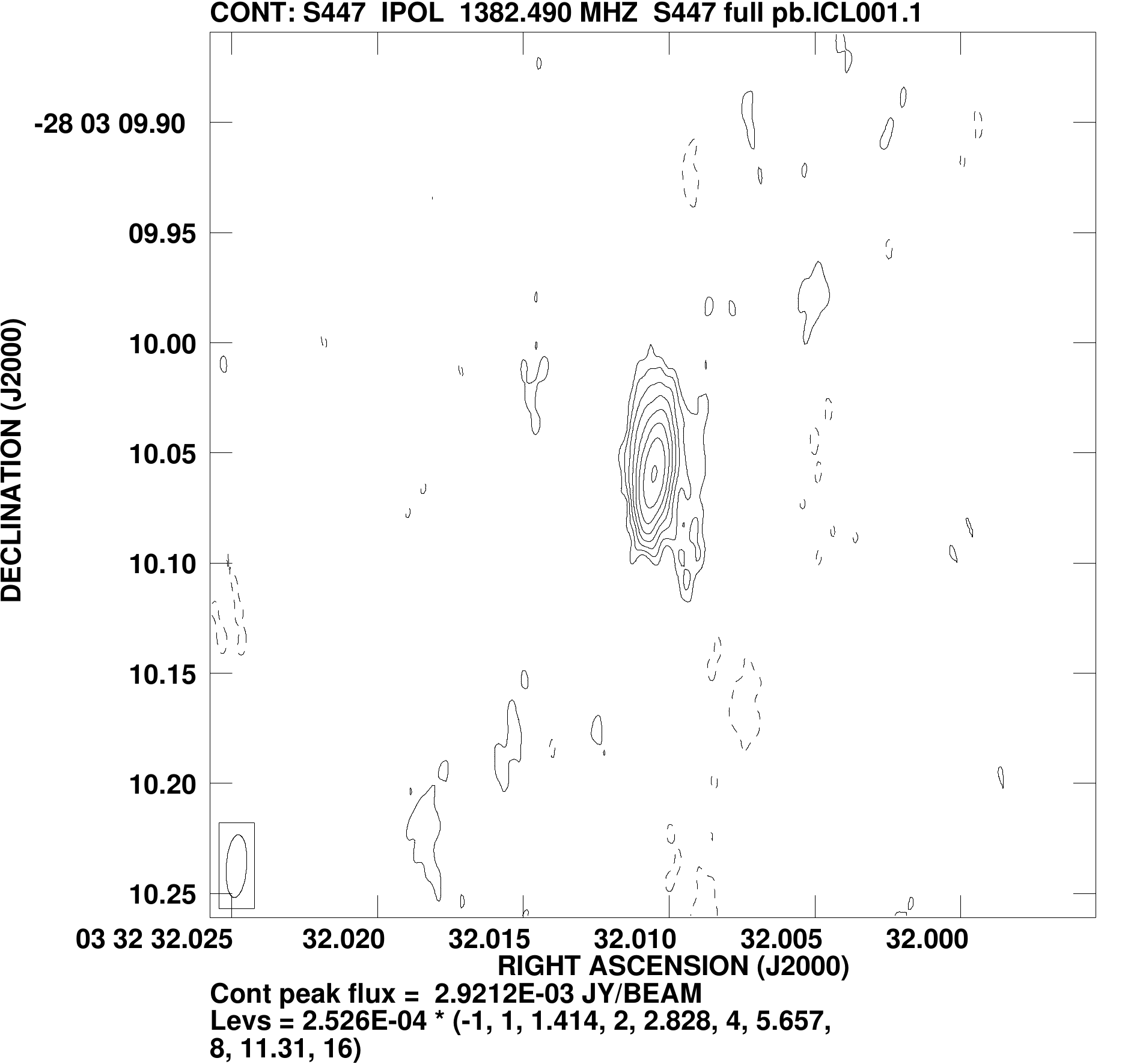}
\includegraphics[width=0.3\linewidth]{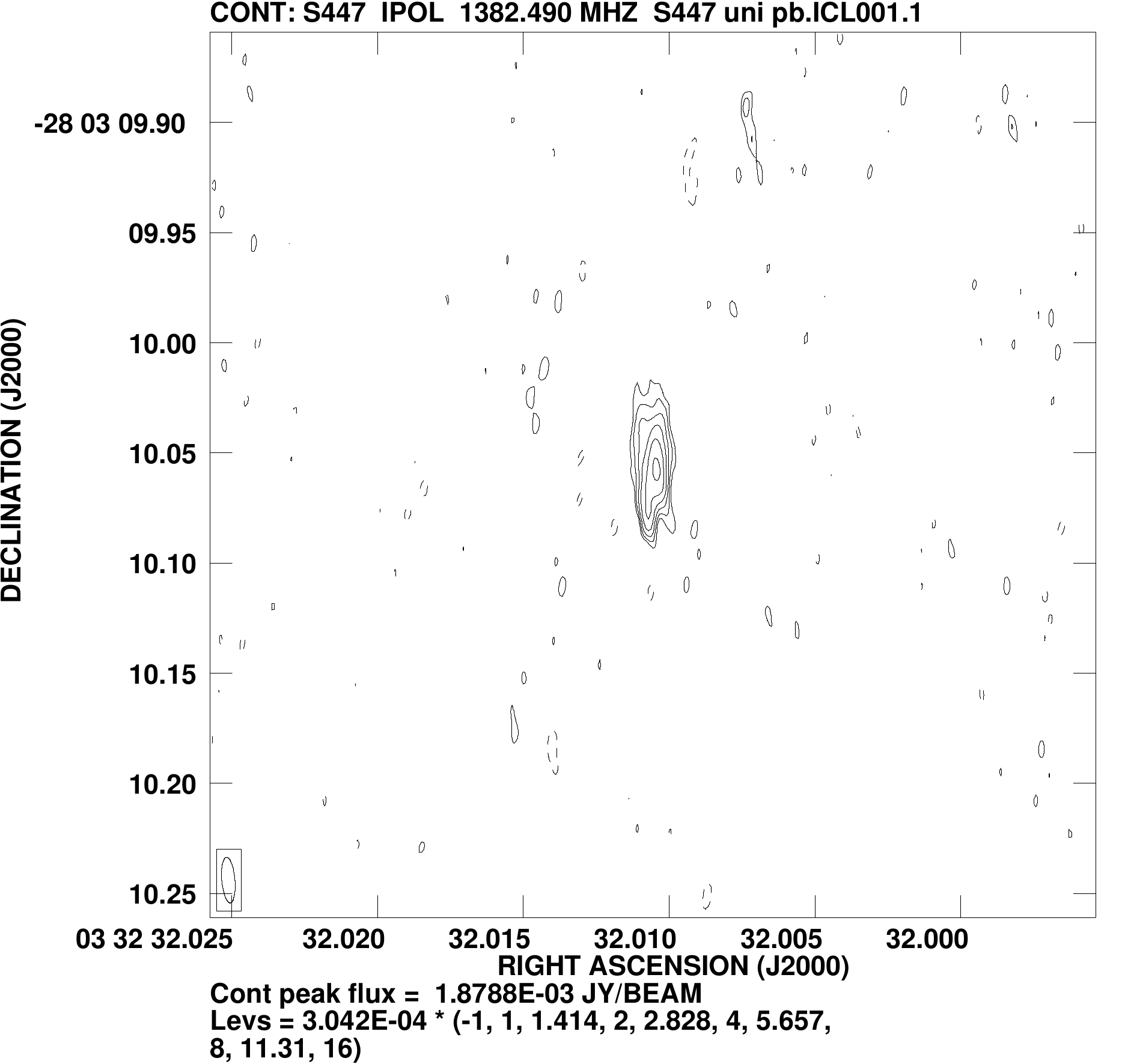} \\
\caption{Continued}
\end{figure*}

\addtocounter{figure}{-1}

\begin{figure*}
\includegraphics[width=0.3\linewidth]{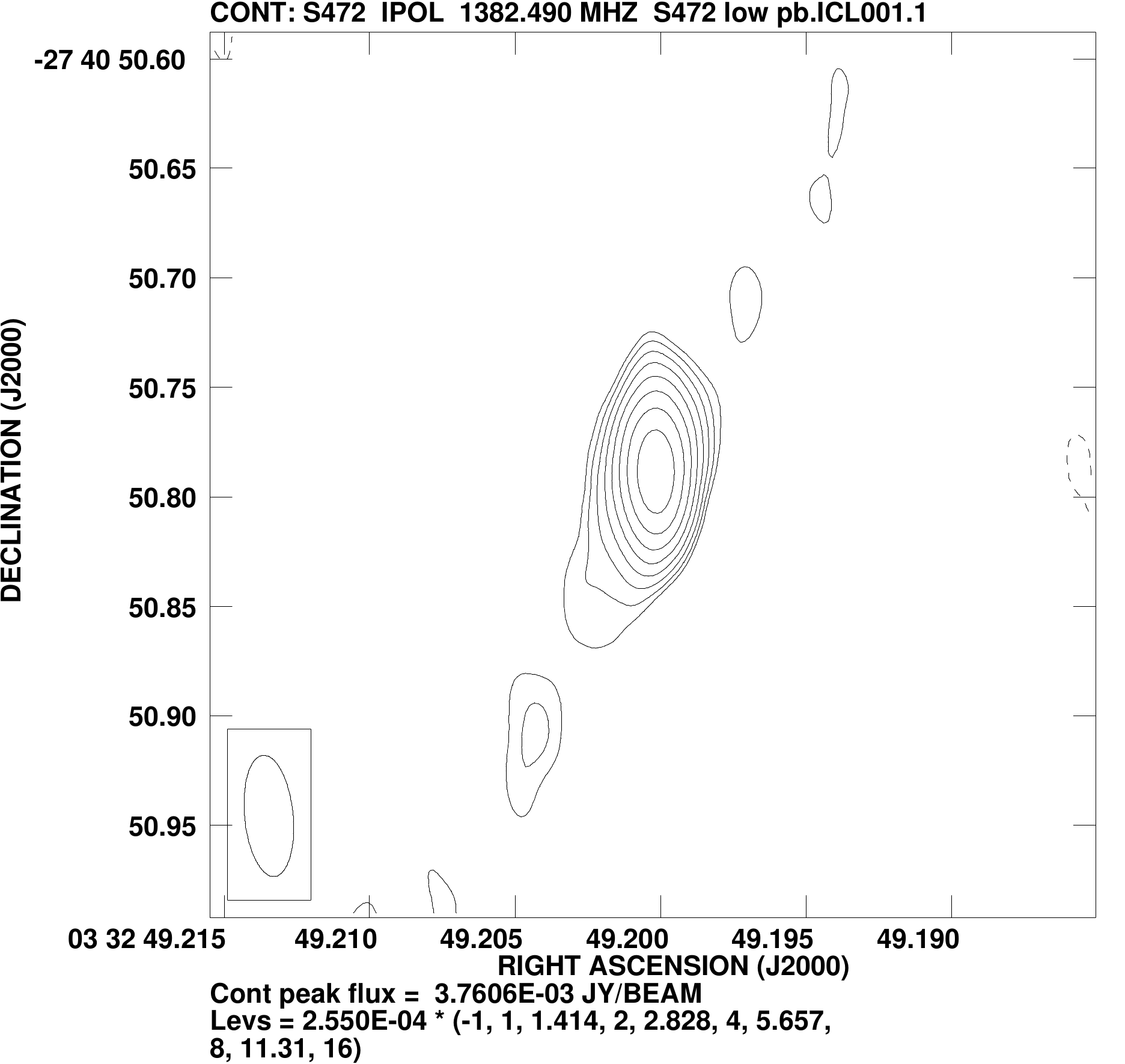} 
\includegraphics[width=0.3\linewidth]{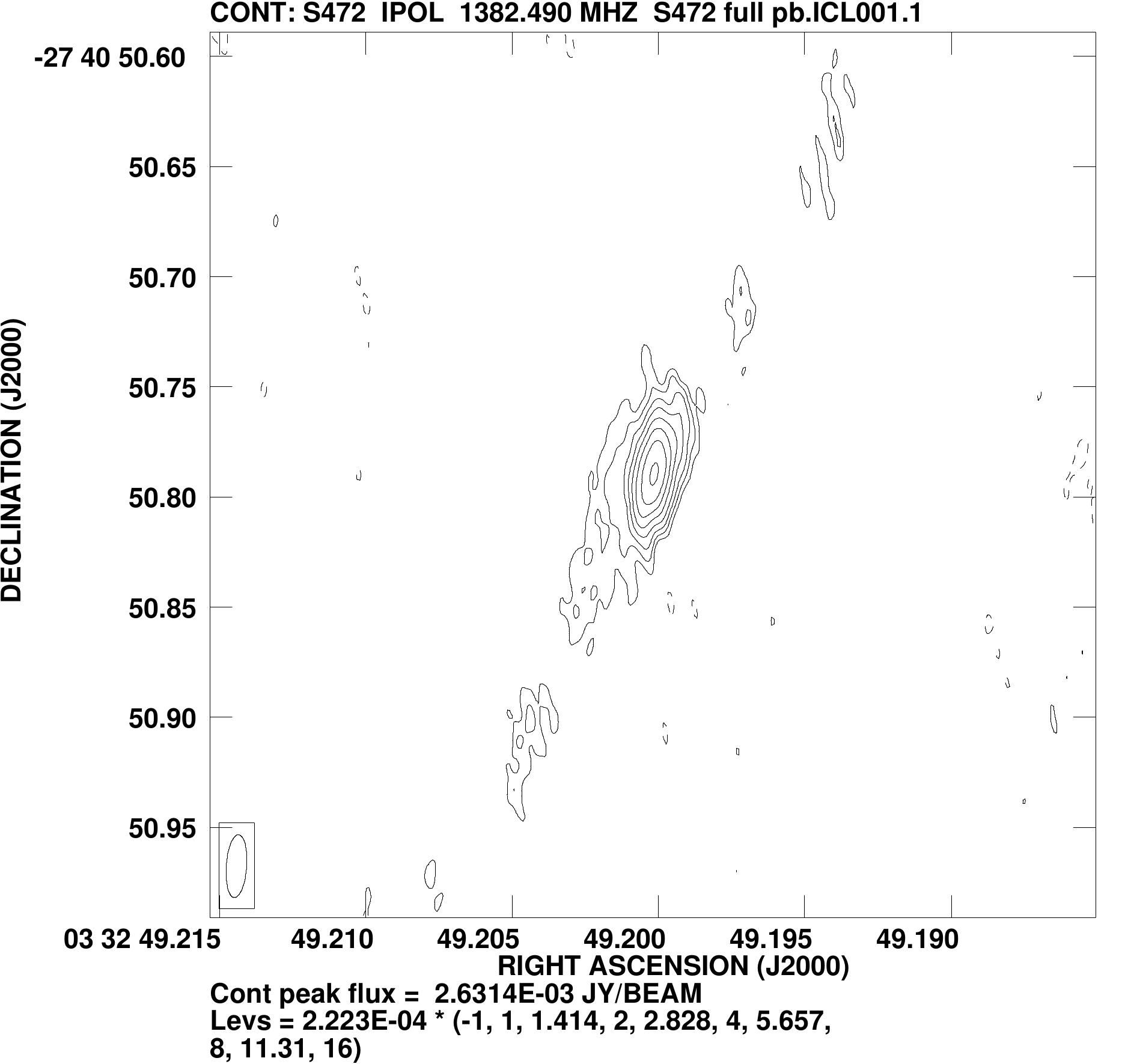}
\includegraphics[width=0.3\linewidth]{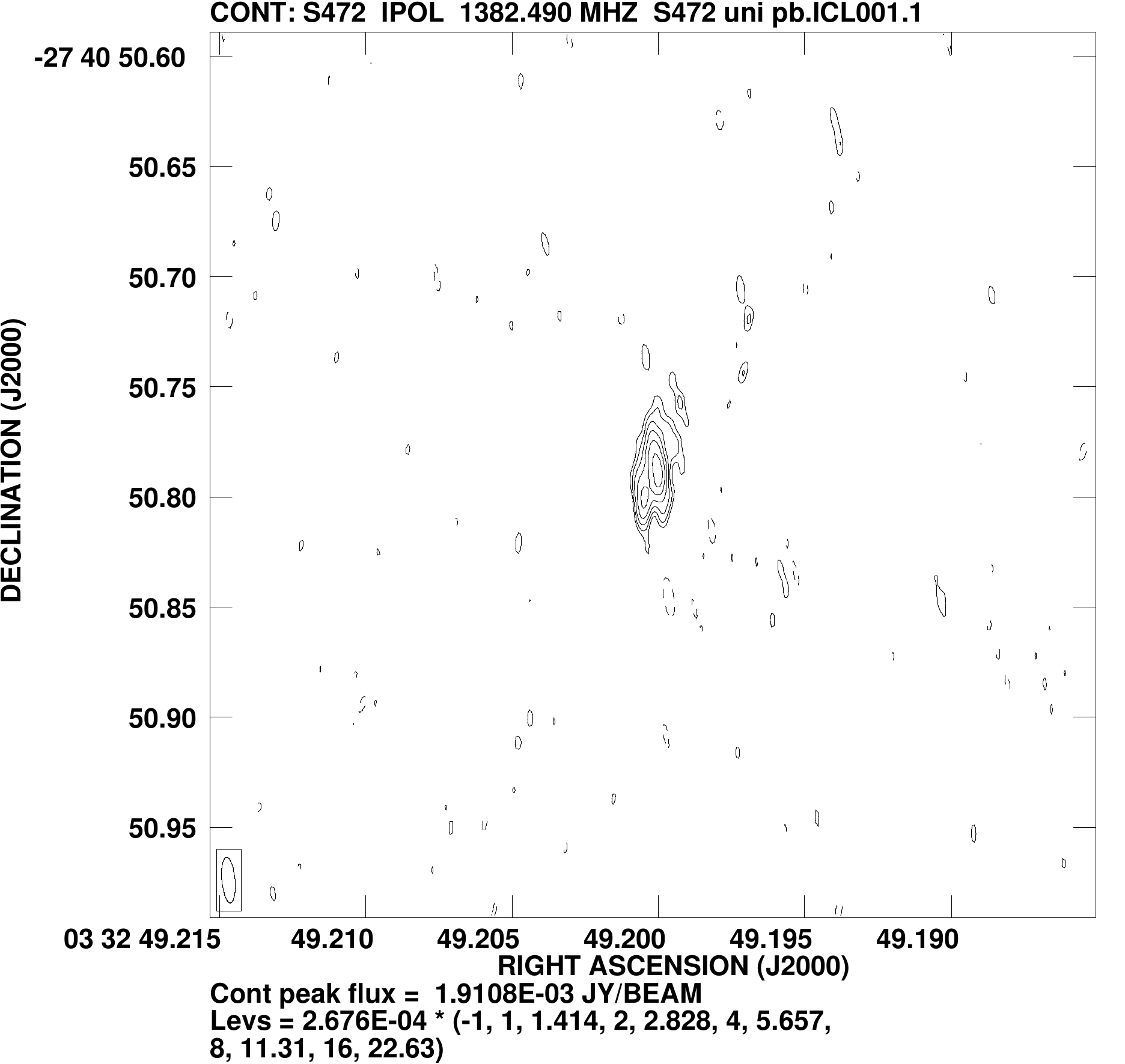} \\
\includegraphics[width=0.3\linewidth]{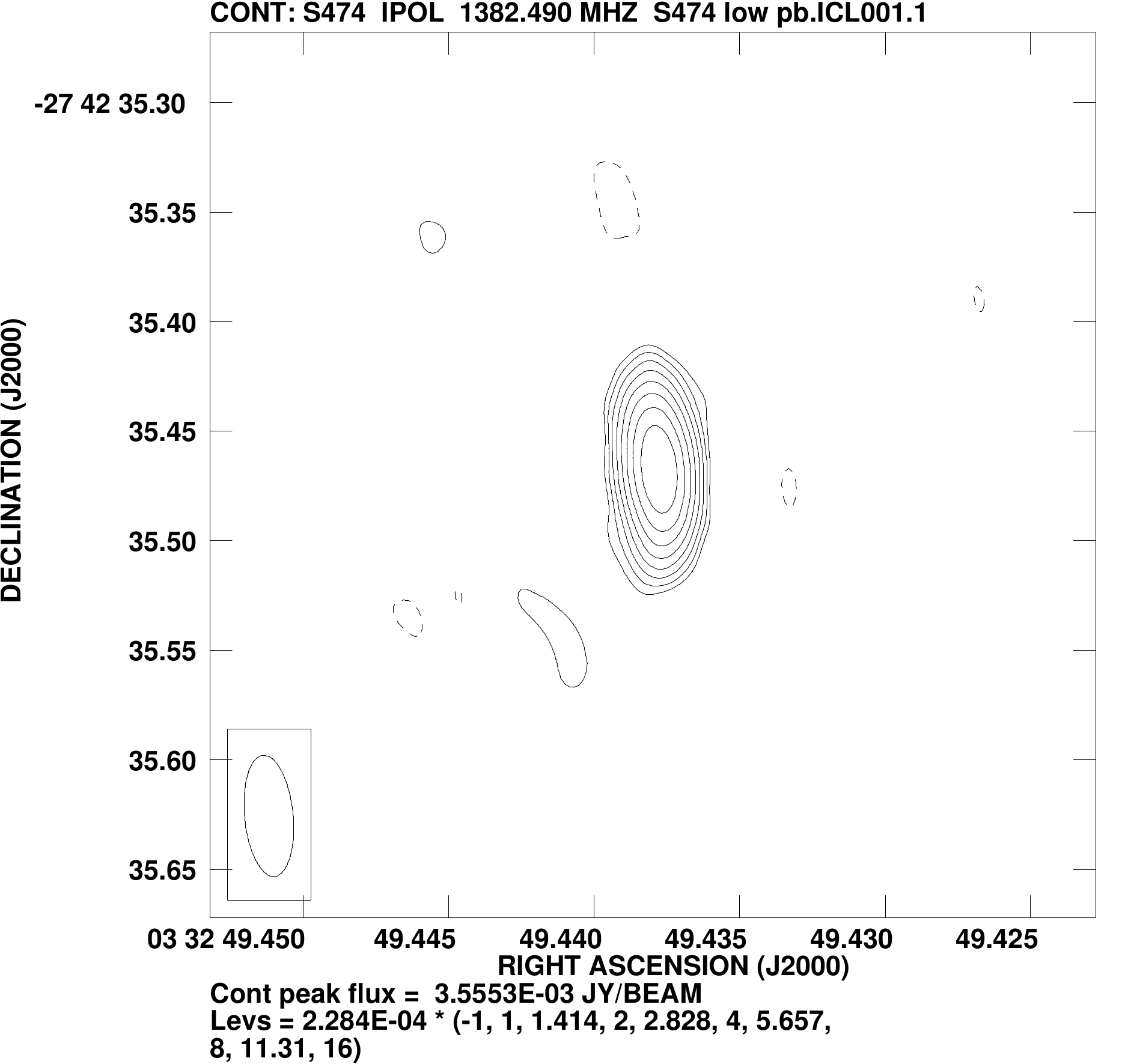} 
\includegraphics[width=0.3\linewidth]{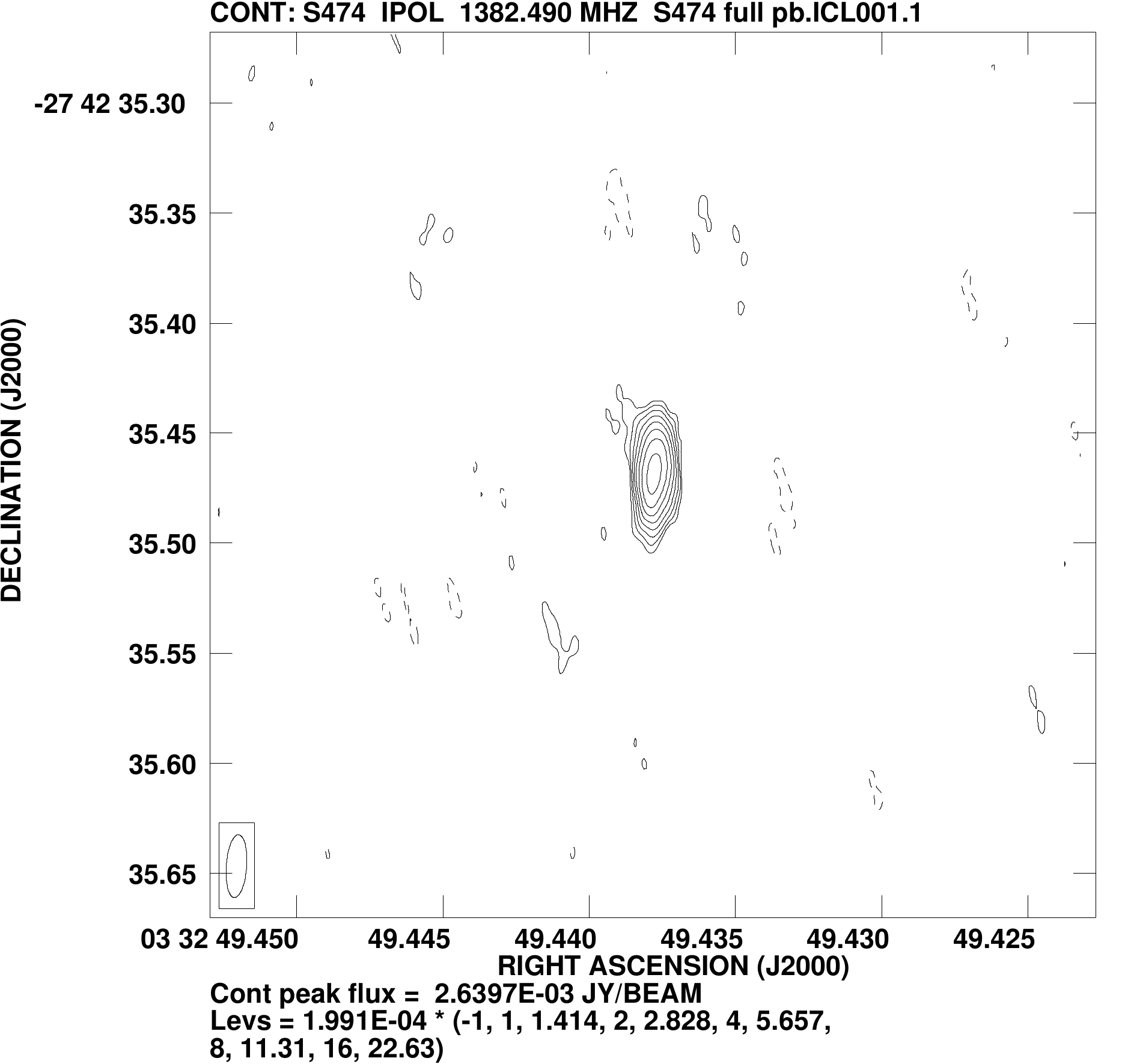}
\includegraphics[width=0.3\linewidth]{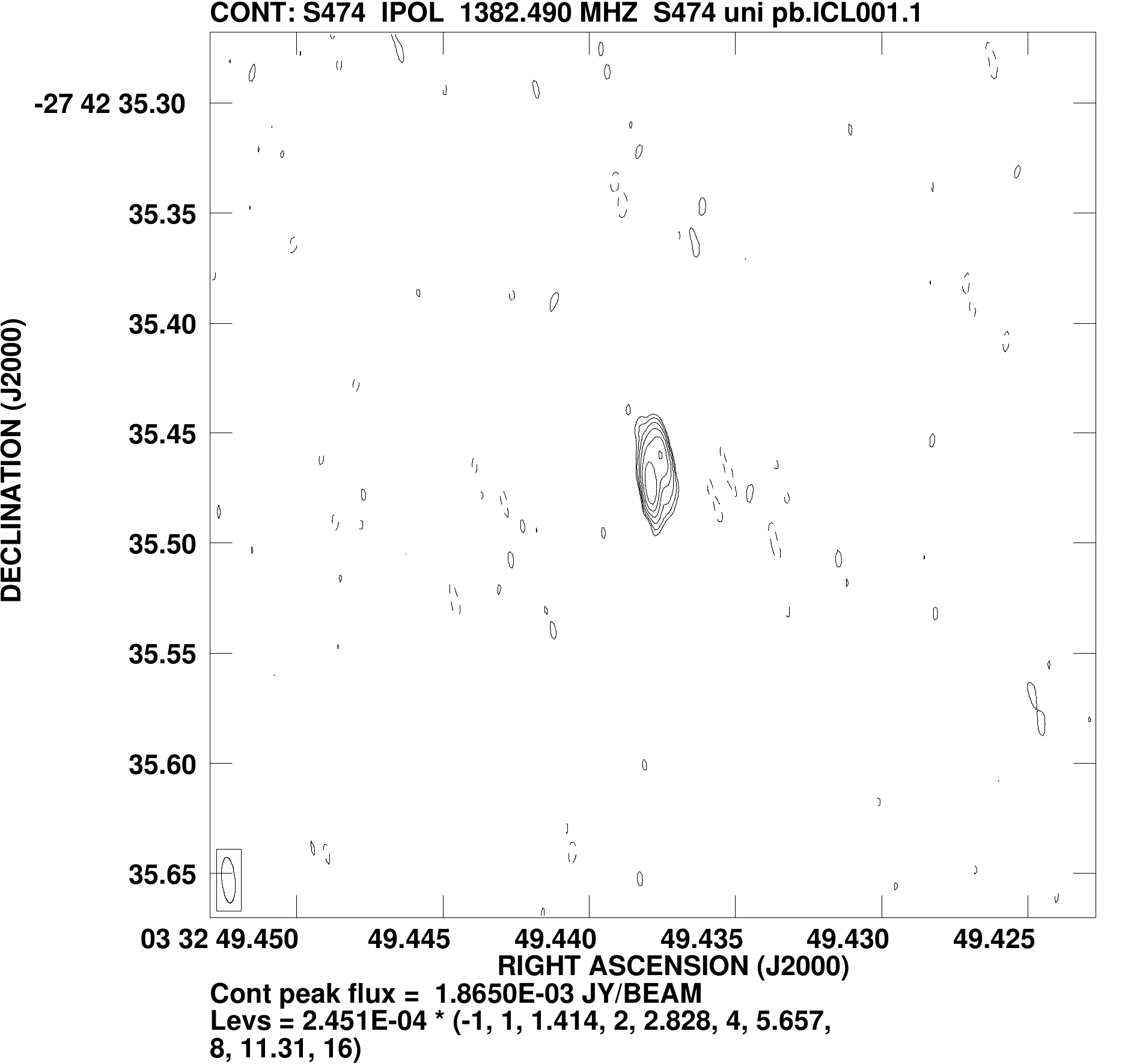} \\
\includegraphics[width=0.3\linewidth]{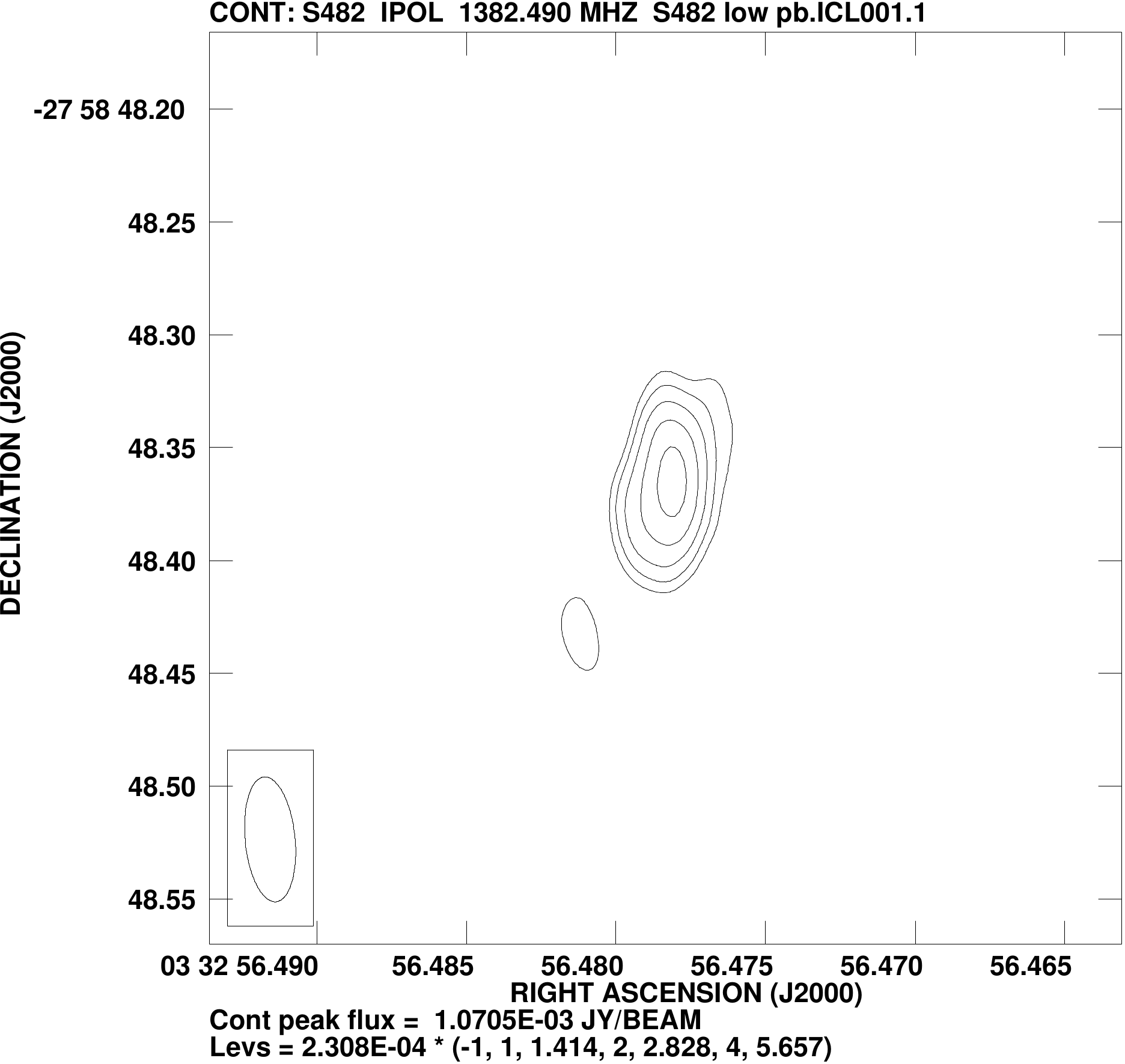} 
\includegraphics[width=0.3\linewidth]{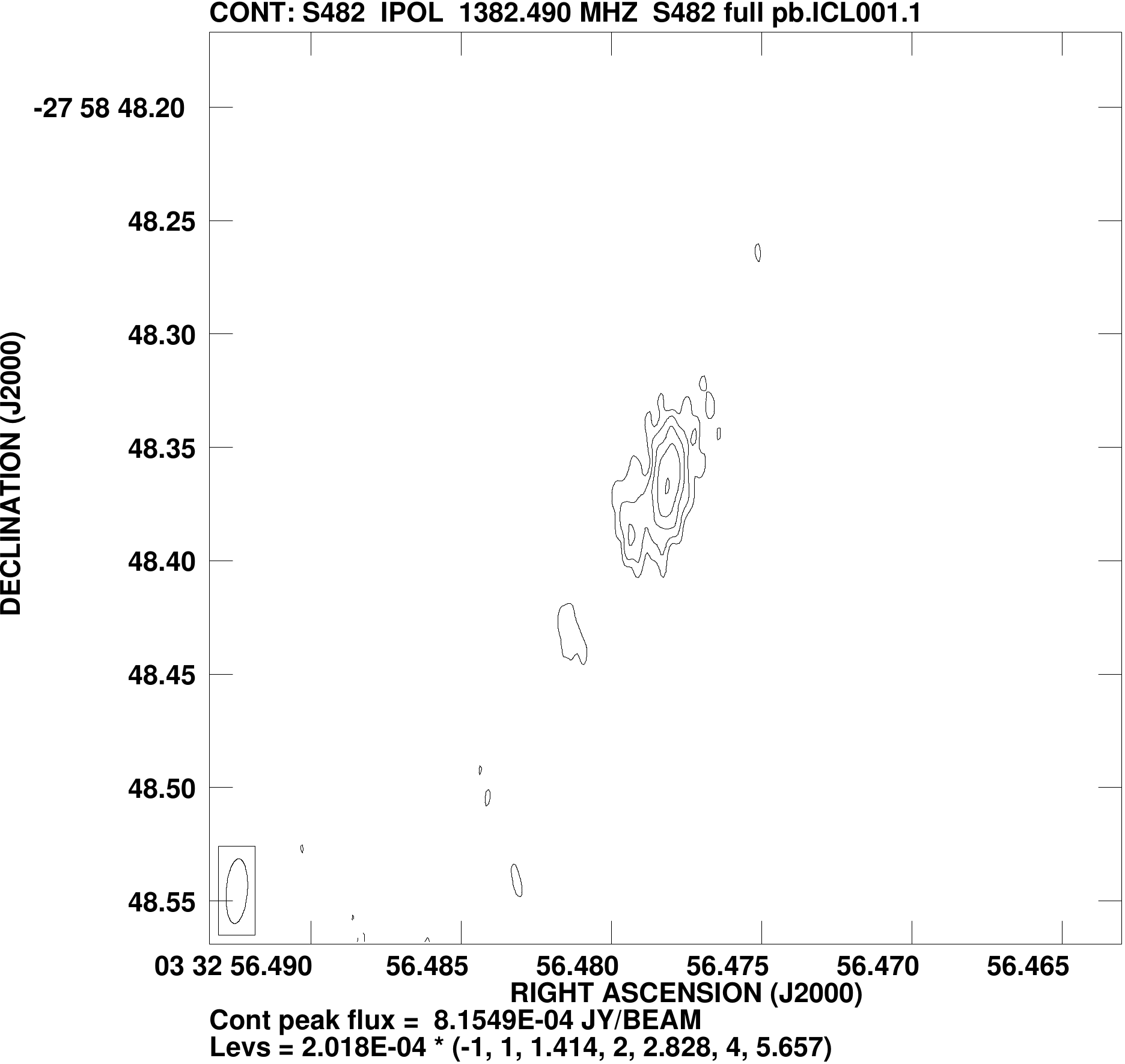}
\includegraphics[width=0.3\linewidth]{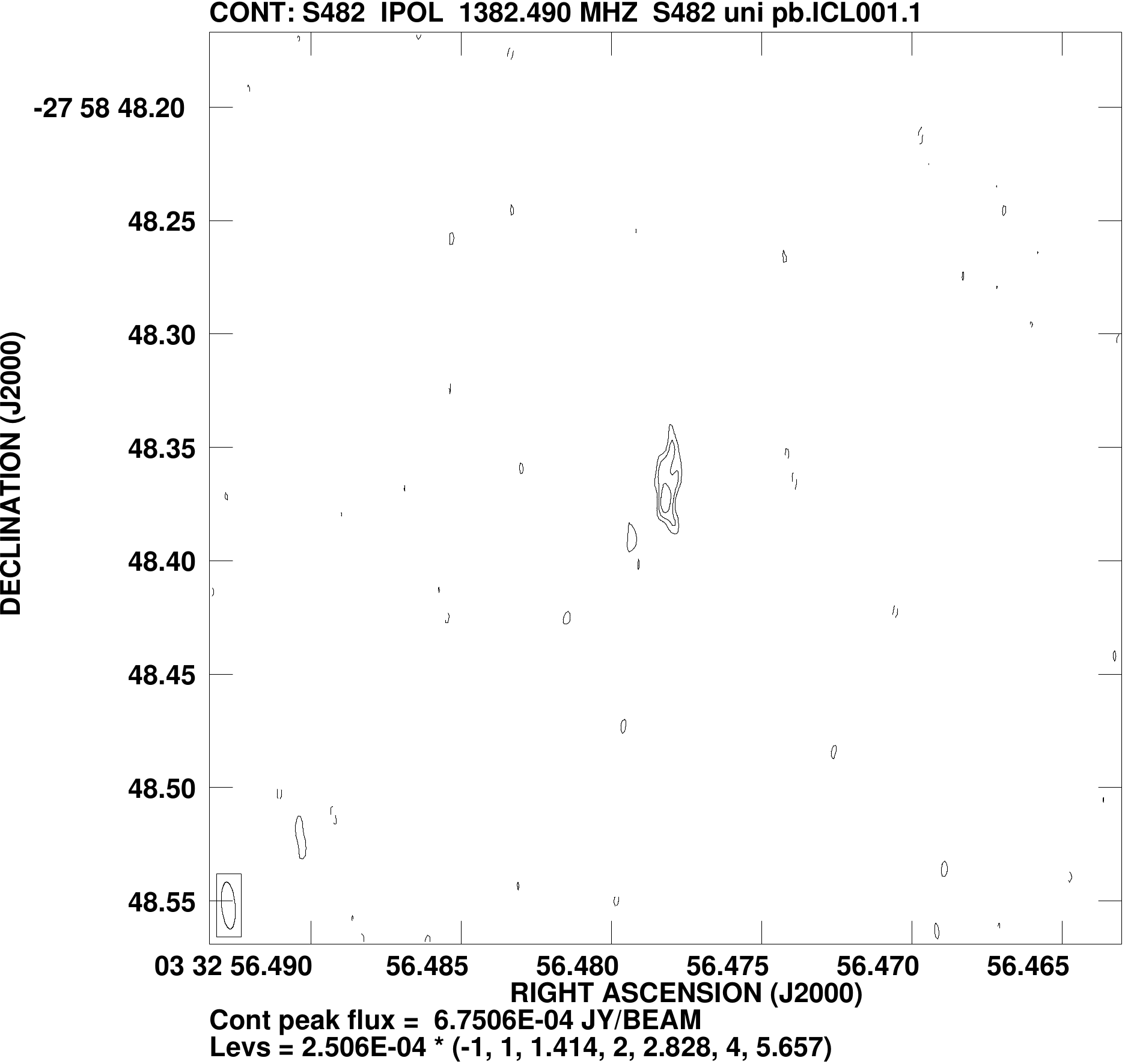} \\
\includegraphics[width=0.3\linewidth]{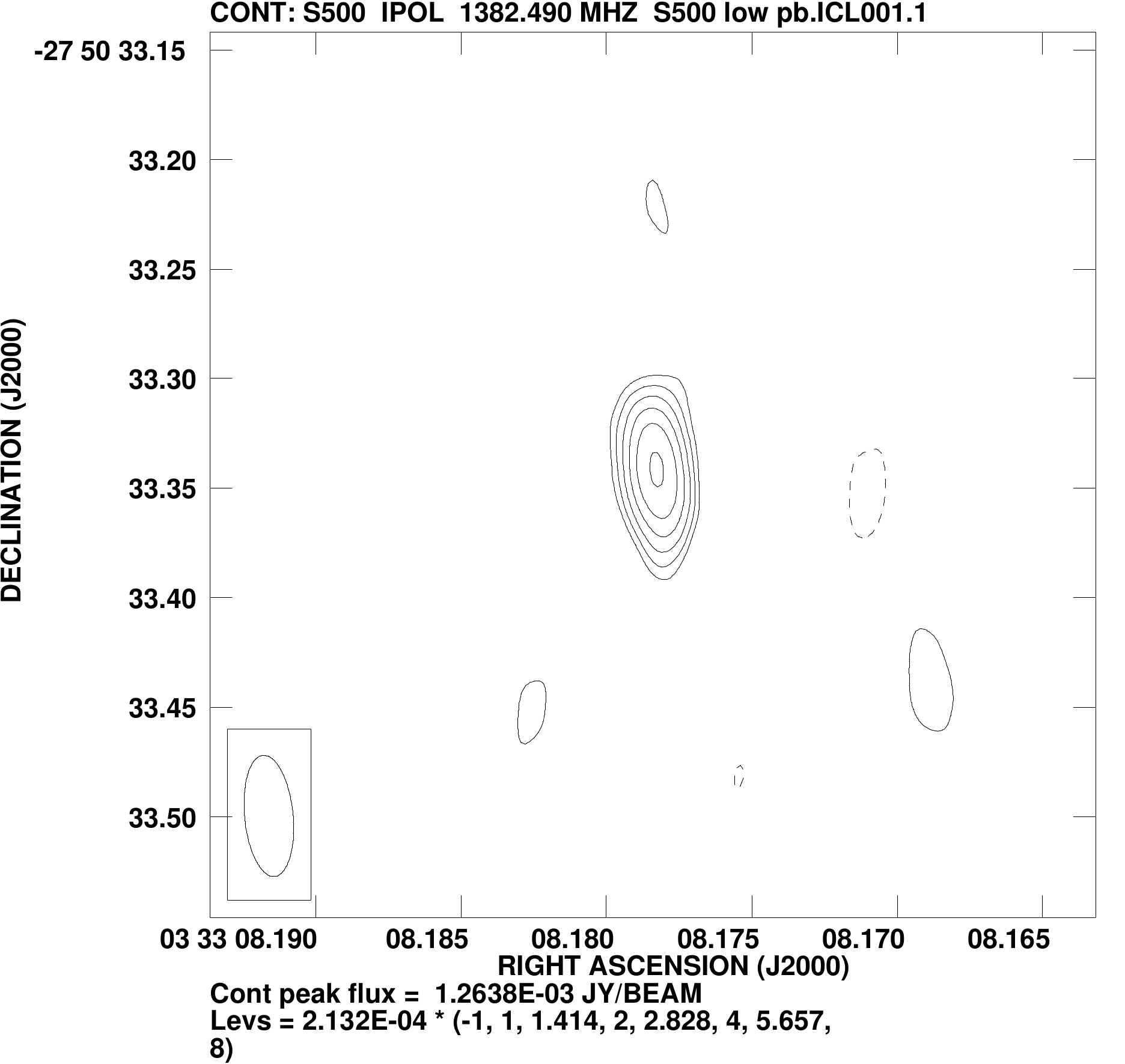} 
\includegraphics[width=0.3\linewidth]{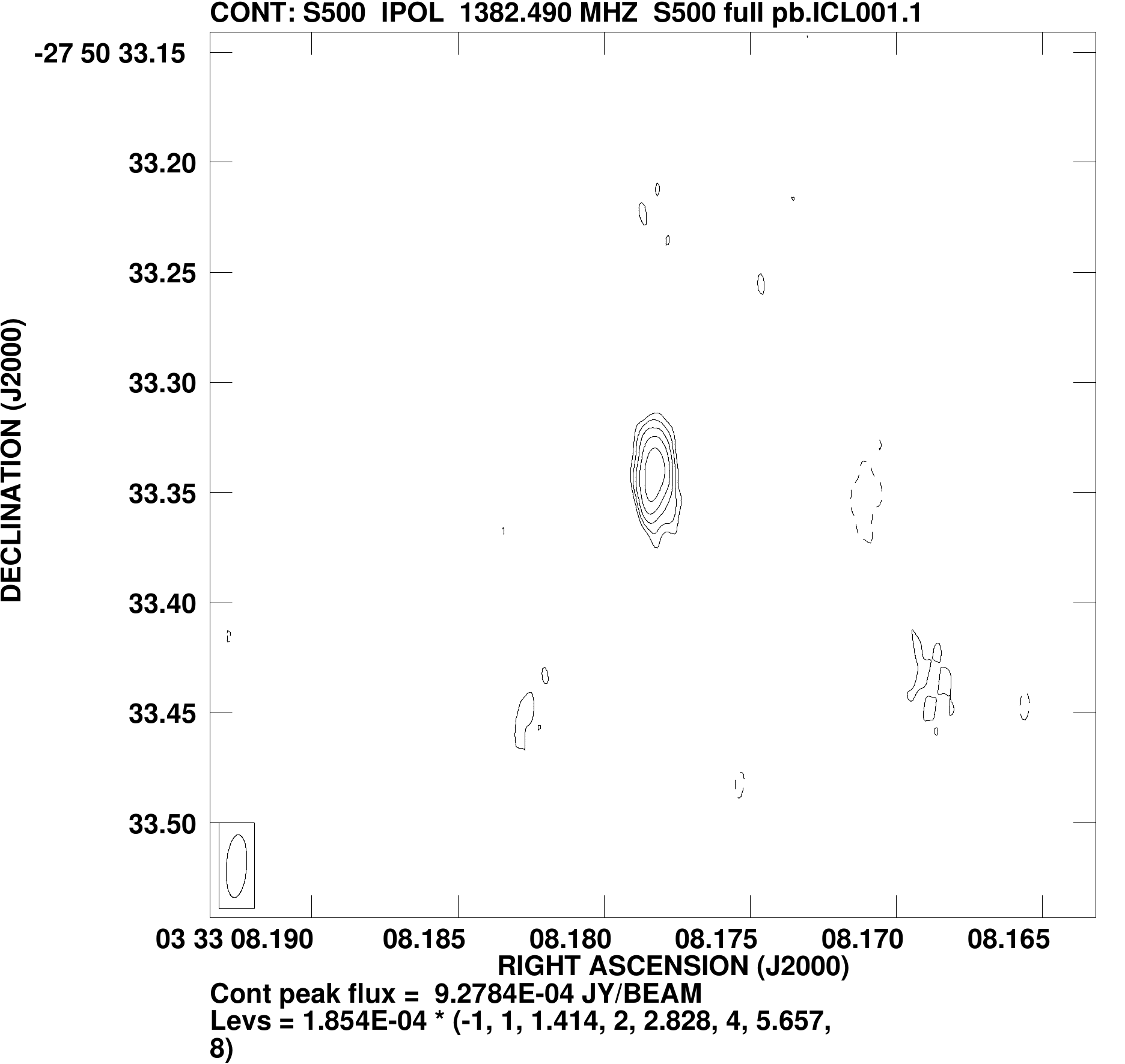}
\includegraphics[width=0.3\linewidth]{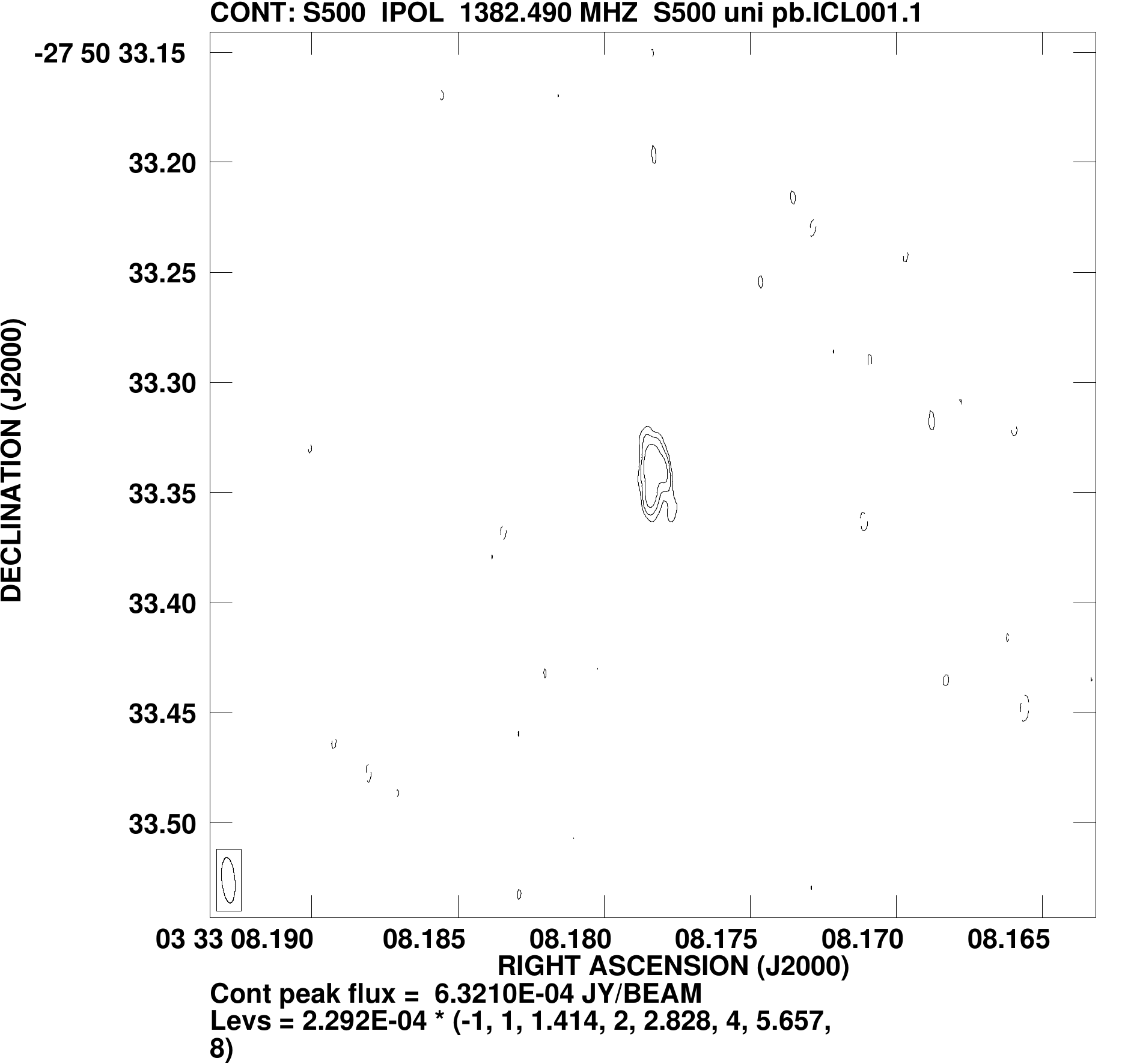} \\
\caption{Continued}
\end{figure*}

\addtocounter{figure}{-1}

\begin{figure*}
\includegraphics[width=0.3\linewidth]{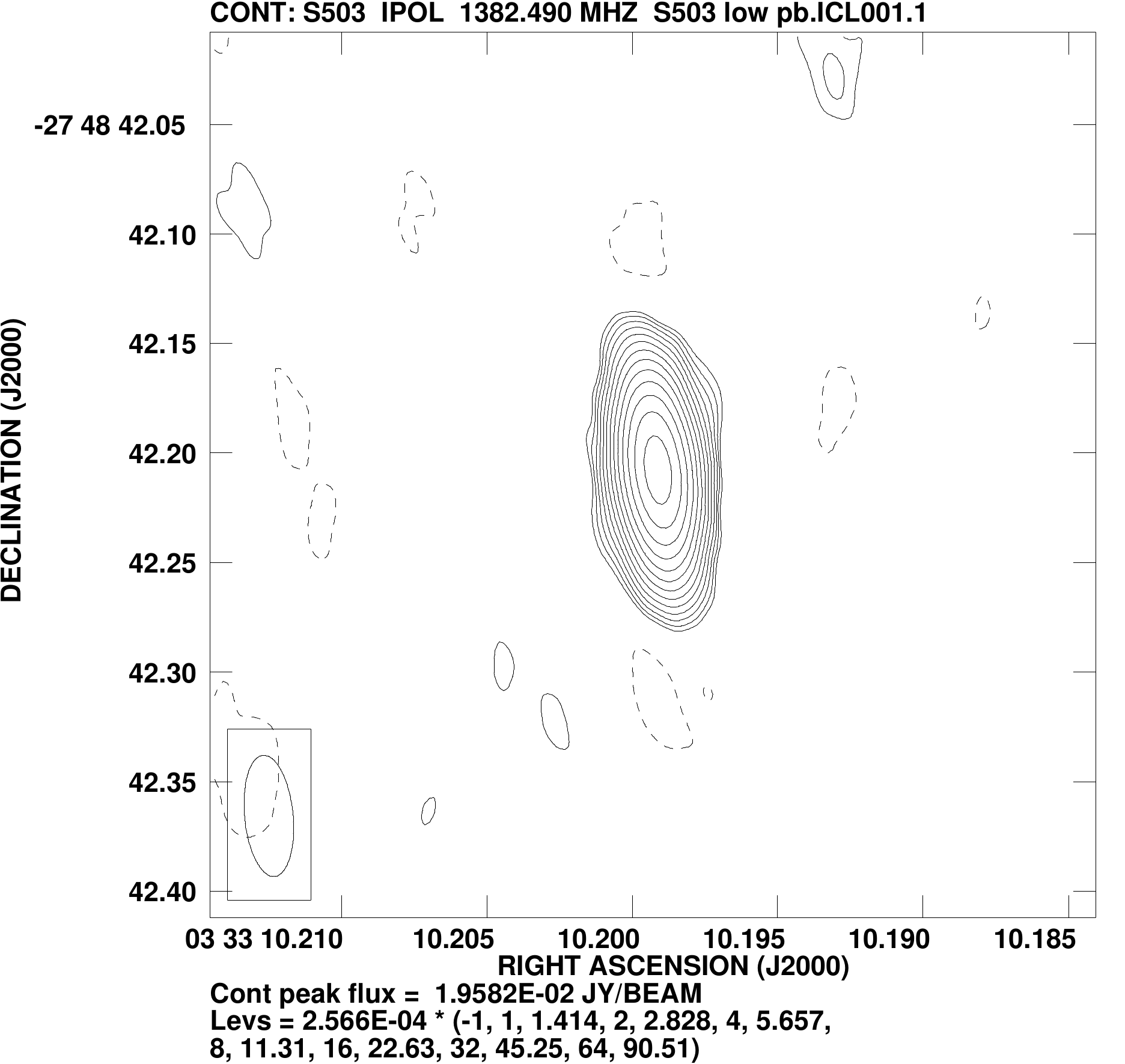} 
\includegraphics[width=0.3\linewidth]{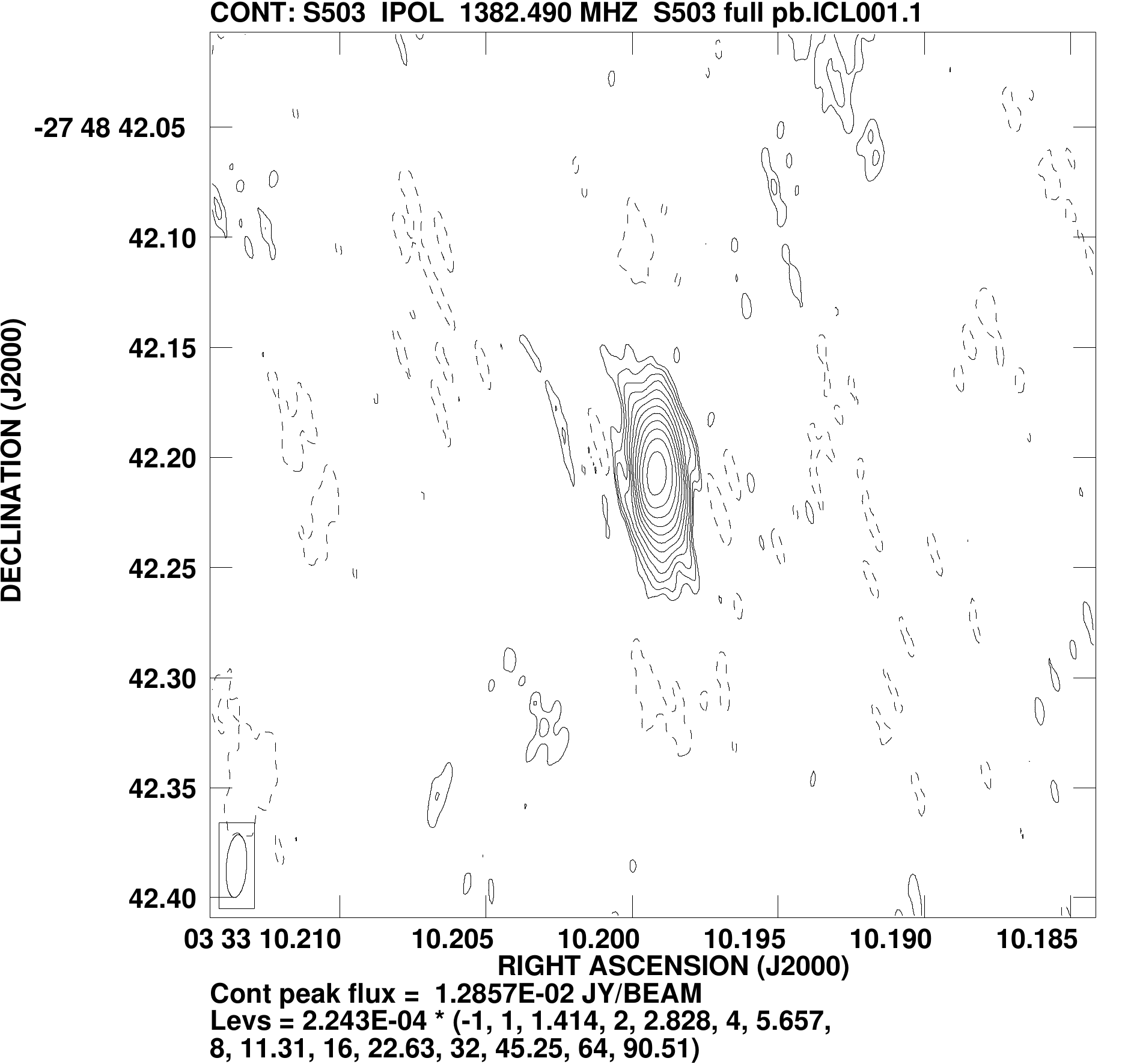}
\includegraphics[width=0.3\linewidth]{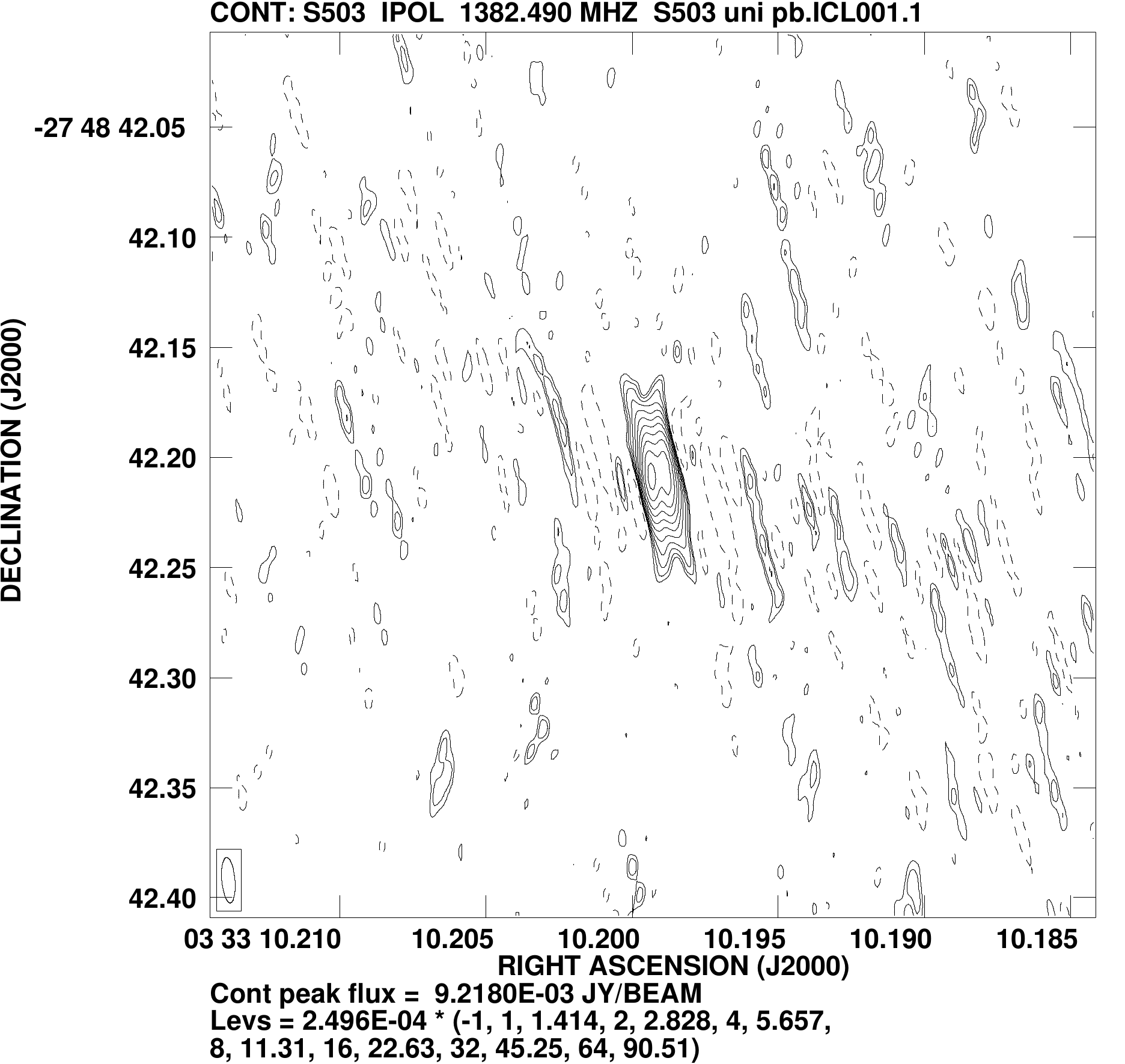} \\
\includegraphics[width=0.3\linewidth]{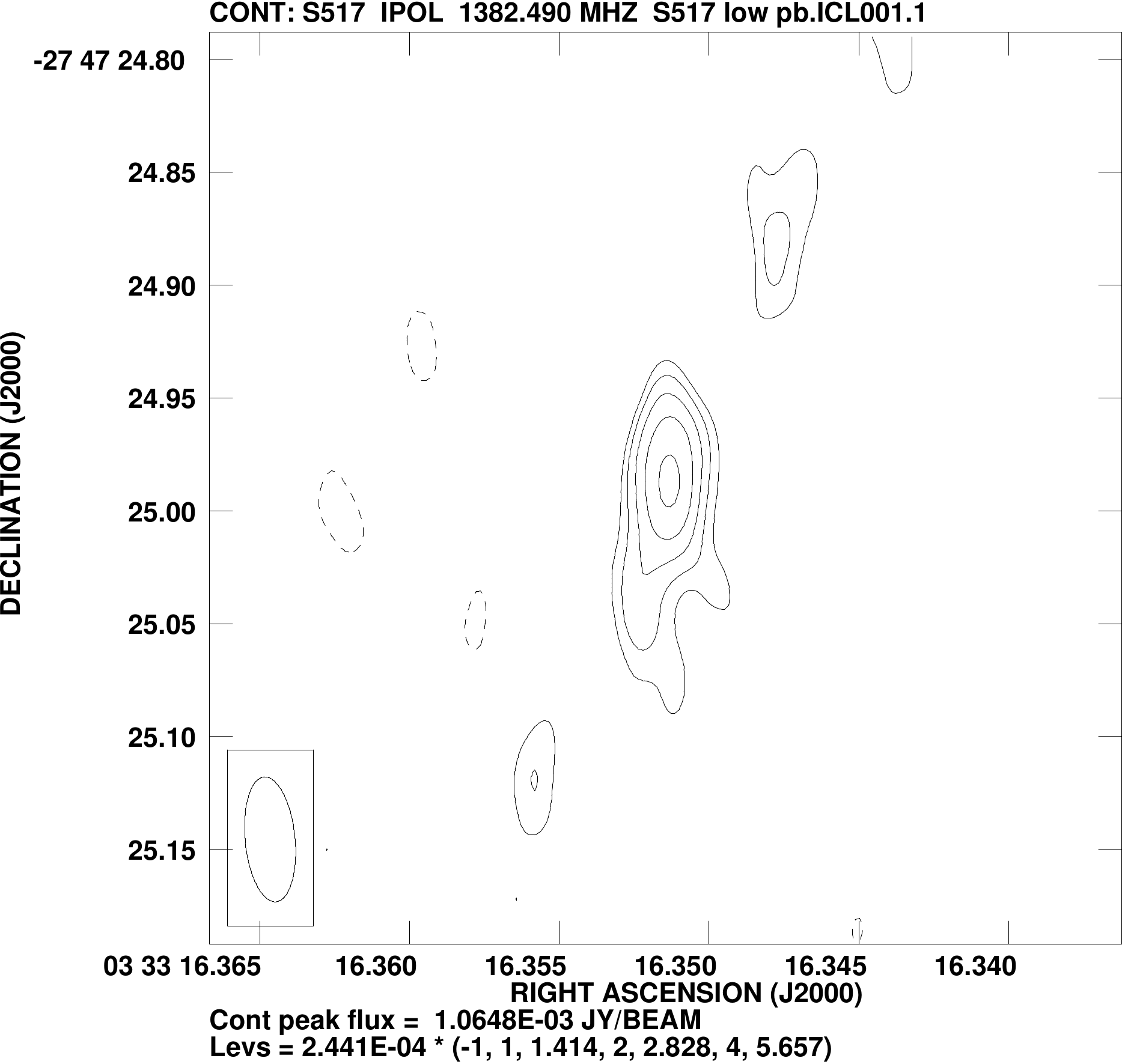} 
\includegraphics[width=0.3\linewidth]{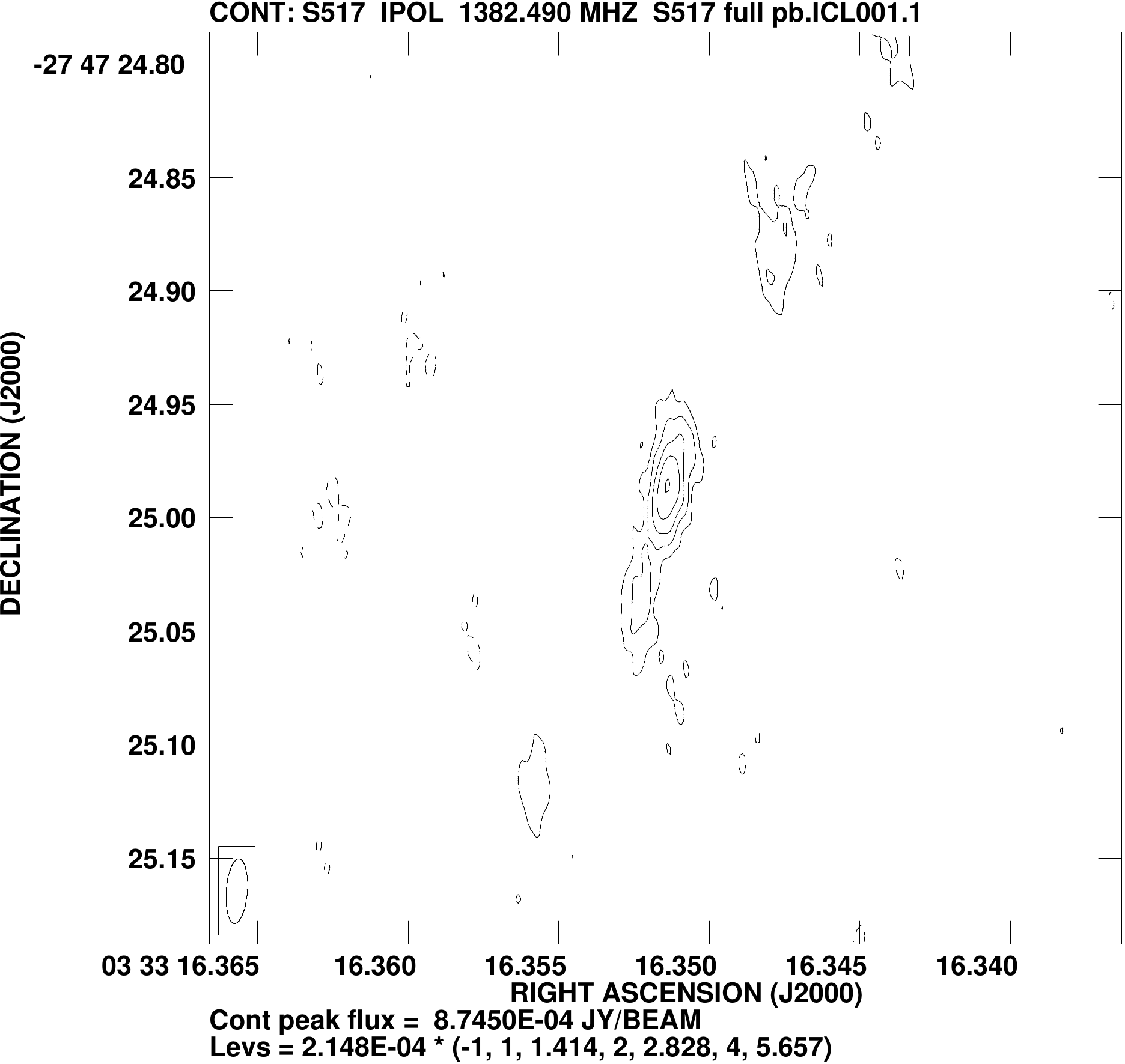}
\includegraphics[width=0.3\linewidth]{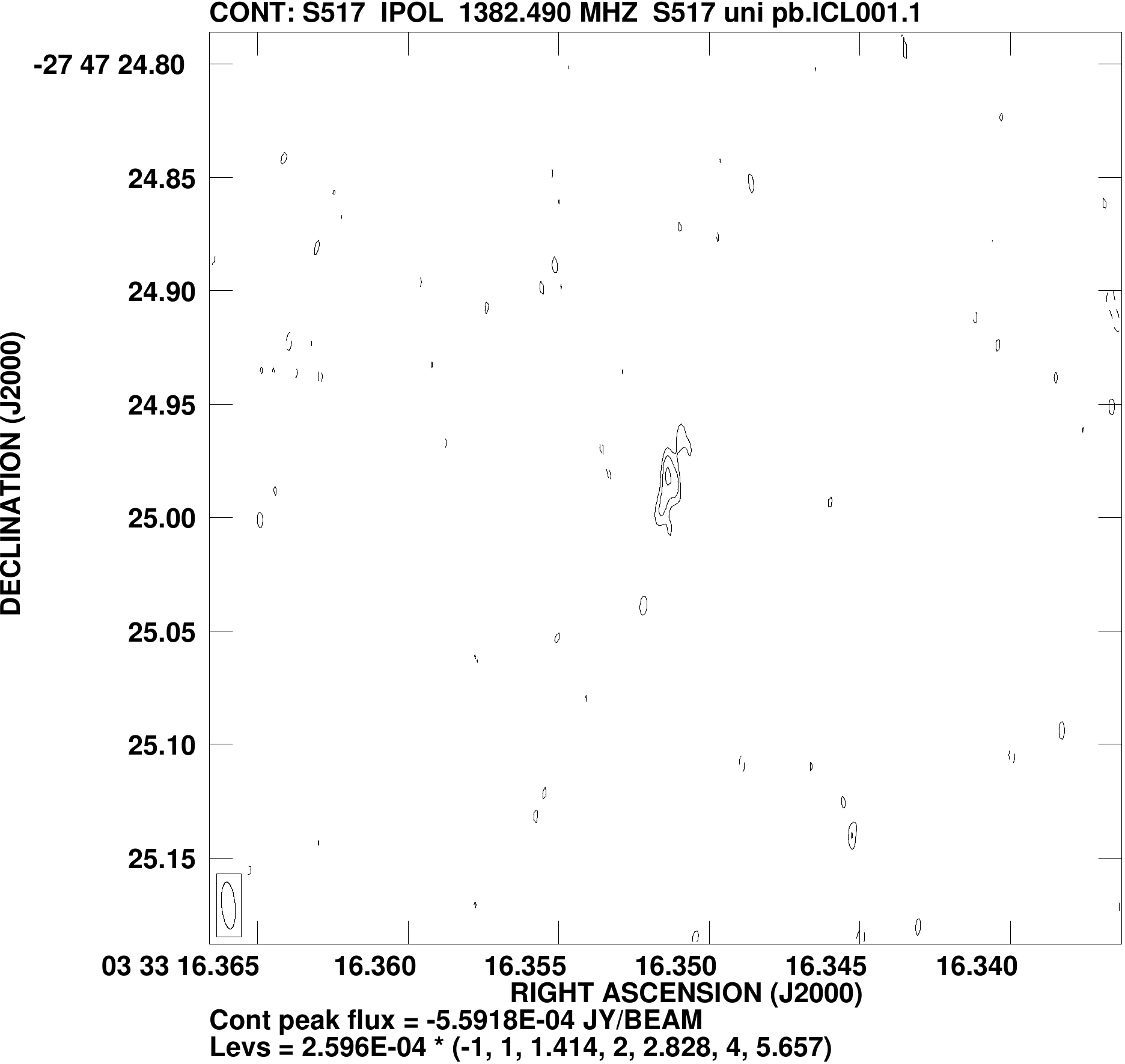} \\
\includegraphics[width=0.3\linewidth]{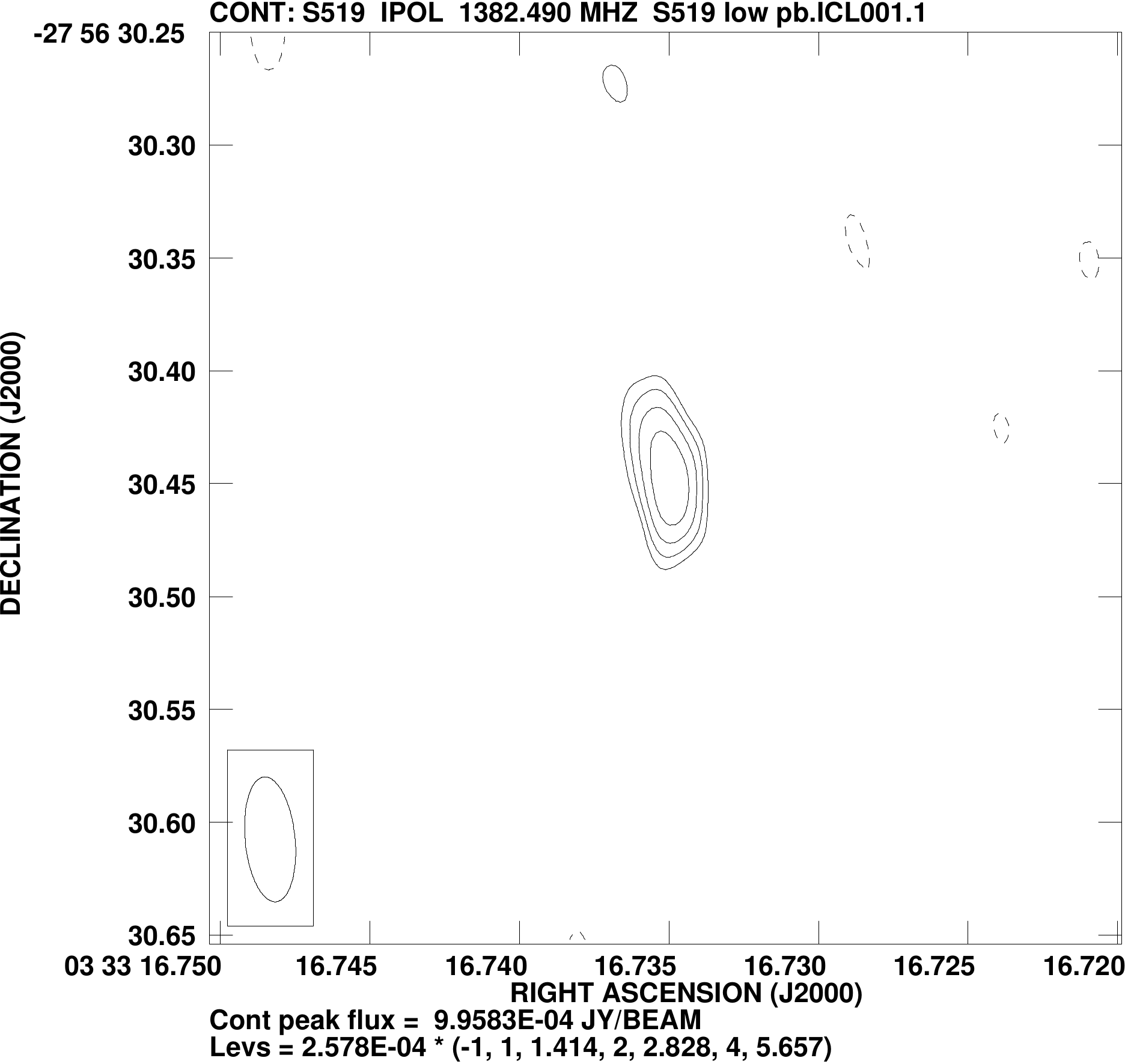} 
\includegraphics[width=0.3\linewidth]{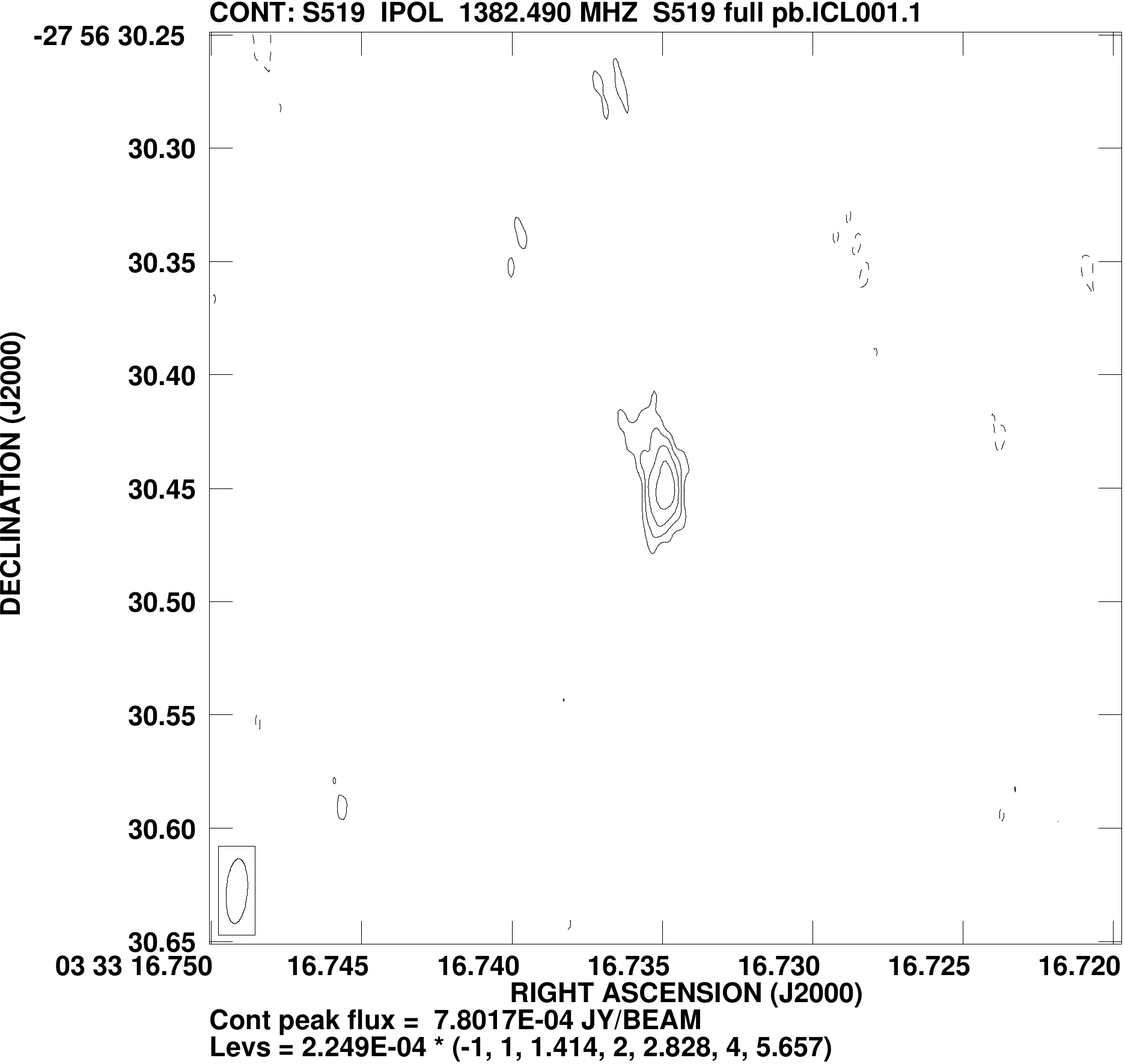}
\includegraphics[width=0.3\linewidth]{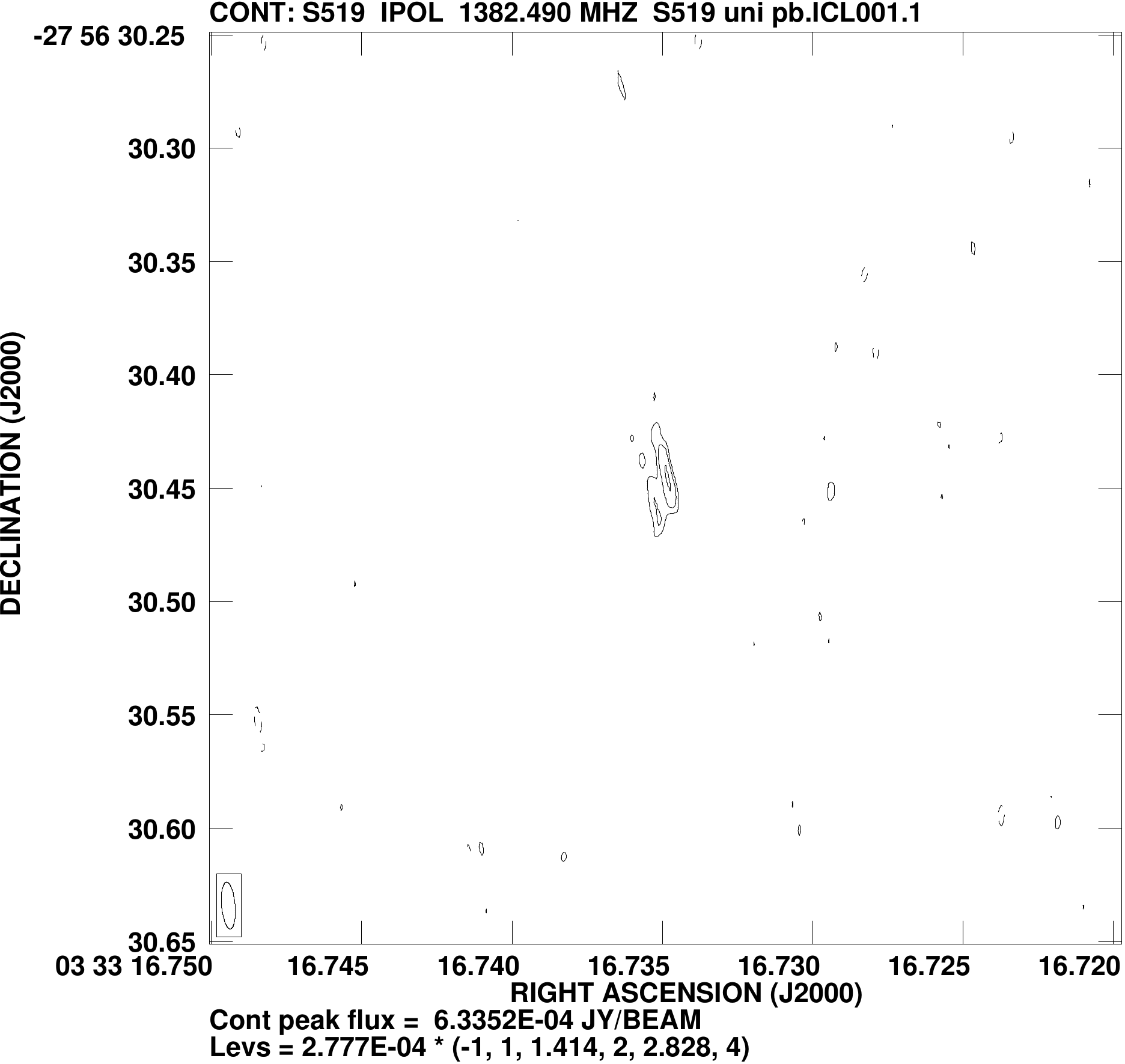} \\
\includegraphics[width=0.3\linewidth]{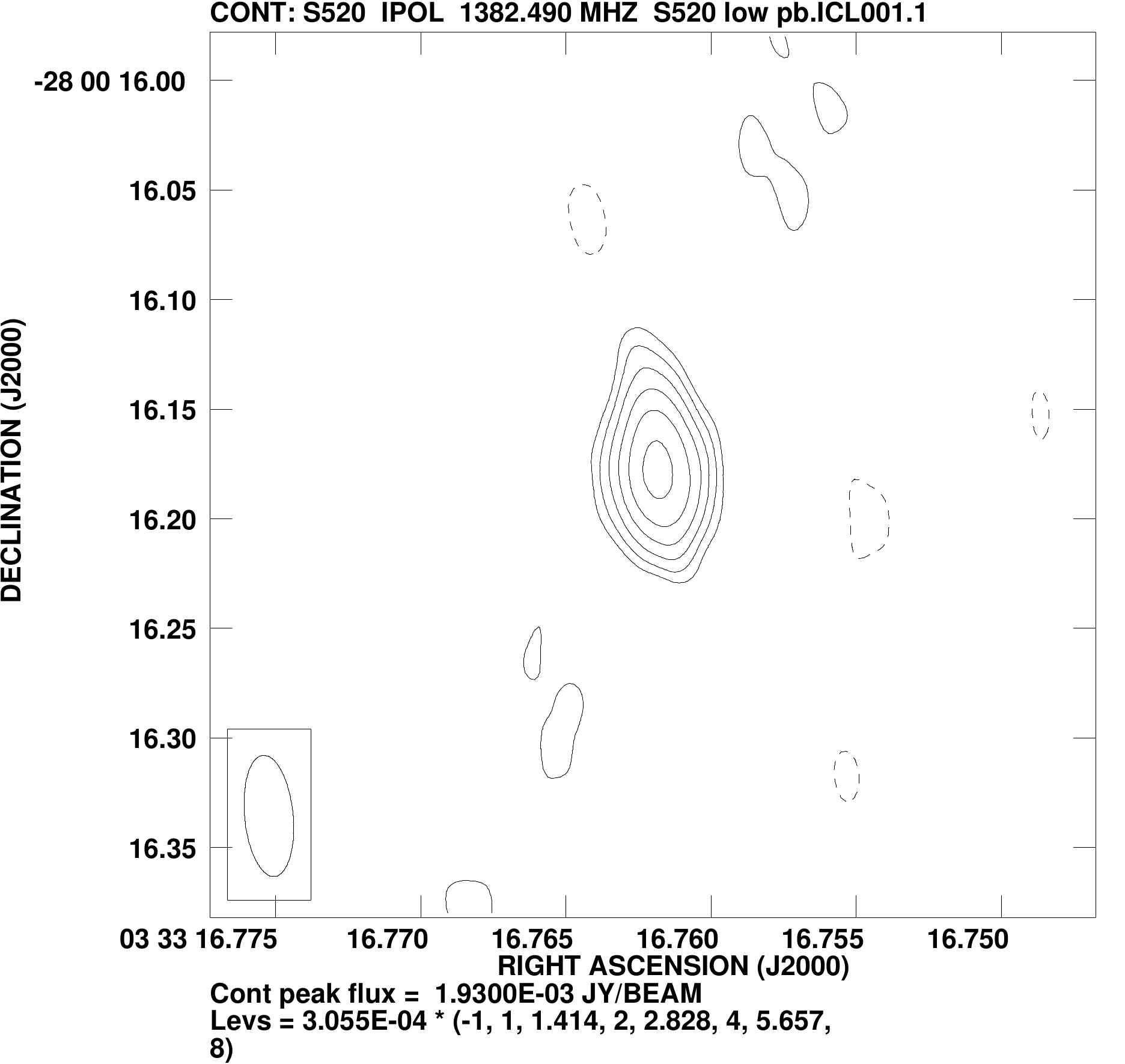} 
\includegraphics[width=0.3\linewidth]{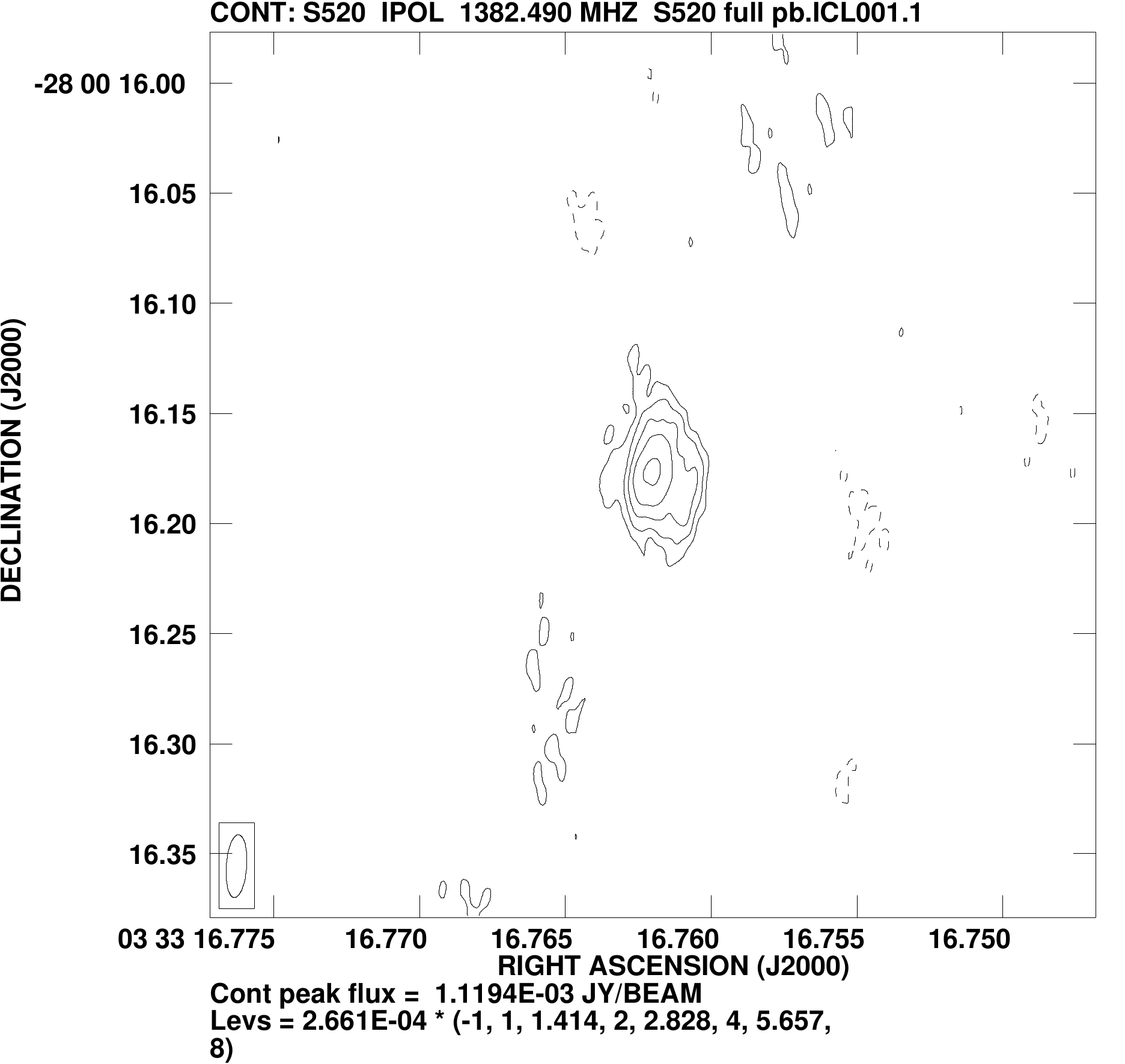}
\includegraphics[width=0.3\linewidth]{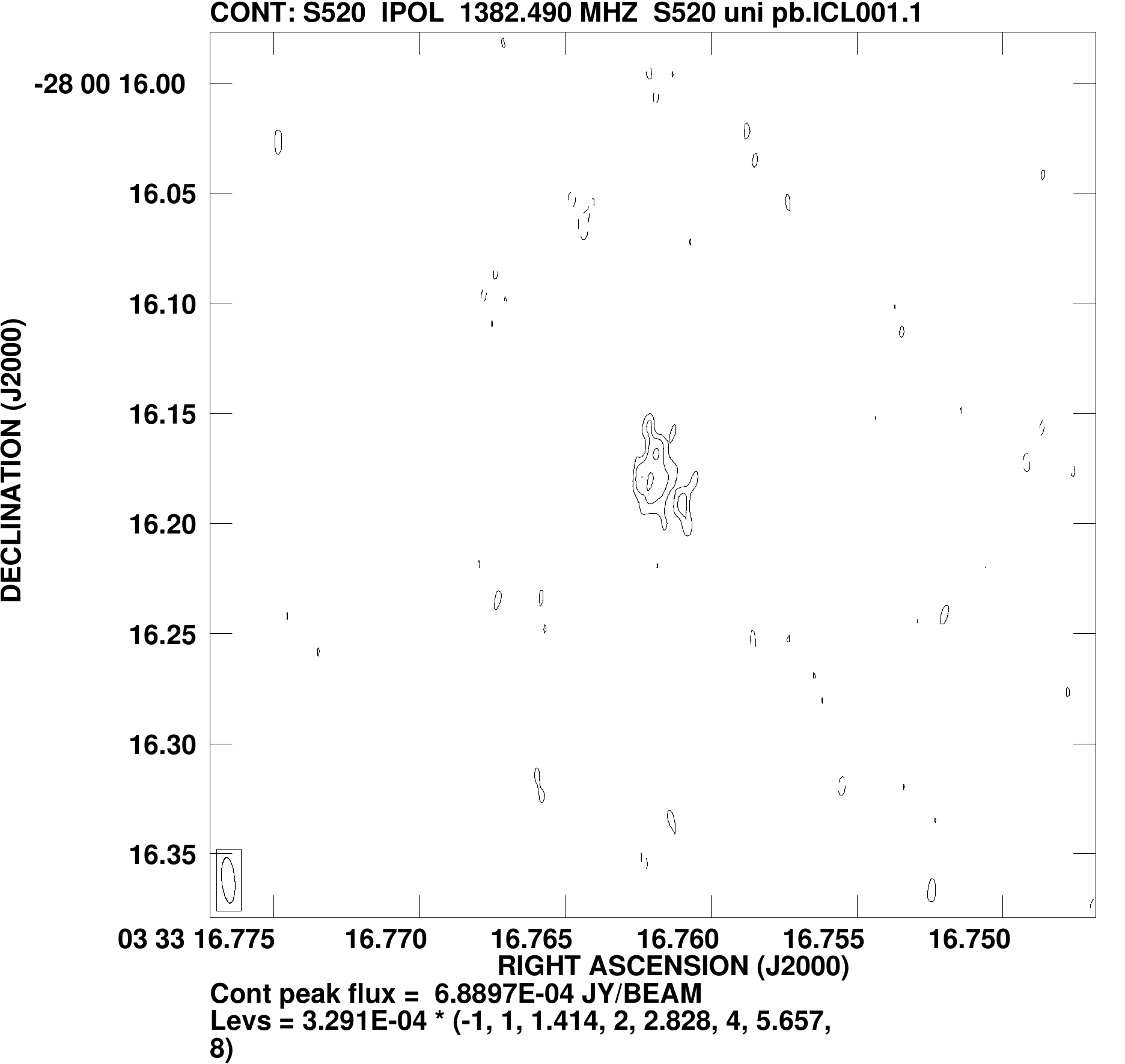} \\
\caption{Continued}
\end{figure*}

\addtocounter{figure}{-1}

\begin{figure*}
\includegraphics[width=0.3\linewidth]{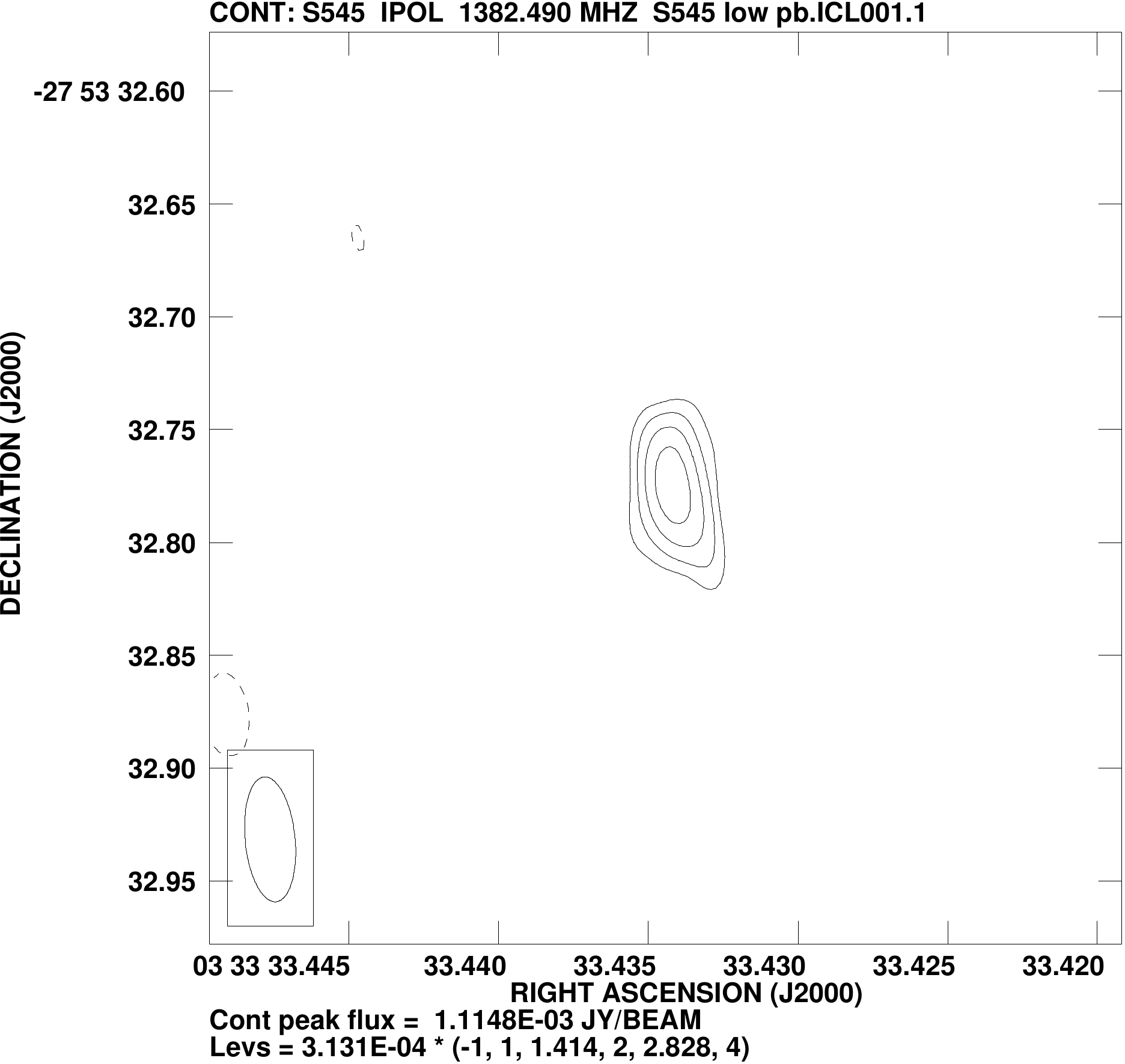} 
\includegraphics[width=0.3\linewidth]{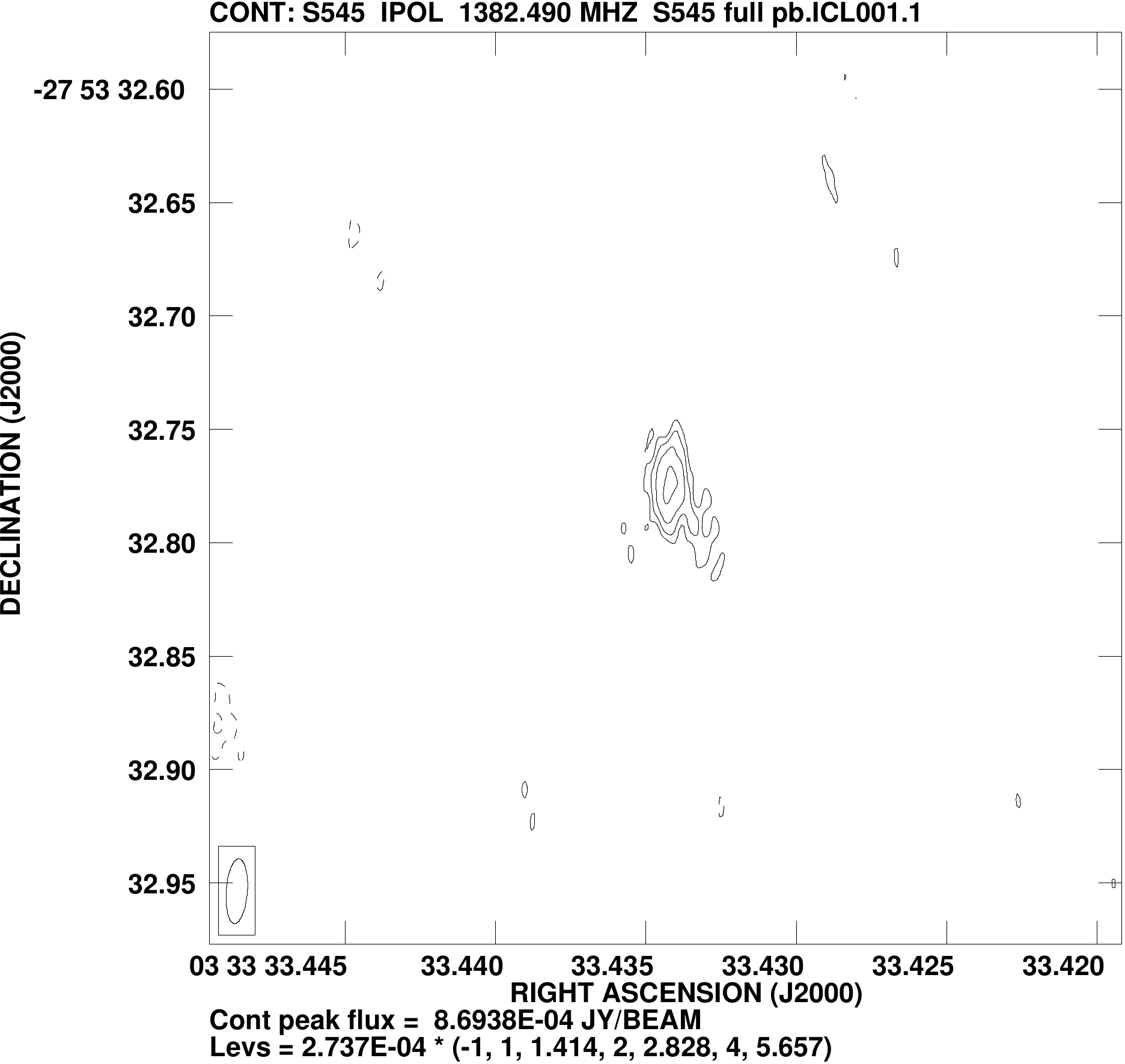}
\includegraphics[width=0.3\linewidth]{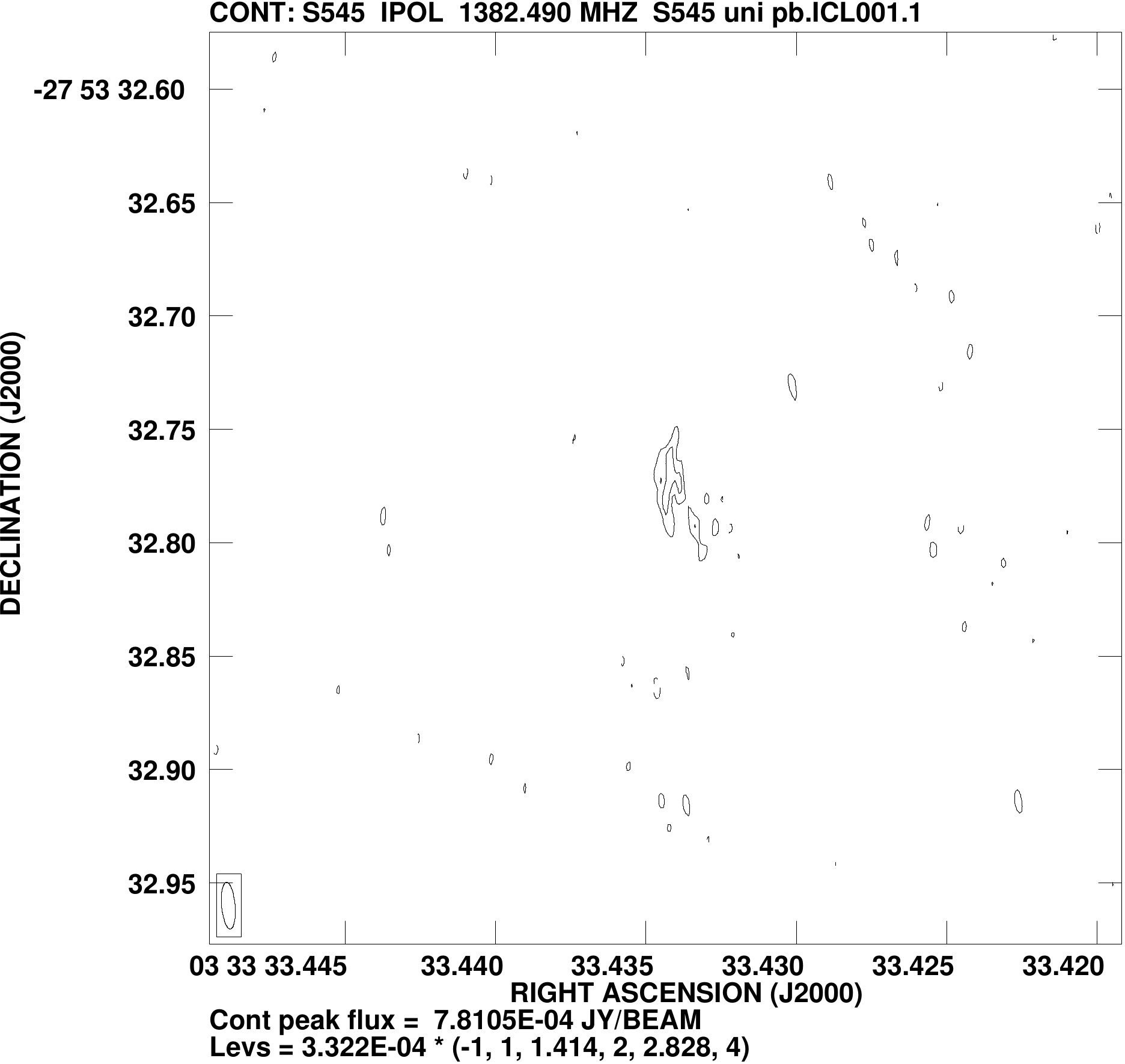} \\
\caption{Continued}
\end{figure*}

\section{Discussion}
\label{sec-5}

\subsection{Statistics}

\subsubsection{Detections}

One of the goals of this project was to identify AGN in a blind survey
of radio sources, i.e., without selecting objects other than by
requiring a radio flux density in the ATCA survey. Given that the
sensitivity of the VLBI observations varies by a factor of more than
two over the observed area, the detection statistics are difficult to
analyse. Also, our observations are biased towards compact objects,
and this bias is a function of flux density. Fainter sources will only
be detected in VLBI observations if they are increasingly compact,
whereas strong sources such as S393 (the second-brightest in our
sample) can be detected if a fraction as low as 10\,\% is contained
within a compact core. This means that the source types that our
observations are sensitive to is a function of flux density.

To provide some analysis of the detection statistics despite these
shortcomings, we have divided our sample into three groups: sources
with ATCA flux densities, $S$, of 10\,mJy or more (bin 1, 7 objects),
sources with $1\,{\rm mJy}\leq S<10\,{\rm mJy}$ (bin 2, 16 objects),
and sources with $S<1\,{\rm mJy}$ (bin 3, 73 objects).

Out of the three bins, 5, 11 and 5 objects have been detected,
respectively, corresponding to percentages of 71\,\%, 69\,\%, and
7\,\% (see Table~\ref{tab:detections}). Most of the VLBA-detected
sources (14 sources) are compact, which is indicated by the ratio of
VLBI-scale flux density to the ATCA-scale flux density being
consistent with unity (see Table~\ref{tab:results}). We note, however,
that the ATCA flux densities by \cite{Norris2006a} are probably
affected by clean bias, which tends to underestimate the flux
densities of weak sources. For a very similar observation,
\cite{Middelberg2008a} show that clean bias can decrease the flux
density by up to 5\,\% for sources with SNR smaller than 20. Hence the
ratio of the VLBI-scale flux density to ATCA-scale flux density is
biased towards higher values, in particular for sources with an ATCA
flux density of less than 1\,mJy, and even exceeds unity in some
cases.

\begin{table}[htbp]
\centering
\caption{The number of observed radio sources grouped into three flux
  density bins. $N_{\rm ATCA}$ is the number of radio sources found
  with the ATCA in a given flux density range, $N_{\rm VLBA}$ is the
  number of sources detected out of these, $\%_{\rm VLBA}$ is this
  number expressed as a percentage of $N_{\rm ATCA}$, AGN is the
  number of known AGN prior to our observations, and nAGN is the
  number of new AGN we have identified.}
\begin{tabular}{|l|r|r|r|r|r|}
\hline
\hline
flux density & $N_{\rm ATCA}$ & $N_{\rm VLBA}$ & $\%_{\rm VLBA}$ & AGN & nAGN\\
\hline
$\geq$10\,mJy                       & 7  & 5  & 71\,\%  &  5   & 0 \\
$1\,{\rm mJy}\leq S<10\,{\rm mJy}$  & 16 & 11 & 69\,\%  &  10  & 5 \\
$S<1\,{\rm mJy}$                    & 73 & 5  & 7\,\%   &  19  & 2 \\
\hline
\end{tabular}
\label{tab:detections}
\end{table}

\subsubsection{New AGN}

It was noted in the introduction that a detection in a VLBI
observation is a strong indicator for AGN activity. To determine
whether a source is already a known AGN we have used the information
from various sources listed in \cite{Norris2006a} and the criteria by
\cite{Szokoly2004} (using the X-ray hardness ratio and X-ray
luminosity, see Sec.~\ref{sec:hardness}). 

Without the VLBI data presented here, the combined radio, IR, optical,
and X-ray data were able to classify 36 sources (38\,\%) as AGN. There
are 7 sources however, which have not been identified as AGN and are
detected in the VLBI observations. This number is 7\,\% of the total
number of radio objects observed, but 35\,\% of the sources detected
with the VLBA.

All VLBI-detected bright sources in bin 1 were known to be AGN before,
but out of the 16 medium-bright sources in bin 2, only 10 (63\,\%)
were previously known to be AGN, and our VLBA observations have added
5 (31\,\%). Among the 73 faint sources in bin 3, 19 sources were known
AGN (26\,\%), and we have added 2 (4\,\%). These statistics suggest
that VLBI observations can make a significant contribution to the
classification of galaxies below a flux density of 10\,mJy, because
brighter objects are already known AGN, and objects fainter than
1\,mJy are only occasionally detected. This ability, of course, is a
function of the sensitivity of the VLBI observations, which will soon
improve significantly as the bandwidths of radio telescopes increase.

Two sources, S423 and S443, have not been classified as AGN based on
their optical and X-ray properties, but are detected in our VLBI
observations. In these two cases the ancilliary data indicated a
starburst or elliptical and a spiral galaxy, respectively. S423 is
detected marginally in the uniformly-weighted image, with a resolution
of 20.8$\times$5.9\,mas$^2$. Its peak flux density is
300\,$\mu$Jy/beam, which indicates a brightness temperatures of
$2.2\times10^6$\,K. Hence it displays characteristics of a normal
galaxy while at the same time exhibiting a brightness temperature
clearly indicating an AGN. The VLBI detection of S443 is offset from
the centre of the galaxy and can therefore not be taken as evidence
for an AGN; for further discussion on possible origins, see
Section~\ref{sec:sources}. It has also been ignored in the following
analyses which compare properties of VLBI-detected and VLBI-undetected
sources.

\subsubsection{Radio spectral index}

Radio spectral index can be used as an indicator for AGN
activity. When the spectral index is larger than $-0.3$ ($S\propto
\nu^\alpha$) then the emission is deemed to come from self-absorbed
synchrotron emission such as that found in the very compact regions of
AGN and their jets. When the spectral index is smaller than $-1.2$
then the emission is thought to come from extended structures formed
by AGN in the high-redshift Universe. Between these values, a
separation between AGN and star-forming activity can not be made. We
have taken the data measured by \cite{Kellermann2008} and found that
the median spectral index among the VLBI-detected sources is $-0.7$,
whereas the undetected sources have a median of $-0.8$. In both cases,
the scatter of the distribution as characterised by the median average
difference (MAD) is 0.2. A Kolmogorov-Smirnov test (K-S-test) returns
a $p$ of 0.80 that the two samples are drawn from the same parent
population, indicating that there is no significant difference in
spectral index between sources detected and undetected with VLBI.

One might expect that the VLBI-detected sources show higher spectral
indices than the undetected sources, because the VLBI observations
require the emission to come from smaller regions, which then have a
tendency to have smaller optical depths. We argue that since the
linear resolution of our observations is of the order of 280\,pc (VLBI
flux densities were measured from images with a resolution of
55$\times$22\,mas and a typical redshift is 1) the emission from the
observed regions is still dominated by optically thin synchrotron
radiation. Unfortunately, the number of VLBI-detected objects is too
small to test this hypothesis statistically.

\subsubsection{Redshifts}

We have searched the literature for redshift measurements of our
sample. In total, we have found redshifts for 82 sources, of which 19
were detected in our VLBA observations. The median redshift of the
detected sources is 0.98 with a MAD of 0.29, whereas the undetected
sources have a median redshift of 0.67 with a MAD of 0.41. Again the
samples are not significantly different, with a $p$ of 0.18 in a
K-S-test.

\subsubsection{Infrared-Faint Radio Sources (IFRS)}

IFRS are mysterious objects which can be strong (tens of mJy) radio
sources but do not have near-infrared counterparts detected by the
{\it Spitzer} observatory as part of the {\it Spitzer} Wide-area
InfraRed Extragalactic Survey (SWIRE). IFRS are unexpected because all
known classes of galaxy at $z<2$ are expected to be detectable given
the IR survey sensitivities. See \cite{huynh2010}, \cite{Norris2010}
and \cite{Middelberg2010} and their references for recent work on
these sources. Previous VLBI observations of six IFRS have resulted in
two detections (\citealt{Norris2007b}, \citealt{Middelberg2008c}),
supporting the idea that IFRS are AGN-driven. There are three radio
sources in our sample which have been classified as IFRS: S415, S446,
and S506, which have catalogued arcsec-scale flux densities of
1.2\,mJy, 0.3\,mJy, and 0.2\,mJy, but none was detected.

\subsection{Co-located X-ray observations}

X-ray observations are claimed to be a very direct tracer of AGN
activity (e.g., \citealt{Mushotzky2004}). The radiation originates
very close to the supermassive black hole, it can leave these regions
rather unabsorbed, in particular at energies of a few keV, and there
is little contamination from other sources, such as stars.

The surveyed region has been observed intensively with the Chandra
X-ray observatory. In particular, a 940\,ks exposure of a
0.11\,deg$^2$ region as indicated in Figure~\ref{fig:overview}, known
as ``the'' Chandra Deep Field South (CDFS), has resulted in the
detection of 346 sources (\citealt{Giacconi2002}). The region has been
reobserved by \cite{Luo2008}, adding more than 1\,Ms to the total
exposure time and resulting in a central sensitivity of
$7.1\times10^{-17}\,{\rm erg\,cm^{-2}\,s^{-1}}$. The total number of
X-ray sources found in this observation was 578. Furthermore, a
240\,ks exposure of a 0.3\,deg$^2$ region, known as the Extended
Chandra Deep Field South (ECDFS), covers entirely the area observed
here (\citealt{Lehmer2005}). This observation, with a maximum
sensitivity of $3.5\times10^{-16}\,{\rm erg\,cm^{-2}\,s^{-1}}$,
yielded the detection of 762 point sources. All observations were
carried out in the energy range 0.5\,keV$-$8\,keV and provide
information about source flux densities in a soft band
(0.5\,keV$-$2\,keV) and in a hard band (2\,keV$-$8\,keV). We have
cross-matched our detections to these catalogues.

\subsubsection{Identifications in the CDFS}

\begin{figure}[htpb]
\includegraphics[width=0.95\linewidth,angle=270]{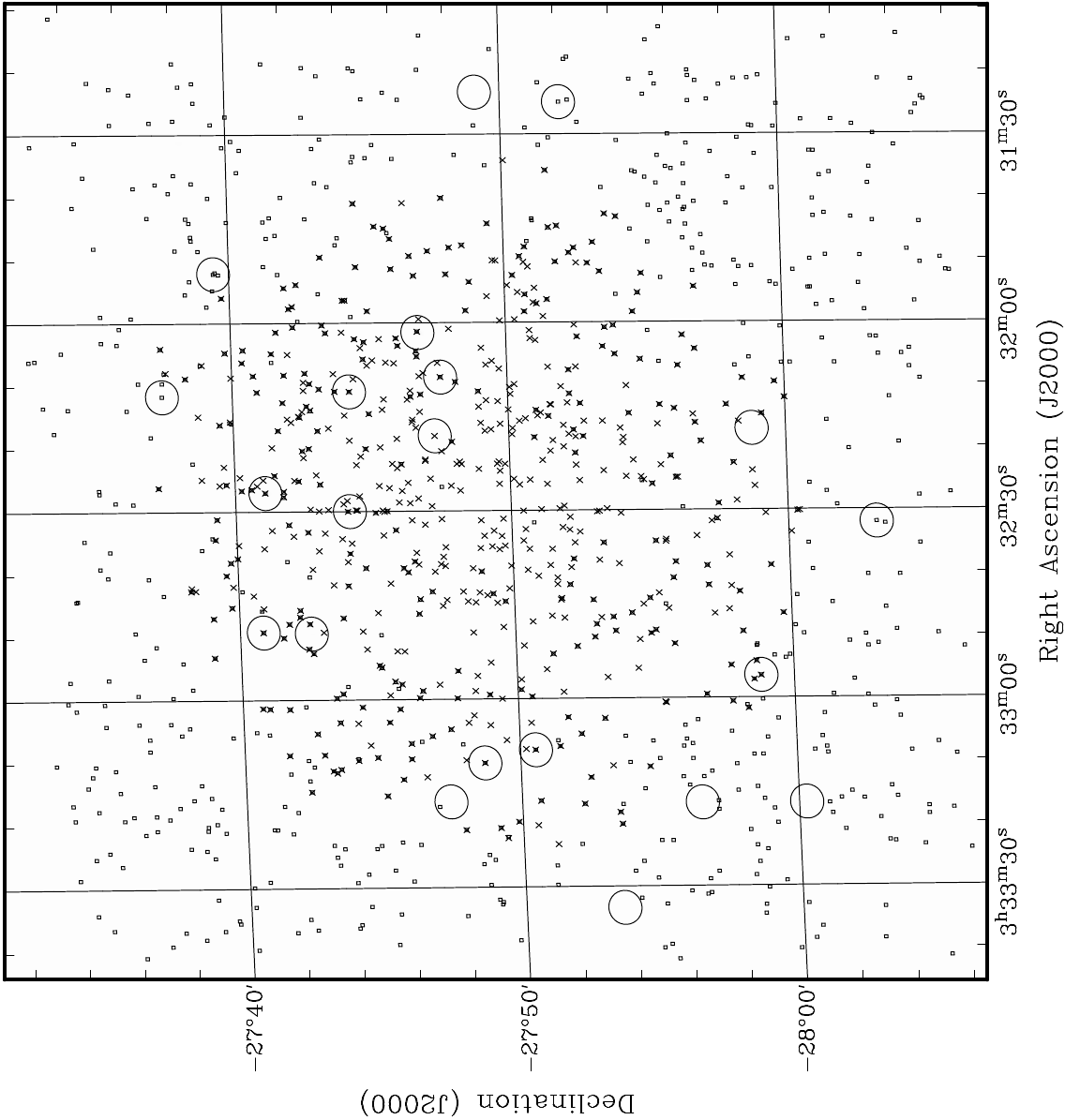}
\caption{Diagram illustrating the locations of X-ray sources in the
  CDFS (crosses), in the ECDFS (squares) and the sources detected in our
  VLBI observations (circles).}
\label{fig:xray}
\end{figure}

The region indicated with the medium circle in Figure~\ref{fig:overview}
contains 23 radio sources, of which 6 were detected by our VLBI
observations. A search for counterparts of the VLBI-detected sources
in the catalogues by \cite{Luo2008} returned 5 matches. There is only
one VLBI-detected source, S474, which clearly lies in the area covered
by the X-ray observations, but is not detected in the X-ray. Whilst it
lies towards the edge of the area covered by the 2\,Ms exposure, the
integration times of the sources surrounding S474 suggest that its
location has been observed for around 1.3\,Ms. A further 4
VLBI-detected sources have X-ray counterparts outside the most
sensitive region (see Figure~\ref{fig:xray}).

\subsubsection{Identifications in the ECDFS}

Repeating the search in the catalogue by \cite{Lehmer2005}, which
covers an area that includes all sources observed by us, returned 12
sources, 8 of which were identical to sources discovered in the search
of the CDFS data. Hence only one VLBI source was detected in the
deep CDFS observations but not in the shallower ECDFS observations. In
total 7 VLBI sources have no X-ray counterpart at all.

This result sheds an interesting light on X-ray observations as a means
of filtering AGN: at a sensitivity level corresponding to integration
times of between 240\,ks and 2\,Ms the fraction of VLBI-detected AGN
with X-ray counterparts increases from $12/20=60$\,\% to $5/6=$83\,\%.

\subsubsection{Hardness ratios}
\label{sec:hardness}

The hardness ratio

\begin{equation}
HR=\frac{H{\rm cts}-S{\rm cts}}{H{\rm cts}+S{\rm cts}},
\end{equation}

where $H{\rm cts}$ and $S{\rm cts}$ are the net counts in the hard and
soft band, can be used as an indicator for the degree of obscuration
of an object. Soft X-ray spectra (with $HR<-0.2$) generally indicate
that objects are unobscured and that, in the case of AGN, one is
looking at a type 1 object. Harder spectra ($HR>-0.2$) indicate
obscured lines of sight as would be caused by a dusty circumnuclear
torus and can therefore be taken as indicators for type 2 objects. The
classification based on hardness ratio alone can be improved when the
total X-ray luminosity is also taken into account, as has been
demonstrated by \cite{Szokoly2004}. Sources with X-ray luminosities
below $10^{35}$\,W ($10^{42}\,{\rm erg\,s^{-1}}$) are treated as
non-AGN, sources with luminosities between $10^{35}$\,W and
$10^{37}$\,W are considered as AGN and sources with X-ray luminosities
in excess of $10^{37}$\,W are classified as QSO. The hardness ratio is
then used to further distinguish between type 1 and type 2 objects.

\begin{figure}
\includegraphics[width=0.95\linewidth]{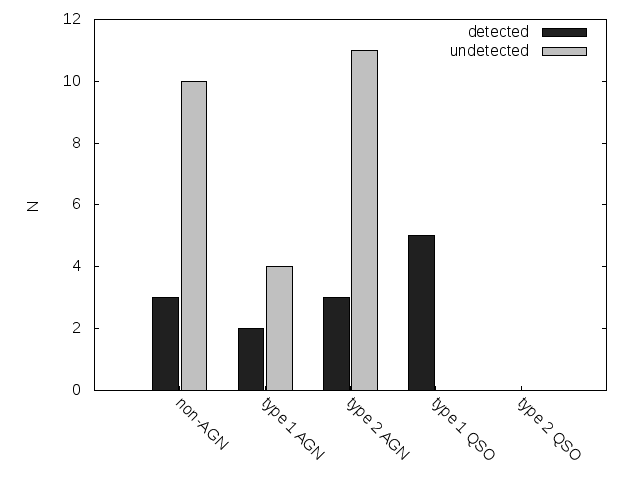}
\caption{X-ray classification statistics. Black bars represent sources
  with a VLBI detection and grey bars sources without. The lack of
  undetected type 1 QSOs is striking (see discussion in text).}
\label{fig:x-ray-stats}
\end{figure}

We have calculated the hardness ratio for all 13 VLBI-detected sources
with X-ray counterparts. Where both \cite{Luo2008} and
\cite{Lehmer2005} had made a detection we have adopted the
measurements of the former observation since it is more sensitive and
its errors are smaller. Also note that in the case of sources with
very few counts a rigorous error estimate essentially requires a
re-analysis of the raw data (\citealt{Park2006}), which is beyond the
scope of this paper. We therefore do not give an error for the
hardness ratios. For all 14 objects redshifts were available in the
literature, hence the total X-ray luminosity could be calculated and
the \cite{Szokoly2004} criteria were used for classification. The
scheme resulted in 3 non-AGNs (23\,\%), 2 and 3 type 1/2 AGNs (15\,\%
and 23\,\%), respectively, and 5 type 1 QSOs (39\,\%, no type 2
QSOs). Out of the VLBI-undetected sources, 25 had sufficient data
available for this classification, resulting in 10 non-AGN (40\,\%), 4
and 11 type 1/2 AGNs (16\,\% and 44\,\%), and no QSOs. See
Figure~\ref{fig:x-ray-stats} for an illustration of these
statistics. It is difficult to read a trend off this distribution,
however, the lack of undetected type 1 QSOs is striking. Whilst only
$\sim10$\,\% of optically selected type 1 QSOs are radio-loud, a much
larger fraction of X-ray selected type 1 QSOs appear to be radio-loud.

Part of the cause of this effect is that all 5 type 1 QSOs are
brighter than 1\,mJy, where the fraction of VLBI detections is
generally high (69\,\%).

\subsubsection{Potential causes of the X-ray non-detections}

Since X-ray observations have been claimed to be a very good tracer of
AGN activity (e.g., \citealt{Mushotzky2004}, \citealt{Brandt2005}),
the lack of X-ray counterparts to almost 1/3 of the VLBI-detected
sources deserves a closer look.

First, it must be noted that the VLBI-detected source without X-ray
counterpart in general have lower flux densities in other bands, too:
only one of these objects, S329, does have a measured optical
magnitude, and only one other, S519, has a measured $24\,\mu{\rm m}$
flux density. One object, S545, does have no optical or infrared flux
density at all.

\begin{figure}
\includegraphics[width=0.95\linewidth]{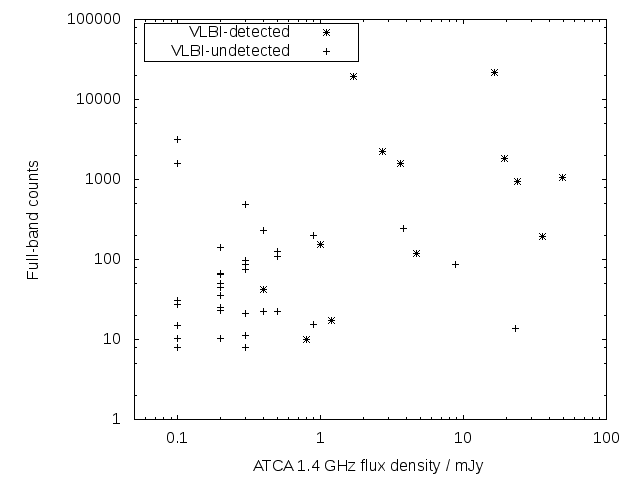}
\caption{Full-band X-ray counts as a function of 1.4\,GHz flux
  density. Neither the VLBI-detected (asterisks) nor the
  VLBI-undetected radio sources (pluses) show a correlation with X-ray
  counts.}
\label{fig:radio_vs_xray}
\end{figure}

Second, there appears to be no correlation between X-ray and radio
flux densities (Figure~\ref{fig:radio_vs_xray}). Hence some objects
will unavoidably have no X-ray counterpart, just as some sources with
X-ray detections have no VLBI detection.

Third, some sources could be Compton-thick and be so obscured that
they remained undetected in the X-ray observations. This happens at
column densities of $N_{\rm H}\gg1.5\times10^{24}\,{\rm cm^{-2}}$
(\citealt{Brandt2005}), which reduces the X-ray flux by two orders of
magnitude, even at high ($>10\,{\rm keV}$) energies.

\subsection{Notes on individual sources}
\label{sec:sources}

\begin{itemize}
\item{S329} This source appears to be mildly resolved in our VLBI
  image, and is undetected at 0.5\,keV -- 8\,keV by
  \cite{Lehmer2005}. Its redshift is 1.00 (\citealt{Mainieri2008}).

\item{S331} This source has been classified as AGN because of its more
  than tenfold radio excess over the radio-infrared relation. Its
  value of $q_{24}$ is -1.96 which clearly puts it into the AGN
  regime. Its X-ray properties (\citealt{Lehmer2005},
  \citealt{Szokoly2004}) suggest that it is a type 1 AGN.

\item{S380} This compact radio object was classified as AGN based on
  its $q_{24}$ value of $-0.21$ (\citealt{Norris2006a}). Its X-ray
  properties qualify it as a galaxy because its X-ray luminosity in
  the 0.5\,keV -- 8\,keV band is only $L_X=4.6\times10^{31}$\,W
  (\citealt{Lehmer2005}, \citealt{Szokoly2004}). It has a spectral
  index of $-0.6$ (\citealt{Kellermann2008}) and a redshift of 0.02
  (\citealt{Wolf2004}), making it the closest object in our sample.

\item{S393} This is the only VLBI-detected source with a pronounced
  extension in the VLBI observations. It is an X-ray source at a
  redshift of 1.07 (\citealt{Zheng2004}), and it has been identified
  as AGN by \cite{Norris2006a} because of its X-ray hardness ratio.
  The \cite{Szokoly2004} criteria result in a type 2 AGN
  classification, and its radio spectral index is -1.2
  (\citealt{Kellermann2008}). Only a very small fraction (5\,\%) of
  its arcsec-scale flux density is recovered in the VLBI
  observations. At its redshift, the angular separation of 32\,mas of
  the two components seen in our VLBI images translate to a projected
  distance of 263\,pc, and is therefore in the regime of GPS sources
  (and is of a size comparable to the narrow-line region).

\item{S404} The VLBI images of this source display almost all the flux
  density found in the ATCA observations. It has a compact morphology
  and its X-ray properties classify it as a type 1 QSO
  (\citealt{Luo2008}, \citealt{Szokoly2004}). It has a spectral index
  of 0.0 (\citealt{Kellermann2008}), indicating a compact,
  self-absorbed radio source, and its redshift is 0.54
  (\citealt{Norris2006a}). The HST images show an unresolved object
  with diffraction spikes, indicative of a point-like optical
  source. Its $q_{24}$, however, is 0.31 which is only a mild radio
  excess over the radio-infrared relation.

\item{S411} This is a compact source -- all arcsec-scale emission is
  recovered in the VLBI image. Its hardness ratio of -0.44 and X-ray
  luminosity qualify it as a type 1 QSO (\citealt{Luo2008},
  \citealt{Szokoly2004}) and it is located at a redshift of 1.61
  (\citealt{Afonso2006}). It has a spectral index of -0.5
  (\citealt{Kellermann2008}), and it is a featureless, round object in
  the HST ACS images (\citealt{Giavalisco2004}).

\item{S414} This is a compact source -- the VLBI observations recover
  95\,\% of the flux density measured with the ATCA
  (\citealt{Norris2006a}). It has $q_{24}=-0.29$ which makes this an
  AGN. S414 has a redshift of 1.57 (\citealt{Mainieri2008}) and is the
  brightest X-ray source in our sample, with $L_X=1.5\times10^{38}$\,W
  (0.5\,keV -- 8\,keV), qualifying it as a type 1 QSO
  (\citealt{Lehmer2005}, \citealt{Szokoly2004}).

\item{S421} The radio data for this source are consistent with a point
  source since it is unresolved in our VLBI image. It is not an X-ray
  source, and it has a spectral index of -0.6
  (\citealt{Kellermann2008}) and a redshift of 0.13
  (\citealt{Mainieri2008}).

\item{S423} This compact radio source is classified as a normal galaxy
  using the scheme by \cite{Szokoly2004}. It has a spectral index of
  -0.2 (\citealt{Kellermann2008}) and a redshift of 0.73
  (\citealt{Afonso2006}). Afonso et al. report spectroscopic evidence
  for star-forming activity in this galaxy, potentially merging with a
  nearby ($0.5''$) object. They identify five objects closer than
  $30''$ with similar redshifts. \cite{Mobasher2004} classify this
  object as an elliptical galaxy, using photometry in 17 bands. Hence
  there is no evidence except the VLBI detection that this object
  harbours an AGN. Given its redshift it has a 5\,GHz radio luminosity
  of $7.5\times10^{23}\,{\rm W\,Hz^{-1}}$, which is about an order of
  magnitude too low to classify it as a radio-loud object, and is more
  typical of Seyferts.

\item{S437} This source has been classified as AGN based on $q_{24}$
  and the literature (\citealt{Norris2006a}). Only 4\,\% of its ATCA
  flux density were recovered, and the VLBI images show an unresolved
  source. Yet it is a powerful X-ray source exceeding a luminosity of
  $10^{37}$\,W (0.5\,keV -- 8\,keV) and it has a hardness ratio of
  -0.48, indicating a type 1 QSO (\citealt{Luo2008},
  \citealt{Szokoly2004}). The redshift of this object is 0.73
  (\citealt{Afonso2006}). It is one of four objects covered by the
  Great Observatories Origins Deep Survey (GOODS,
  \citealt{Giavalisco2004}) with the {\it Hubble} Space Telescope. It
  is an unresolved object in all four bands, and diffraction spikes
  indicate a strong point-like component.

\item{S443} This compact object is the faintest VLBI-detected radio
  source in our survey, with an arcsec-scale flux density of only
  0.4\,mJy (\citealt{Norris2006a}). It is associated with a spiral
  galaxy, with a spectral index of -0.5 (\citealt{Kellermann2008}) and
  a redshift of 0.076 (\citealt{Afonso2006}).  Its $q_{24}$ value of
  1.05 is broadly consistent with the value of 0.84 found by
  \cite{Appleton2004} for the radio-infrared relation, and also its
  X-ray properties suggest that it is a galaxy without
  AGN. \cite{Afonso2006} report that the X-ray emission comes from
  north of the galaxy nucleus, with indications of star-forming
  activity. \cite{Mobasher2004} classify it as an Sbc object based on
  18-band photometry.
  
  We consider this a marginal detection, partly because the VLBI
  position places the VLBI source at the very edge of the visible
  perimeter of the galaxy, which can be treated as evidence against a
  detection (see Figure~\ref{fig:S443}). On the other hand, a radio
  transient cannot be ruled out (see \citealt{Lenc2008}). We also note
  that the radio luminosity of this source, $5\times
  10^{21}$\,W\,Hz$^{-1}$, is similar to one of the brightest radio
  SNe, SN1986J. We therefore consider it possible that S443 is a radio
  supernova.

\item{S447} This source has been classified as AGN based on $q_{24}$
  and also is an X-ray source. It has a spectroscopic redshift of 1.96
  (\citealt{Norris2006a}), where 1\,mas corresponds to 8.5\,pc. It is
  unresolved in our images, hence most of the VLBI flux density
  (20\,\% of the arcsec-scale flux density) comes from a region
  smaller than $\sim$100\,pc. It is the second most luminous X-ray
  source in our sample of 96 sources, with $L_X=1.3\times10^{38}$\,W
  (0.5\,keV -- 8\,keV, \citealt{Lehmer2005}). Nevertheless it has a
  soft spectrum with $HR=-0.37$ which qualifies it as a type 1 QSO
  (\citealt{Szokoly2004}).

\item{S472} This is a very compact source since 93\,\% of the ATCA
  flux density are seen in the VLBI image. Based on its $q_{24}=-0.31$
  this is an AGN. There is conflicting information about its redshift:
  \cite{Zheng2004} give a photometric redshift of 1.22, but
  \cite{LeFevre2004} measured a spectroscopic redshift of 0.55, which
  we deem more reliable. Its X-ray spectrum is hard between 0.5\,keV
  -- 8\,keV and indicates an absorbed type 2 AGN
  (\citealt{Lehmer2005}, \citealt{Szokoly2004}).

\item{S474} This is a compact source which displays all of its ATCA
  flux density in a low-resolution VLBI image. Its redshift is 0.98
  (\citealt{LeFevre2004}) and its spectral index is -1.4
  (\citealt{Kellermann2008}). It is a faint infrared source and is
  undetected in the X-ray. This is the only VLBI-detected radio source
  which remained undetected by the 2\,Ms Chandra exposure
  (\citealt{Luo2008}).

\item{S482} This is a compact radio source. It is also an X-ray source
  (\citealt{Lehmer2005}) with hardness ratio -0.33, but its luminosity
  is only $L_X=5.2\times10^{33}$\,W (0.5\,keV -- 8\,keV), which then
  qualifies it as a galaxy (\citealt{Szokoly2004}). It was classified
  by \cite{Norris2006a} as an AGN based on its $q_{24}$ of -0.61. It
  has a spectral index of -0.5 (\citealt{Kellermann2008}) and redshift
  of 0.15 (\citealt{Norris2006a}).

\item{S500} This compact radio object has an absorbed X-ray spectrum
  with hardness ratio 0.6 (\citealt{Lehmer2005}) , and it is
  classified as a type 2 AGN by the \cite{Szokoly2004} criteria. It
  has a spectral index of -0.9 (\citealt{Kellermann2008}) and a
  redshift of 0.73 (\citealt{Mainieri2008}). Based on $q_{24}=-0.21$,
  this is an AGN.

\item{S503} This is the brightest source in our VLBI observations,
  with a flux density of $(21.2\pm3.0)\,{\rm mJy}$ (compared to an
  ATCA flux density of 19.3\,mJy). Hence this source is very compact,
  but our uniformly-weighted image indicates that its core is resolved
  into two components with a separation of around 10\,mas (76\,pc at
  the distance of S503; $z=0.81$, \citealt{Mainieri2008}). It is an
  X-ray source indicating a type 1 AGN (\citealt{Lehmer2005},
  \citealt{Szokoly2004}).

\item{S517} This compact radio source is not a detected X-ray source,
  and it has a spectral index of -0.9 (\citealt{Kellermann2008}) and a
  redshift of 1.03 (\citealt{Mainieri2008}). The extensions towards
  the south-east seen in the VLBI images are imaging artifacts and do
  not indicate a true extension.

\item{S519} This compact radio source has previously been classified
  as AGN based on its $q_{24}=-0.42$. It is not an X-ray source, and
  it has a redshift of 0.69 (\citealt{Norris2006a}).

\item{S520} This source is mildly resolved. From its arcsec-scale flux
  density of 3.7\,mJy $(2.5\pm0.4)\,{\rm mJy}$ could be recovered in
  our lowest-resolution image. There is a hint of a $\sim$25\,mas
  extension towards the south-west in the naturally and
  uniformly-weighted image which at the redshift of the source of 1.4
  (\citealt{Wolf2004}) corresponds to around 200\,pc. It has only a
  very faint IR counterpart in the co-located {\it Spitzer} image, and
  is not detected in the ECDFS survey by \cite{Lehmer2005} (it is
  outside the GOODS region).

\item{S545} This compact radio source also does not exhibit X-ray
  emission, and neither a spectral nor a redshift are available.

\end{itemize}

\begin{figure}
\includegraphics[width=0.95\linewidth]{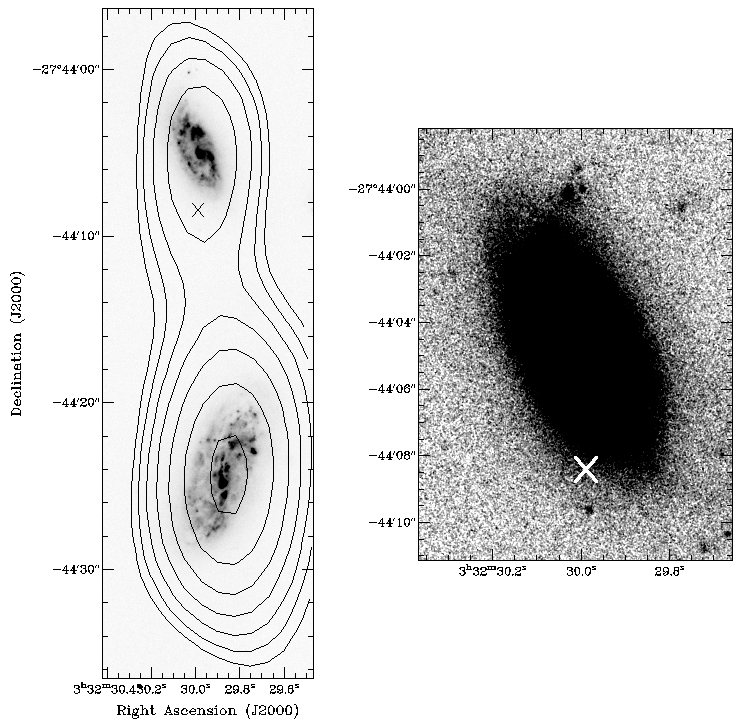}
\caption{{\it Left panel:} HST B-band image of S443 (at the top) and
  S442 (bottom). Black contours (starting at 0.1\,mJy and increasing
  by factors of $\sqrt{2}$) represent the ATCA 1.4\,GHz flux density
  and the cross indicates the S443 VLBI position. {\it Right panel:} An
  enlarged region of the HST image of S443 with enhanced contrast and
  a cross indicating the VLBI position, which is at the very edge of
  the galaxy, but not outside it.}
\label{fig:S443}
\end{figure}

\section{Conclusions}

Wide-field VLBI observations of 96 sources in the Chandra Deep Field
South have been carried out using a new correlation technique on a
software correlator. The correlation was carried out only once, with
high temporal and spectral resolution to avoid averaging effects, and
the visibilities were phase-rotated to the 96 target fields inside the
correlator, and then averaged. This approach is equivalent to a
multi-pass correlation, but requires only a small computational
overhead. It results in many small, normal-sized data sets, which can
be calibrated using standard techniques. The results of this project
are as follows.

\begin{itemize}
\item We have imaged all sources and have detected a fraction of
  21\,\%. The detection of one additional source is in the outer
  regions of the optical host galaxy, and could be a radio
  supernova. Most sources have flux densities of the same order as the
  arcsec-scale flux densities, and the radio-emitting regions
  therefore must be smaller than hundreds of pc (or less in some
  cases). Given their redshifts we can confidently interpret nearly
  all our VLBI detections as AGN.

\item A search of co-located, sensitive X-ray observations revealed
  that VLBI observations can identify AGN which have previously been
  missed, even though X-ray observations are expected to be a very
  good tracer of AGN activity. A total of 7 sources were classified as
  AGN for the first time.  Using X-ray data alone, only 10 out of the
  13 detected sources with available X-ray counterparts were
  identified as AGN, missing about one-quarter.

\item The VLBI core in the star-forming galaxy S443, which is one of
  the lowest redshift galaxies in our sample, has a radio luminosity
  of $5\times 10^{21}$\,W\,Hz$^{-1}$, consistent with that of a
  supernova. The VLBI core in the galaxy S423 at a redshift of 0.73 is
  surprising, because multiband photometry and X-ray data do not
  indicate the presence of an AGN.

\item Surprisingly, the VLBI detections include every X-ray detected
  type 1 QSO in our field, whereas only about 10\,\% of
  optically-selected type 1 QSOs are radio-loud. Therefore, either
  X-ray observations preferentially select radio-loud QSOs, or even
  radio-quiet QSOs have compact cores detectable with VLBI.
\end{itemize}

\section{Appendix A: source finding using images of separate polarisations}

In large images, the probability of finding a random noise peak
exceeding a given threshold can become significant. However, if the
data are split in half and searched for emission independently, and if
the positions are then compared between these two images, then the
probability of finding a random noise peak can be lower than in the
first case. We assume that the noise in the images has a Gaussian
distribution, and that sources are unpolarised -- these assumptions
are good approximations in our case.

Suppose that the rms of a Stokes I $=$(RCP+LCP)/2 image is
$\sigma$. Because the two orthogonal polarisations RCP and LCP are
processed by independent antenna electronics the noise in images using
only RCP or LCP is independent and larger than in Stokes I by a factor
of $\sqrt{2}$. Hence locations in the Stokes I image which have an SNR
of 5 have an SNR in the RCP and LCP image of $5/\sqrt{2}=3.54$.

The number of random noise peaks, $N'$, in an image above a threshold,
$z$, is

\begin{equation}
N'=N\times p(z)
\end{equation}

where

\begin{equation}
p(z)=1-\frac{1}{\sqrt{2\pi}}\int_{-\infty}^{z}{\rm exp}(-\frac{1}{2}t^2)dt
\end{equation}

and $N$ is the number of resolution elements in the image. For
example, for $z=5$, $p(z)$ evaluates to $2.87\times10^{-7}$, and in
images which have $8192^2$\,pix and a resolution element consists of
209\,pix, there are $N'=8192^2{\rm pix}/209\,{\rm
  pix}\times2.87\times10^{-7}=0.092$ resolution elements above a
$5\,\sigma$ cutoff.

To search for sources with the same absolute flux density in images
made from RCP and LCP only, one has to lower the search threshold to
$z/\sqrt{2}$, which results in a much greater number of random
noise peaks above the threshold. However, an additional constraint
can be imposed on the search now, because one requires that peaks in
the RCP and LCP images occur at the same location.

Randomly selecting one resolution element out of $N$ from RCP and LCP
each results in $1/N$ chance coincidences. Selecting one resolution
element in RCP and $N'$ from the LCP image (the number of resolution
elements above the $\sigma/\sqrt{2}$ threshold) therefore results in
$N'\times 1/N$ chance coincidences. Repeating that $N'$ times then
results in $N'^2/N$ chance coincidences. Hence the number of
random noise peaks occurring independently at the same location in RCP
and LCP when a detection threshold is set at $z=5\,\sigma$ in the
Stokes I image is

\begin{equation}
N''=N'^2/N=(N\,p(z/\sqrt{2}))^2/N=N\,p(z/\sqrt{2})^2
\end{equation}

Using the numbers from the example above, $p(z/\sqrt{2})$ evaluates to
$2.05\times 10^{-4}$, and $N''=8192^2{\rm pix}/209\,{\rm pix}\times
(2.05\times 10^{-4})^2=0.014$.

In our case the number of random noise peaks matching these criteria
is $0.092/0.014=6.6$ times lower than a simple $5\,\sigma$ cutoff.

\section{Acknowledgements}
\label{sec-6}

We thank Craig Walker for providing detailed beam width and squint
measurements, and for discussions on a primary beam attenuation
scheme. We also thank the anonymous referee who has helped to improve
this paper considerably. This research has made use of the NASA/IPAC
Extragalactic Database (NED) which is operated by the Jet Propulsion
Laboratory, California Institute of Technology, under contract with
the National Aeronautics and Space Administration.

\cleardoublepage
~\\

\begin{sidewaystable}[htbp]
\centering
\caption{Results of our observations. Source S443 has been included
  here but is shown separately from the other sources to indicate that
  its detection does not imply the presence of an AGN. Column (1): ID
  we use in this paper; column (2): the rms of the low-resolution
  image in mJy/beam; columns (3)-(4): the flux density extracted from
  the low-resolution image and its error in mJy; columns (5)-(6): the
  coordinates of the source in the low-resolution image; columns
  (7)-(8): the fraction of the ATCA flux density recovered in the
  low-resolution image and its error; columns (9)-(10): the
  classification given by \cite{Norris2006a} and based on the
  \cite{Szokoly2004} criteria; column (11): a 1 here indicates that
  this source had not been previously known to contain an AGN; column
  (12): the value of $q_{24}$; columns (13)-(15): the soft-band
  (0.5\,keV to 2\,keV) and hard-band (2\,keV to 8\,keV) X-ray counts
  listed by \cite{Luo2008} and \cite{Lehmer2005}, and the derived
  hardness ratio; columns (16)-(17): the spectral index $\alpha$
  listed by \cite{Kellermann2008}, multiplied with $-1$ to convert to
  our convention of $S\propto\nu^\alpha$, and its error; column
  (17)-(18): the redshift and its reference: M08 -
  \cite{Mainieri2008}, N06 - \cite{Norris2006a}, A06 -
  \cite{Afonso2006}, Z04 - \cite{Zheng2004}, W04 - \cite{Wolf2004},
  L04 - \cite{LeFevre2004}} {\tiny
\begin{tabular}{|l|r|r|r|r|r|r|r|l|c|l|r|r|r|r|r|r|r|c|}
\hline
\multicolumn{1}{|c|}{ID} & \multicolumn{1}{|c|}{rms} & \multicolumn{1}{|c|}{$S_{\rm VLBI}$} & \multicolumn{1}{|c|}{$\Delta S_{\rm VLBI}$} & \multicolumn{1}{|c|}{RA} & \multicolumn{1}{|c|}{Dec} & \multicolumn{1}{|c|}{$\frac{S_{\rm VLBI}}{S_{\rm ATCA}}$} & \multicolumn{1}{|c|}{$\Delta(\frac{S_{\rm VLBI}}{S_{\rm ATCA}})$} & \multicolumn{1}{|c|}{class} & \multicolumn{1}{|c|}{xclass} & \multicolumn{1}{|c|}{new} & \multicolumn{1}{|c|}{$q_{24}$} & \multicolumn{1}{|c|}{$S_{\rm cts}$} & \multicolumn{1}{|c|}{$H_{\rm cts}$} & \multicolumn{1}{|c|}{$HR$} & \multicolumn{1}{|c|}{$\alpha$} & \multicolumn{1}{|c|}{$\Delta\alpha$} & \multicolumn{1}{|c|}{$z$} & \multicolumn{1}{|c|}{ref}\\
\multicolumn{1}{|c|}{(1)} & \multicolumn{1}{|c|}{(2)} & \multicolumn{1}{|c|}{(3)} & \multicolumn{1}{|c|}{(4)} & \multicolumn{1}{|c|}{(5)} & \multicolumn{1}{|c|}{(6)} & \multicolumn{1}{|c|}{(7)} & \multicolumn{1}{|c|}{(8)} & \multicolumn{1}{|c|}{(9)} & \multicolumn{1}{|c|}{(10)} & \multicolumn{1}{|c|}{(11)} & \multicolumn{1}{|c|}{(12)} & \multicolumn{1}{|c|}{(13)} & \multicolumn{1}{|c|}{(14)} & \multicolumn{1}{|c|}{(15)} & \multicolumn{1}{|c|}{(16)} & \multicolumn{1}{|c|}{(17)} & \multicolumn{1}{|c|}{(18)} & \multicolumn{1}{|c|}{(19)}\\
\hline
S329 & 0.12 & 0.84 & 0.25 & 03:31:23.309 & -27:49:05.878 & 0.76 & 0.27 &  &  & \multicolumn{1}{r|}{1} & \multicolumn{1}{l|}{} & \multicolumn{1}{l|}{} & \multicolumn{1}{l|}{} & \multicolumn{1}{l|}{} & \multicolumn{1}{l|}{} & \multicolumn{1}{l|}{} & 1.00 & M08 \\ \hline
S331 & 0.12 & 1.69 & 0.32 & 03:31:24.912 & -27:52:08.038 & 0.05 & 0.01 & AGN & AGN1 &  & -1.96 & 144.21 & 51.78 & -0.47 & \multicolumn{1}{l|}{} & \multicolumn{1}{l|}{} & 1.26 & M08 \\ \hline
S380 & 0.11 & 1.00 & 0.24 & 03:31:52.127 & -27:39:26.545 & 1.25 & 0.39 & AGN & Gal &  & -0.21 & 7.89 & 8.64 & 0.05 & -0.6 & 0.10 & 0.02 & W04 \\ \hline
S393 & 0.07 & 2.47 & 0.42 & 03:32:01.548 & -27:46:48.302 & 0.05 & 0.01 & AGN & AGN2 &  & \multicolumn{1}{l|}{} & 253.27 & 821.3 & 0.53 & -1.2 & 0.20 & 1.07 & Z04 \\ \hline
S404 & 0.07 & 1.40 & 0.23 & 03:32:08.674 & -27:47:34.649 & 0.83 & 0.21 & AGN & QSO1 &  & 0.31 & 14202.81 & 5220.74 & -0.46 & 0.0 & 0.10 & 0.54 & N06 \\ \hline
S411 & 0.07 & 2.73 & 0.41 & 03:32:10.925 & -27:44:15.248 & 1.01 & 0.25 & AGN & QSO1 &  & \multicolumn{1}{l|}{} & 1592.43 & 625.48 & -0.44 & -0.5 & 0.10 & 1.62 & A06 \\ \hline
S414 & 0.11 & 3.41 & 0.52 & 03:32:11.654 & -27:37:26.282 & 0.95 & 0.24 & AGN & QSO1 &  & -0.29 & 1033.6 & 553 & -0.3 & -0.4 & 0.10 & 1.57 & M08 \\ \hline
S421 & 0.08 & 2.55 & 0.38 & 03:32:17.061 & -27:58:46.619 & 0.98 & 0.25 &  &  & \multicolumn{1}{r|}{1} & \multicolumn{1}{l|}{} & \multicolumn{1}{l|}{} & \multicolumn{1}{l|}{} & \multicolumn{1}{l|}{} & -0.6 & 0.10 & 0.13 & W04 \\ \hline
S423 & 0.06 & 0.53 & 0.13 & 03:32:18.024 & -27:47:18.751 & 1.32 & 0.42 & SF & Gal & \multicolumn{1}{r|}{1} & \multicolumn{1}{l|}{} & 40.92 & 13.17 & -0.51 & -0.2 & 0.10 & 0.73 & A06 \\ \hline
S437 & 0.08 & 0.67 & 0.17 & 03:32:27.014 & -27:41:05.366 & 0.04 & 0.01 & AGN & QSO1 &  & -1.19 & 15929.72 & 5664.26 & -0.48 & -0.8 & 0.20 & 0.74 & A06 \\ \hline
S447 & 0.09 & 4.85 & 0.71 & 03:32:32.010 & -28:03:10.058 & 0.20 & 0.05 & AGN & QSO1 &  & -1.7 & 651.26 & 300.11 & -0.37 & \multicolumn{1}{l|}{} & \multicolumn{1}{l|}{} & 1.96 & N06 \\ \hline
S472 & 0.08 & 4.36 & 0.63 & 03:32:49.200 & -27:40:50.788 & 0.93 & 0.23 & AGN & AGN2 &  & -0.31 & 29.39 & 83.54 & 0.48 & -0.8 & 0.10 & 0.55 & L04 \\ \hline
S474 & 0.08 & 3.77 & 0.55 & 03:32:49.438 & -27:42:35.467 & 1.18 & 0.29 &  &  & \multicolumn{1}{r|}{1} & \multicolumn{1}{l|}{} & \multicolumn{1}{l|}{} & \multicolumn{1}{l|}{} & \multicolumn{1}{l|}{} & -1.4 & 0.30 & 0.98 & L04 \\ \hline
S482 & 0.08 & 1.33 & 0.23 & 03:32:56.478 & -27:58:48.366 & 1.10 & 0.29 & AGN & Gal &  & -0.61 & 14 & 7.11 & -0.33 & -0.5 & 0.10 & 0.15 & N06 \\ \hline
S500 & 0.07 & 1.34 & 0.23 & 03:33:08.178 & -27:50:33.342 & 1.34 & 0.35 & AGN & AGN2 &  & -0.21 & 33.45 & 135.15 & 0.6 & -0.9 & 0.10 & 0.73 & M08 \\ \hline
S503 & 0.09 & 21.18 & 3.00 & 03:33:10.199 & -27:48:42.208 & 1.10 & 0.27 &  & AGN1 &  & \multicolumn{1}{l|}{} & 1238.67 & 617.45 & -0.33 & -0.9 & 0.20 & 1.03 & M08 \\ \hline
S517 & 0.08 & 1.31 & 0.23 & 03:33:16.351 & -27:47:24.988 & 0.55 & 0.15 &  &  & \multicolumn{1}{r|}{1} & \multicolumn{1}{l|}{} & \multicolumn{1}{l|}{} & \multicolumn{1}{l|}{} & \multicolumn{1}{l|}{} & -0.9 & 0.10 & 1.03 & M08 \\ \hline
S519 & 0.09 & 1.10 & 0.22 & 03:33:16.735 & -27:56:30.446 & 1.22 & 0.34 & AGN &  &  & -0.42 & \multicolumn{1}{l|}{} & \multicolumn{1}{l|}{} & \multicolumn{1}{l|}{} & \multicolumn{1}{l|}{} & \multicolumn{1}{l|}{} & 0.69 & N06 \\ \hline
S520 & 0.10 & 2.46 & 0.39 & 03:33:16.762 & -28:00:16.177 & 0.66 & 0.17 &  &  & \multicolumn{1}{r|}{1} & \multicolumn{1}{l|}{} & \multicolumn{1}{l|}{} & \multicolumn{1}{l|}{} & \multicolumn{1}{l|}{} & \multicolumn{1}{l|}{} & \multicolumn{1}{l|}{} & 1.40 & W04 \\ \hline
S545 & 0.10 & 1.29 & 0.26 & 03:33:33.434 & -27:53:32.777 & 1.43 & 0.40 &  &  & \multicolumn{1}{r|}{1} & \multicolumn{1}{l|}{} & \multicolumn{1}{l|}{} & \multicolumn{1}{l|}{} & \multicolumn{1}{l|}{} & \multicolumn{1}{l|}{} & \multicolumn{1}{l|}{} & \multicolumn{1}{l|}{} &  \\ \hline
\hline
S443 & 0.07 & 0.44 & 0.13 & 03:32:29.989 & -27:44:08.400 & 1.11 & 0.40 & SF & Gal & \multicolumn{1}{r|}{1} & 1.05 & 170.43 & 73.12 & -0.4 & -0.5 & 0.10 & 0.08 & A06 \\ \hline
\hline
\end{tabular}
}
\label{tab:results}
\end{sidewaystable}

\end{document}